\documentclass[twocolumn]{aastex62}

\newcommand\referee{}

\usepackage{comment}
\usepackage{multirow}
\definecolor{darkgreen}{rgb}{0,0.5,0}
\usepackage{makecell}
\usepackage{fourier} 
\usepackage{array}
\usepackage[graphicx]{realboxes}

\usepackage{threeparttablex, tablefootnote}
\usepackage{appendix}
\graphicspath{{./}{figures/}}

\usepackage[all]{nowidow}

\submitjournal{ApJ}

\shorttitle{VAST TDE Search}
\shortauthors{Dykaar et al.}
\begin{document}
\title{An Untargeted Search for Radio-Emitting Tidal Disruption Events in the VAST Pilot Survey}

\author[0009-0008-6396-0849]{Hannah~Dykaar}
\affiliation{Dunlap Institute for Astronomy and Astrophysics, University of Toronto, 50 St. George St., Toronto, ON M5S 3H4, Canada}
\affiliation{David A. Dunlap Department of Astronomy and Astrophysics, University of Toronto, 50 St. George St., Toronto, ON M5S 3H4, Canada}

\author[0000-0001-7081-0082]{Maria~R.~Drout}
\affiliation{David A. Dunlap Department of Astronomy and Astrophysics, University of Toronto, 50 St. George St., Toronto, ON M5S 3H4, Canada}

\author[0000-0002-3382-9558]{B.~M.~Gaensler}
\affiliation{Dunlap Institute for Astronomy and Astrophysics, University of Toronto, 50 St. George St., Toronto, ON M5S 3H4, Canada}
\affiliation{David A. Dunlap Department of Astronomy and Astrophysics, University of Toronto, 50 St. George St., Toronto, ON M5S 3H4, Canada}
\affiliation{Department of Astronomy and Astrophysics, University of California Santa Cruz, 1156 High Street, Santa Cruz, CA 95064, USA}

\author[0000-0001-6295-2881]{David~L.~Kaplan}
\affiliation{Department of Physics, University of Wisconsin-Milwaukee, P.O. Box 413, Milwaukee, WI 53201, USA}

\author[0000-0002-2686-438X]{Tara~Murphy}
\affiliation{Sydney Institute for Astronomy, School of Physics, University of Sydney, Sydney, New South Wales 2006, Australia}

\author[0000-0002-5936-1156]{Assaf~Horesh}
\affiliation{Racah Institute of Physics. The Hebrew University of Jerusalem. Jerusalem 91904, Israel}

\author[0000-0001-6295-2881]{Akash~Anumarlapudi}
\affiliation{Department of Physics, University of Wisconsin-Milwaukee, P.O. Box 413, Milwaukee, WI 53201, USA}

\author[0000-0003-0699-7019]{Dougal~Dobie}
\affiliation{Centre for Astrophysics and Supercomputing, Swinburne University of Technology, Hawthorn, Victoria, Australia}
\affiliation{ARC Centre of Excellence for Gravitational Wave Discovery (OzGrav), Hawthorn, Victoria, Australia}

\author[0000-0002-4405-3273]{Laura~N.~Driessen}
\affiliation{Sydney Institute for Astronomy, School of Physics, University of Sydney, Sydney, New South Wales 2006, Australia}

\author[0000-0002-9994-1593]{Emil~Lenc}
\affiliation{ CSIRO Space and Astronomy, PO Box 76, Epping, NSW, 1710, Australia}

\author[0000-0001-8026-5903]{Adam~J.~Stewart}
\affiliation{Sydney Institute for Astronomy, School of Physics, University of Sydney, Sydney, New South Wales 2006, Australia}

\begin{abstract}

We present a systematic search for tidal disruption events (TDEs) using radio data from the Variables and Slow Transients (VAST) Pilot Survey conducted using the Australian Square Kilometre Array Pathfinder (ASKAP). Historically, TDEs have been identified using observations at X-ray, optical, and ultraviolet wavelengths. After discovery, a few dozen TDEs have been shown to have radio counterparts through follow-up observations. With systematic time-domain radio surveys becoming available, we can now identify new TDEs in the radio regime. A population of radio-discovered TDEs has the potential to provide several key insights including an independent constraint on their volumetric rate. We conducted a search to select variable radio sources with a single prominent radio flare and a position consistent within 2\,$\sigma$ of the nucleus of a known galaxy. While TDEs were the primary target of our search, sources identified in this search may also be consistent with active galactic nuclei exhibiting unusual flux density changes at the timescales probed, uncharacteristically bright supernovae, or a population of gamma-ray bursts. We identify a sample of 12 radio-bright candidate TDEs. The timescales and luminosities range from $\sim$6 to 230 days and $\sim$10$^{38}$ to 10$^{41}$\,erg\,s$^{-1}$, consistent with models of radio emission from TDEs that launch relativistic jets. After calculating the detection efficiency of our search using a Monte Carlo simulation of TDEs, and assuming all 12 sources are jetted TDEs, we derive a volumetric rate for jetted TDEs of 0.80$^{+0.31}_{-0.23}$\,Gpc$^{-3}$\,yr$^{-1}$, consistent with previous empirically estimated rates.
\end{abstract}

\keywords{Tidal disruption;
Time domain astronomy; 
Galaxy nuclei;
Supermassive black holes; 
Sky Surveys}

\section{Introduction} \label{sec:intro}
A tidal disruption event (TDE) occurs when a star gets sufficiently close to a supermassive black hole (SMBH) that the tidal forces overcome the star’s self gravity, breaking it apart \citep{hills_75}. The subsequent transient accretion can result in an electromagnetic flare \citep{Rees}. The identification of this electromagnetic radiation from TDEs is useful for multiple reasons. For example, TDEs are capable of probing quiescent SMBHs that would otherwise be invisible to detection \citep{Duran}. They can also be used to help understand the galactic nuclei they reside in, including their stellar dynamics, circumnuclear material and accretion history \citep{TDE_review}.

Historically, TDEs have been discovered using observations at soft X-ray \citep[e.g.,][]{soft_xray_TDE}, optical \citep[e.g.,][]{SDSS_TDEs}, and ultraviolet (UV) wavelengths \citep[e.g.,][]{UV_TDE}. The soft X-ray emission is thought to be produced primarily by a hot accretion disk that forms after the stellar debris from the disruption circularizes \citep{Rees, komossa_1999}. Proposed emission mechanisms for the optical and UV emission include shocks from the stellar stream collisions that convert the kinetic energy of the streams into thermal energy \citep{UV_TDE, stream_collisions, formation_vs_accretion} and reprocessing of X-ray emission from the accretion disk \citep{Wevers_2019, optical_TDEs}. \cite{Dai_2018} presented a model that unifies both soft X-ray and UV/optical observations, where the optical depth of scattered electrons depends on the viewing angle due to an optically thick wind from a super-Eddington accretion disk. A class of jetted TDEs was later discovered using hard X-ray observations, along with infrared and radio followup. The observed relativistic, non-thermal radiation was shown to be the result of a relativistic jet launched by a TDE \citep{1644_bloom,cenko_2012,brown_2015}.

Approximately 30 TDEs discovered over the past decade have been detected in follow up radio observations \citep[e.g.,][]{1644_z,arp299,goodwin_2023} and shown to have radio counterparts that are well described by synchrotron emission produced by outflowing material \citep[e.g.,][]{1644_z}. Notably, this radio emission can persist for months to years \citep{J2058, Ravi_2021}. In some cases, there is a delayed radio flare appearing months to years after discovery, either as the only detectable radio emission or as rebrightening after an initial flare \citep{TDE_repeater, delayed_flares, repeaters}.

Of the TDEs with observed radio emission, some show emission from a relativistic outflow shown to be the result of a jet \citep[e.g.,][]{J2058, 1644_z}, while some show non-relativistic ejecta \citep[e.g.,][]{14li_a, at2019dsg_neutrino}. The nature of this non-relativistic radio emission is still debated, with possible explanations including a sufficiently decelerated jet interacting with the surrounding medium \citep{velzen}, shocks inside of a relativistic jet \citep{internal_jet}, a wind produced during a period of super-Eddington accretion \citep{14li_a}, \referee{an outflow induced by the self-intersection of the fallback stream \citep{Lu_stream_collisions}}, or emission from the unbound debris of the leftover star \citep{K16}.

Our understanding of the emission mechanisms that govern radio emission from TDEs is still not complete. More observations of radio-bright TDEs are required to understand the emission mechanisms of these sources. In general, understanding the jets and outflows that emerge from TDEs offers insight into the accretion processes of SMBHs and can place constraints on the fraction of TDEs that launch jets. In all cases, radio observations are uniquely capable of probing the density of surrounding material, as well as the size and velocity of the outflow \citep{radio_tde_review}. With systematic time-domain radio surveys now becoming available, we have an unprecedented opportunity to discover TDEs in this regime. 

While a few dozen TDEs have been targeted with radio follow up observations, TDEs have only recently being discovered independent of other wavelength detections in the radio regime \referee{\citep{Anderson_2019,Ravi_2021,somalwar_pop}}. A \emph{population} of radio-discovered TDEs has the potential to provide several key insights, particularly an independent constraint on the volumetric rates of TDEs. There is a discrepancy between theoretical rates of TDEs and those inferred from observations. X-ray, optical, and UV TDEs imply a rate of $\sim$10$^{2}$\,Gpc$^{-3}$\,yr$^{-1}$ \citep[e.g.,][]{TDE_demographics}.
The theoretical rates based on two-body relaxation are significantly higher, with the most conservative estimates at $\sim$3\,$\times$\,10$^{3}$\,Gpc$^{-3}$\,yr$^{-1}$ \citep{stone_metzger}. Many other TDEs could be occurring in galaxies with high levels of extinction, but are presumably being missed by optical surveys, demonstrating a possible advantage of performing a search in the radio regime.

A radio population of TDEs would also provide a unique perspective on TDE host galaxy types. The current population of TDEs, largely discovered at optical, X-ray, and UV wavelengths, shows an overabundance of TDEs occurring in E+A, or post-starburst galaxies \citep{French2020}. 
While the reason for this overabundance is debated (with options including disturbed stellar orbits or a binary SMBH from a previous merger) \citep{French2020}, it is notable that the two TDEs discovered in the radio regime so far, independent of observations at other wavelengths \citep{Anderson_2019, Ravi_2021}, did \emph{not} occur in post-starburst galaxies. A larger population of radio-discovered TDEs could tell us whether this over-representation is a physical effect or is due to an observational bias. 

Notably, \cite{somalwar_pop} presented a population of six radio-selected TDEs using the Very Large Array (VLA) Sky Survey \citep[VLASS;][]{VLASS} with transient optical counterparts from the Zwicky Transient Facility \citep[ZTF;][]{ztf}. They first identified a population of nuclear radio flares in nearby galaxies that show no signs of an active galactic nuclei (AGN), and then cross-matched this population with catalogues of optically-discovered TDEs from ZTF \citep{TDE_demographics, hammerstein_ZTF}. Their population of radio-discovered, optically-identified TDE hosts occurred in E+A galaxies at the same rate as the optically-discovered TDE hosts as a whole. While this may indicate that the overabundance of TDEs in post-starburst galaxies is indeed a physical effect, further studies are warranted.

In addition, some TDEs should occur in galaxies with SMBHs that show evidence of active accretion \citep{obs_review}. Due to extinction, optical observations may miss a TDE associated with an AGN with an optically thick torus, demonstrating a possible advantage of conducting a search in the radio band. However, while the radio emission wouldn't suffer from extinction, the intrinsic radio variability of AGN makes \emph{classifying} a radio flare from a TDE difficult, demonstrating the need for studies of the viability of this approach. 

The Australian Square Kilometre Array Pathfinder \citep[ASKAP;][]{ASKAP} survey for Variables and Slow Transients \citep[VAST;][]{VAST} is an untargeted radio time-domain survey. VAST is designed to be sensitive to slowly evolving ($\sim$days to years) extragalactic transients and variable sources --- ranging from AGN and radio supernovae to gamma-ray burst (GRB) afterglows and TDEs. The full VAST survey commenced in December 2022 and has been allocated over 2,100 hours of observing time over 5 years. It consists of 329 fields covering $\sim$ 8\,000 square degrees of the southern sky. In addition, in preparation for the full survey, a pilot version of the VAST survey was conducted between 2019 and 2021. This pilot survey observed a smaller portion of the sky, 5\,131 square degrees, approximately a dozen times over a two year period \citep{vast_pilot} for a total of $\sim$162 hours of observing time. With their large area and long time baseline, both VAST and its Pilot Survey provide an unprecedented opportunity to discover radio transients. The VAST Pilot has already been used to further our understanding of classical novae \citep{vast_novae} and GRB radio afterglows \citep{vast_pilot_grbs}, among other sources.

In this paper, we present an untargeted search for TDEs using the VAST Pilot Survey. Our methods are distinct and complementary to those of \cite{RACS_TDEs}, who use the Rapid ASKAP Continuum Survey \citep[RACS;][]{RACS} and VAST to perform a targeted search for radio emission at the location of known TDEs. In Section~\ref{sec:appearance}, we discuss how we expect the emission of TDEs to appear in the VAST Pilot Survey. In Section~\ref{sec:candidate_selection}, we outline the criteria used to select the sample of TDE candidates and In Section~\ref{sec:properties}, we present the results of the search and describes the properties of the sources in this sample. In Section~\ref{sec:rates}, we use this search to place constraints on the volumetric rate of TDEs. Finally, in Section~\ref{sec:discussion}, we discuss our results and the nature of the sources in our sample. 
Throughout this work, we adopt the following cosmological parameters: $H_{0}=67.7$\,km\,s$^{-1}$\,Mpc$^{-1}$, $\Omega_{M}=0.310$, and $\Omega_{\Lambda}=0.689$ \citep{Planck18}.

\section{Expected Appearance of TDEs in the VAST Pilot Survey}\label{sec:appearance}
In order to select TDEs from the VAST Pilot data, an understanding of their expected observational properties is necessary. However, only a handful of observed TDEs have published radio counterparts \citep[e.g.,][and references therein]{radio_tde_review} \referee{and only one has observations as low as the VAST Pilot Survey frequency of 888 MHz \citep{goodwin_2023}}. We therefore create a large set of mock TDE lightcurves projected onto the frequency, cadence, and sensitivity of the VAST Pilot data using theoretical models. We use these lightcurves to investigate (i) the expected appearance of TDEs in the VAST data and (ii) the types of TDEs that the VAST Pilot will be sensitive to. In subsequent sections, we will use these mock light curves to help define a set of selection criteria for TDE candidates in the VAST Pilot data (Section~\ref{sec:candidate_selection}) and to estimate the volumetric rate of radio TDEs based on the VAST Pilot detection efficiency (Section~\ref{sec:rates}).

\subsection{Key Parameters of the VAST Pilot}\label{sec:param_vast}
Key properties of the VAST Pilot Survey are necessary to create our set of mock TDE radio lightcurves. The VAST Pilot covered 5,131 square degrees in six distinct regions of the sky. There were 17 epochs obtained at a central frequency of 888 MHz and three additional epochs centered at 1\,296 MHz. The bandwidth of the observations is 288 MHz \citep{vast_pilot}. For our study, we exclude the region covering the Galactic plane, focusing only on observations where we would be able to identify an optical host galaxy, as well as the three epochs at 1\,296 MHz. Each VAST observation consists of 12 minutes of integration, resulting in a typical image RMS of 0.24 mJy beam$^{-1}$ at an angular resolution of 12 to 20$\arcsec$. 

\begin{figure}[t]
  \includegraphics[width=\linewidth]{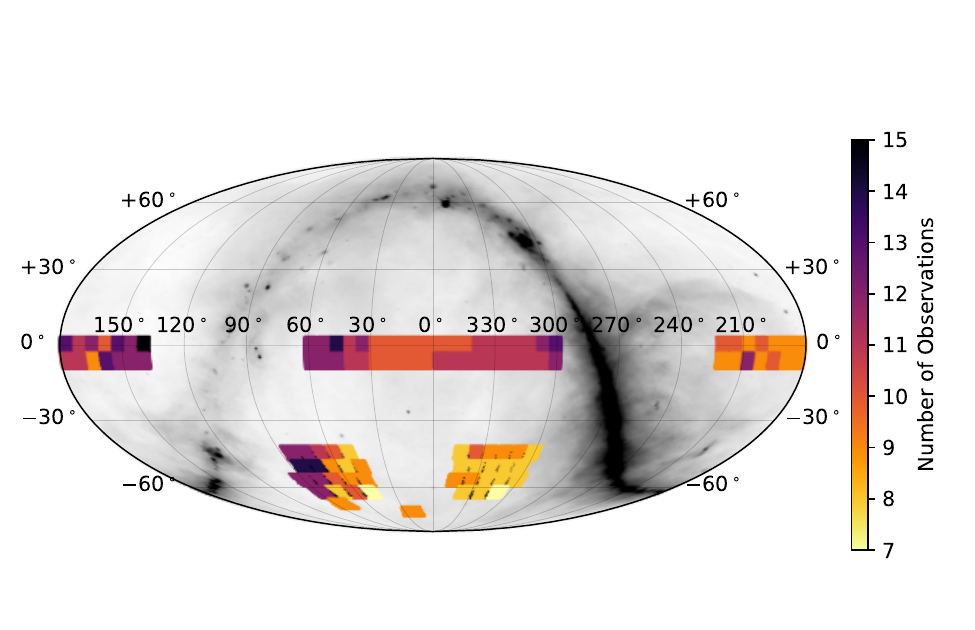}
  \caption{Sky coverage of the VAST Pilot survey observations used in our analysis, this includes all observations that were obtained at a central frequency of 888 MHz and excludes fields observed in the galactic plane.}
  \label{fig:coverage} 
\end{figure}

We also include two epochs from the low-band of RACS that were observed at the same central frequency. The RACS fields were observed for $\sim$15 minutes for a typical image RMS of 0.25 mJy beam$^{-1}$ at an angular resolution of $\sim$15$\arcsec$. In total we consider 19 epochs observed between August 2019 and November 2021, with various portions of the sky observed between 3 and 15 times. There were $\sim$$10^{7}$ individual images observed with the cadence ranging from $\sim$days to $\sim$months. Sky coverage including number of observations per location, is shown in Figure~\ref{fig:coverage}.

\subsection{TDE Models} \label{subsec:models}
We require an approximate theoretical description of expectations for TDE emission at the VAST Pilot frequency in order to broadly understand how their expected luminosities and timescales map onto the VAST Pilot cadence and sensitivity. We consider two models: one for relativistic/jetted radio outflows and one for non-relativistic/quasi-spherical outflows. In both cases, the radio emission is assumed to come from synchrotron emission produced at the interface between expanding material and the ambient medium. 

\emph{Relativistic TDEs:} To approximate the relativistic emission of jetted TDEs, viewed both on- and off-axis, we use the python module \texttt{afterglowpy}\footnote{https://github.com/geoffryan/afterglowpy} \citep{afterglowpy}. This module uses a set of semi-analytic models to numerically compute light curves for \emph{structured} relativistic jet afterglows expanding into constant density media. It has been effectively used to model the afterglows of both short and long-duration gamma-ray bursts \citep[e.g.,][]{afterglowpy_ex} and is able to largely reproduce the results of the \texttt{Boxfit} code (which computes light curves based on the numerical simulations of \citealt{boxfit}) for jets with a ``top-hat'' angular structure. 

\begin{figure*}[t]
\centering
  \includegraphics[width=0.9\linewidth]{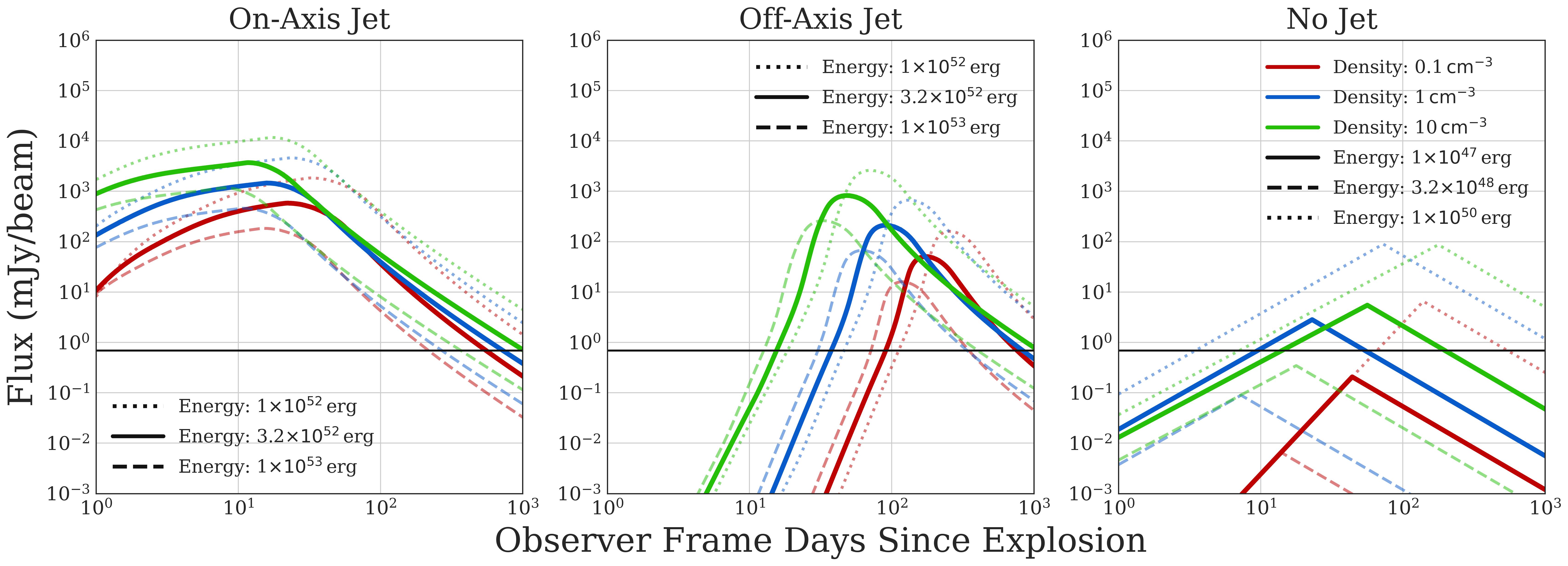}
  \caption{The three types of TDEs modeled in our simulation. From left to right: a relativistic jet viewed on-axis, a relativistic jet viewed off-axis, and no relativistic jet. Each model has nine example simulations plotted. For each model three different simulated densities are shown: 0.1\,cm$^{-3}$ in red, 1\,cm$^{-3}$ in blue, and 10\,cm$^{-3}$in green. For each plotted density, three example outflow energies are shown spanning an order of magnitude. For the non-jetted case these are: $10^{47}$\,erg, as a dashed line, 3$\times$$10^{47}$\,erg as a solid line, and $10^{48}$\,erg as a dotted line. For both the off- and on-axis jetted models these are: $10^{52}$\,erg, as a dashed line, 3$\times$$10^{52}$\,erg as a solid line, and $10^{53}$\,erg as a dotted line. Also shown as a horizontal black line is the 3$\sigma$ flux density limit of VAST. All three models are simulated at a redshift of 0.01.} \label{fig:3_models_lc}
\end{figure*}

The main free parameters of \texttt{afterglowpy} are (i) the structure and half opening angle of the jet, (ii) the fraction of energy in relativistic electrons, $\epsilon_e$, and in magnetic fields, $\epsilon_b$, (iii) the power-law distribution of relativistic electrons, $p$, (iv) the isotropic equivalent energy of the explosion, (v) the density of the circumnuclear density, and (vi) the viewing angle to the observer. For our baseline models, we assume a Gaussian jet with an opening angle of 0.1 radians (5.7 degrees),  and $p$ $=$ 2.5 \citep[e.g.,][]{1644_e}. \referee{We chose $\epsilon_e$ $=$ 0.2 and $\epsilon_b$ $=$ 0.01, which were chosen by \cite{Metzger_2015} to apply the model from \cite{Nakar_Piran_arxiv} to off-axis jetted TDEs. These values are similar to what has been modeled in observed TDEs \citep[e.g.,][]{at2019dsg}. We discuss in Section~\ref{sec:timescale} how the choice of these values affect our models and the interpretation of TDEs in the VAST Pilot.} \referee{The Lorentz factor of the jet has a profile with angular structure, that depends on the input energy, see Equation A1 in \citep{afterglowpy}}. We then compute models for a range of input energies, densities, and viewing angles in order to examine a variety of possible behavior for relativistic radio TDEs. 

Specifically, we consider energies between $10^{52}$ and $10^{54}$\,erg and densities between $10^{-2}$ and $10^{4}$ cm$^{-3}$. These are chosen to span the range of isotropic equivalent energies found for the well-studied relativistic TDEs Sw 1644+57 and AT 2022cmc \citep{1644_b,1644_c,AT2022_xray} and the densities surrounding black holes as calculated from other observed TDEs and SMBHs at various radii \cite[see e.g. Figure 6 from][]{1644_c}, respectively. Finally, we adopt fiducial viewing angles of 10 degrees for on-axis jets and 40 degrees for off-axis jets. Example light curves demonstrating how luminosity and timescale vary with these input parameters are shown in the first two panels of Figure~\ref{fig:3_models_lc}.

\emph{Non-relativistic TDEs:} To approximate the non-relativistic radio emission from TDEs that do not launch jets, we use the model of \cite{Nakar_Piran_arxiv}\footnote{The specific framework that we follow only appears in the Arxiv version and not in the published paper \citep{Nakar_Piran}.}. This provides an analytic framework to calculate the radio synchrotron emission associated with a sub-relativistic spherical outflow interacting with a constant density medium. 

Similar to the relativistic case, key free parameters of this model are the density of the ambient medium, the energy of the outflow, $p$, $\epsilon_e$, and $\epsilon_b$. \referee{We approximate the initial Lorentz factor of the outflow as $\Gamma$$\approx$2, ignoring relativistic effects, as in \cite{Nakar_Piran_arxiv}. The initial outflow velocity may be considerably lower than this \citep[e.g.][]{14li_a}, we discuss how the choice of $\Gamma$ affects the models in Section~\ref{sec:timescale}}. As above, we fix $p=2.5$, \referee{$\epsilon_e$ $=$ 0.2 and $\epsilon_b$ $=$ 0.01}, and compute models for a range of circumnuclear densities and outflow energies. We adopt the same range of densities as above, but consider lower outflow energies ranging between $10^{46}$ and $10^{50}$\,erg. This latter range is based on the estimated energy at early and late times of TDEs with sub-relativistic outflows including ASASSN$-$14li \citep{14li_a} and AT2018hyz which has been interpreted as a non-relativistic outflow \citep{AT2018hyz} and as an off-axis jet \citep{hyz_sfaradi}. With these ranges of parameters, our model non-jetted TDEs have a maximum luminosity of \mbox{$\nu {\rm L}_{\nu}$$\approx$2$\times$10$^{41}$erg\,s$^{-1}$} at 888 MHz.

Finally, using equations 7, 8, and 12 as well as Table 1 of \cite{Nakar_Piran_arxiv}, for a given set of input parameters we calculate (i) the time when the lightcurve will peak at the VAST observing frequency and (ii) the power-law index of the radio lightcurve both before and after the peak. Example light curves for a range energies and densities are shown in the right panel of Figure~\ref{fig:3_models_lc}. 

\emph{Caveats:} We emphasize that the models described above should only be taken as approximate descriptions for the evolution of specific radio TDEs. In particular, both models simulate the evolution of a blast wave into a constant density medium, while the density of the environment surrounding real SMBHs tends to decrease approximately logarithmically with radius (cf.\ Figure 6 from \citealt{1644_c}). \referee{In addition to not being constant, this environment may also not be homogeneous. \cite{Goodwin_2022} proposed that the synchrotron energy index fluctuations seen in the radio-bright TDE AT2019azh, may be due to an inhomogeneous circumnuclear medium.} Furthermore, in recent work, \cite{new_jetted_model} conclude that the viewing angle of a jetted outflow is degenerate with its Lorentz factor. Because of this degeneracy, non-jetted and certain off-axis jets may be indistinguishable\footnote{For example, \cite{new_jetted_model} argue that the TDE AT2019dsg, originally classified as a non-relativistic TDE \citep{at2019dsg}, may actually be a relativistic jet viewed off-axis.}. These factors are important to consider when attempting to derive energies and densities of any specific observed TDE. However, despite these limitations, the models described above are able to reproduce the broad luminosities and timescales observed for jetted and non-jetted TDEs, which are what we require to design a search within (Section~\ref{sec:candidate_selection}), and measure the detection efficiency (Section~\ref{sec:rates}), of the VAST Pilot for TDEs.

\subsection{A Simulated Set of TDEs Observed by VAST}\label{sim_pop}
We use a two-step process to create a set of mock TDEs observations. First, we create a large grid of model lightcurves at a range of redshifts, ambient densities, and outflow energies using the theoretical frameworks outlined in Section~\ref{subsec:models}. We then perform a Monte Carlo simulation to perform mock VAST observations of these models.

For the model grid, we consider three types of radio TDEs: jetted on-axis, jetted off-axis, and non-jetted. For each type of TDE, we create models in 40 redshift bins spaced evenly by $\log_{10} z$ between z=0.05 and z=2. Within each redshift bin, we then create 100 models for each TDE type, sampling 10 energies and 10 ambient densities in the ranges described in Section~\ref{subsec:models} evenly in $\log_{10} n$ and $\log_{10} E$. In all cases, we adjust the frequency sampled such that it corresponds to an \emph{observed} frequency of 888 MHz. We also adjust the observed timescale of the flare according to the simulated redshift.

We then run a Monte Carlo simulation, generating 6000 mock VAST TDE light curves within each redshift bin. For each iteration, we randomly select: (i) a model from within the large grid described above, (ii) an explosion date from within the 1815 days that encompass the duration of the VAST Pilot (815 days) and the 1000 days prior to the first VAST Pilot observation, and (iii) a specific field within the VAST Pilot footprint where the event is located. We allow explosion epochs prior to the commencement of the VAST Pilot, as such objects may be long-lived and detectable as purely fading transients. We specify a location in the sky where the event occurred because the VAST Pilot did not observe each field an equal number of times (Figure~\ref{fig:coverage}). We then project the chosen simulated lightcurve onto the actual observing cadence of the VAST Pilot for the specified field. For each observed epoch, we resample the model flux based on the typical flux density errors of the VAST Pilot.

\subsection{Basic Properties of the Simulated VAST TDEs} \label{sim_pop_properties}
\begin{figure}[t]
  \includegraphics[width=\linewidth]{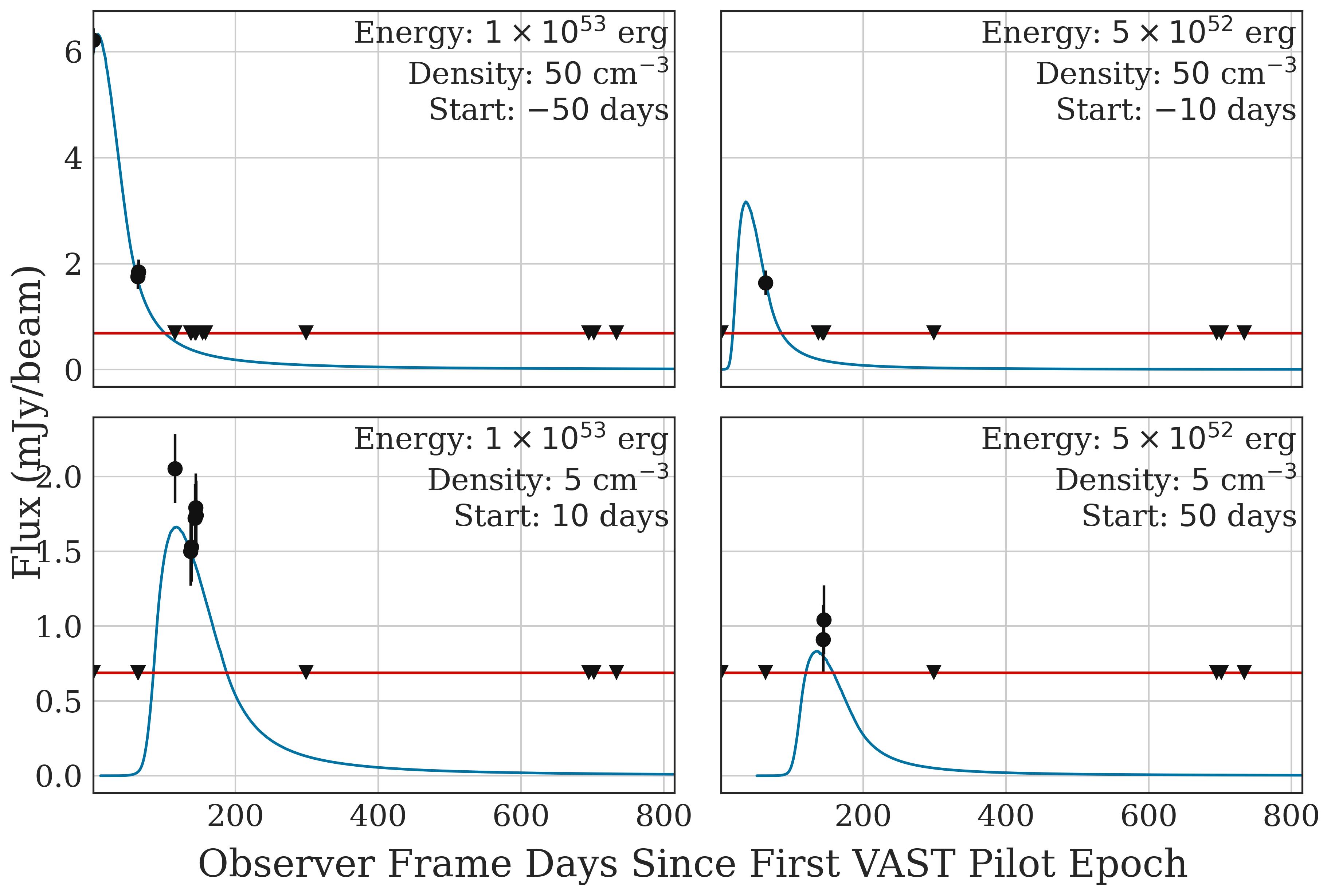}
  \caption{Example lightcurves from the TDE simulated population described in section~\ref{sim_pop}. Each source shown is an off-axis jetted TDE at a redshift of 0.5. The isotropic equivalent energies, circumnuclear densities, and start dates of the disruption relative to the first VAST Pilot epoch are varied to show some of the diversity in flares we expect to see. The red line indicates the sensitivity of the VAST Pilot, the blue curve shows the predicted emission from the TDE, and the black points and triangles show the simulated detections and upper limits in the VAST Pilot respectively. Errors on fluxes in the top two panels are obscured by markers.} \label{fig:model_lc}
\end{figure}

In total we created 2.4$\times10^{5}$ mock VAST Pilot TDE observations spanning 0.05 $<$ z $<$ 2, and the observed appearances are diverse. In Figure~\ref{fig:model_lc} we present four example lightcurves (all off-axis jetted TDEs simulated at z$=$0.5), demonstrating some of this diversity and the typical quality expected. In all plots, the red line marks 0.72 mJy/beam. This is 3 times the typical RMS of the VAST Pilot, and is taken as the detection threshold within our simulations.
All four lightcurves have between one and five observations in which the model flux is above the VAST threshold; one explodes before the beginning of the VAST Pilot such that we only detect the fading emission from the source. 

We now discuss the implications of this simulation for the types of TDEs that the VAST Pilot is sensitive to. In Section~\ref{sec:rate_calculations} we will use these simulated lightcurves to determine our detection efficiency for different classes of TDEs as a function of redshift and implications for their rates. However, if we now adopt a simplified assumption that any lightcurve with a minimum of three detections above the VAST detection threshold is observable, several broad themes are already clear. We find that jetted TDEs in our simulation are observable under this criteria out to z=2, whereas non-jetted TDEs are only observable out to z=0.06. This agrees well with observations, as a jetted source with \mbox{$\nu L_{\nu}\sim3\times10^{41}$\,erg\,s$^{-1}$}, the approximate peak luminosity of Swift J1644+57, would be detectable by VAST in at least \emph{one} epoch out to z=2.4 given VAST's 3$\sigma$ flux density limit of 0.69\,mJy. Similarly, a non-jetted source with \mbox{$\nu L_{\nu}\sim9\times10^{37}$\,erg\,s$^{-1}$}, the approximate peak luminosity of ASASSN$-$14li, would be detectable out to z=0.07.

This simulation provides broad expectations for the timescales and flux densities that we can expect for different types of TDEs at various distances. For example, at a redshift of 0.5, the median flux density of an observable simulated jetted TDE is 5.1\,mJy. The fractional flux change, defined as ($\frac{S_{max}}{S_{min}}$), where $S_{max}$ and $S_{min}$ are the maximum and minimum flux densities of a given lightcurve respectively, has a median value of 4.8. The inferred time in the observer frame that the jetted flares are above half of their peak flux density has a median value of 56 days (see Section~\ref{sec:calculate_param} for further details of how this is calculated). At a redshift of 0.02, the maximum flux density of non-jetted TDEs has a median value of 3.1\,mJy and a median fractional flux change of 2.6. The median time in the observer frame above half of their peak is 90 days.

\section{Candidate Selection}\label{sec:candidate_selection}
We choose a set of criteria to select a sample of transient/variable sources detected in the VAST Pilot that are coincident with galaxy nuclei and broadly consistent with what we expect from TDEs. 

\subsection{Source Identification and Initial Quality Cuts} \label{sec:initial_cuts}
For this work, we take as our starting point a catalog of radio lightcurves previously produced by the VAST Pipeline \citep{vast_pipeline}. This pipeline takes as input a catalogue of source components from the source-finding algorithm \emph{Selavy} \citep{selavy}, produced by \mbox{\emph{ASKAPSoft}} \citep{askapsoft}. It then associates measurements from different epoch with specific sources. An individual source may be detected by \emph{Selavy} in some epochs but not others. In this case, the pipeline uses forced photometry, fitting a Gaussian to the image, in order to estimate the flux density and error at the position where the source was detected in images of other epochs.

We perform the following cuts on the source catalogue to ensure that each source in our sample has sufficient high quality data to be analysed. We require that a source:
\begin{itemize}
\setlength\itemsep{-0.1em}
    \item Be detected above the VAST threshold in at least three epochs, with at least two of these epochs being detected by \emph{Selavy};
    \item Be detected at $\geq$10\,$\sigma$ in at least \emph{one} epoch;
    \item Be at least 20$\arcsec$ away from the nearest neighbour source, consistent with the angular resolution of the VAST Pilot;
    \item Have no other source detected within three times the semimajor axis of the source, as measured by \emph{Selavy};
    \item Have an average compactness, defined as the integrated flux density divided by the peak flux density, of less than 1.4. This selects for point sources as expected for TDEs.
\end{itemize}

After implementing these initial criteria, we are left with a catalogue of $\sim$10$^6$ sources.

\begin{figure}[t]
\includegraphics[width=\linewidth]{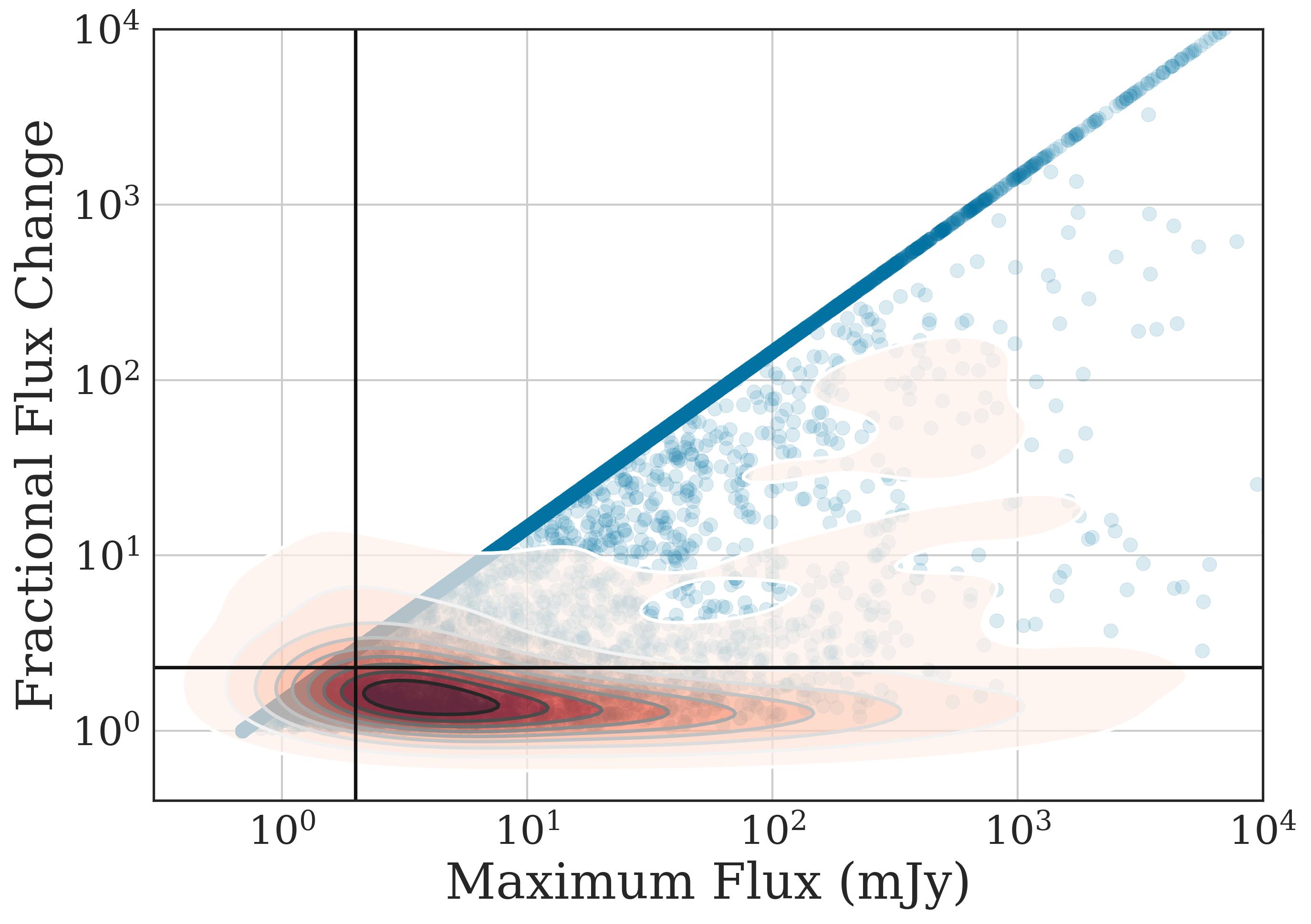}
\caption{Fractional flux change ($S_{max} / S_{min}$) and maximum flux density of simulated TDEs detectable in the VAST Pilot, shown as blue points. Also shown as a red density plot are the variability parameters of AGN in the VAST Pilot catalogue identified in the V\'eron catalogue, with colour scale representing the kernel density estimate of the fractional flux change and maximum flux density for these AGN. The smoothing bandwidth of the Kernel density is estimated using Scott's rule \citep{scott}. The black lines show our selection criteria on the fractional flux change and maximum flux density.} \label{fig:variability_param}
\end{figure}

\subsection{Radio Variability}\label{sec:radio_var}
We next restrict our sample to only include sources that are (i) variable, (ii) display this variability at a high level of significance, and (iii) whose apparent variability is not primarily dominated by a single epoch of observations. For the first point, we note that both TDEs and AGN occur in galactic centers and can be variable on similar timescales \citep{Metzger_2015}. We therefore opt to select a radio variability criteria that will exclude the vast majority of known AGN. This process will likely eliminate some true TDEs with detections in the VAST pilot. However, it will produce a purer sample for examination and will be taken into account when calculating the implication of our final sample for the rates of radio TDEs in Section~\ref{sec:rates}.

To accomplish this, we calculate the fractional flux change---defined as the ratio of the maximum and minimum integrated flux densities of a given lightcurve---for a sample of 798 AGN found by cross-matching the VAST Pilot source catalogue with the V\'eron catalogue of quasars and active nuclei \citep{veron}. We find that requiring a fractional flux change of at least 2 would eliminate 95\% of these known AGN. In Figure~\ref{fig:variability_param} we plot the fractional flux change for both this AGN sample and our mock VAST TDE light curves\footnote{The diagonal line above which no mock TDEs appear is an effect of how the VAST sensitivity is treated in our simulations. Because the minimum measurable flux density is not zero but instead three times the sensitivity of the VAST Pilot, the maximum fractional flux change will also scale linearly with the maximum flux.}. The horizontal line designates a fractional flux of 2, while the vertical line indicates a flux density 10 times the typical VAST sensitivity (which we require for at least one epoch, as described in Section~\ref{sec:initial_cuts}). While the relative number of TDEs found with different properties in this plot are not representative of what would be found by a flux density limited survey (because we simulated an equal number of events in each redshift bin and have not scaled for relative rates of different classes of TDEs) it demonstrates that TDEs are expected to occupy the region of parameter space allowed by these selection criteria (above and to the right of the plotted lines).

Second, we require that all sources have a variability detected at high significance. Specifically, we select sources that have a maximum variability statistic, $V_s$, that is greater than 5. Here, following \cite{Mooley}, $V_s = \frac{\Delta S}{\sigma}$ where $\Delta S$ is the difference between the two flux density measurements and $\sigma$ is the errors on those two flux densities added in quadrature. $V_s$ is calculated for every combination of two measurements in the lightcurve .

\begin{figure}[t]
\includegraphics[width=\linewidth]{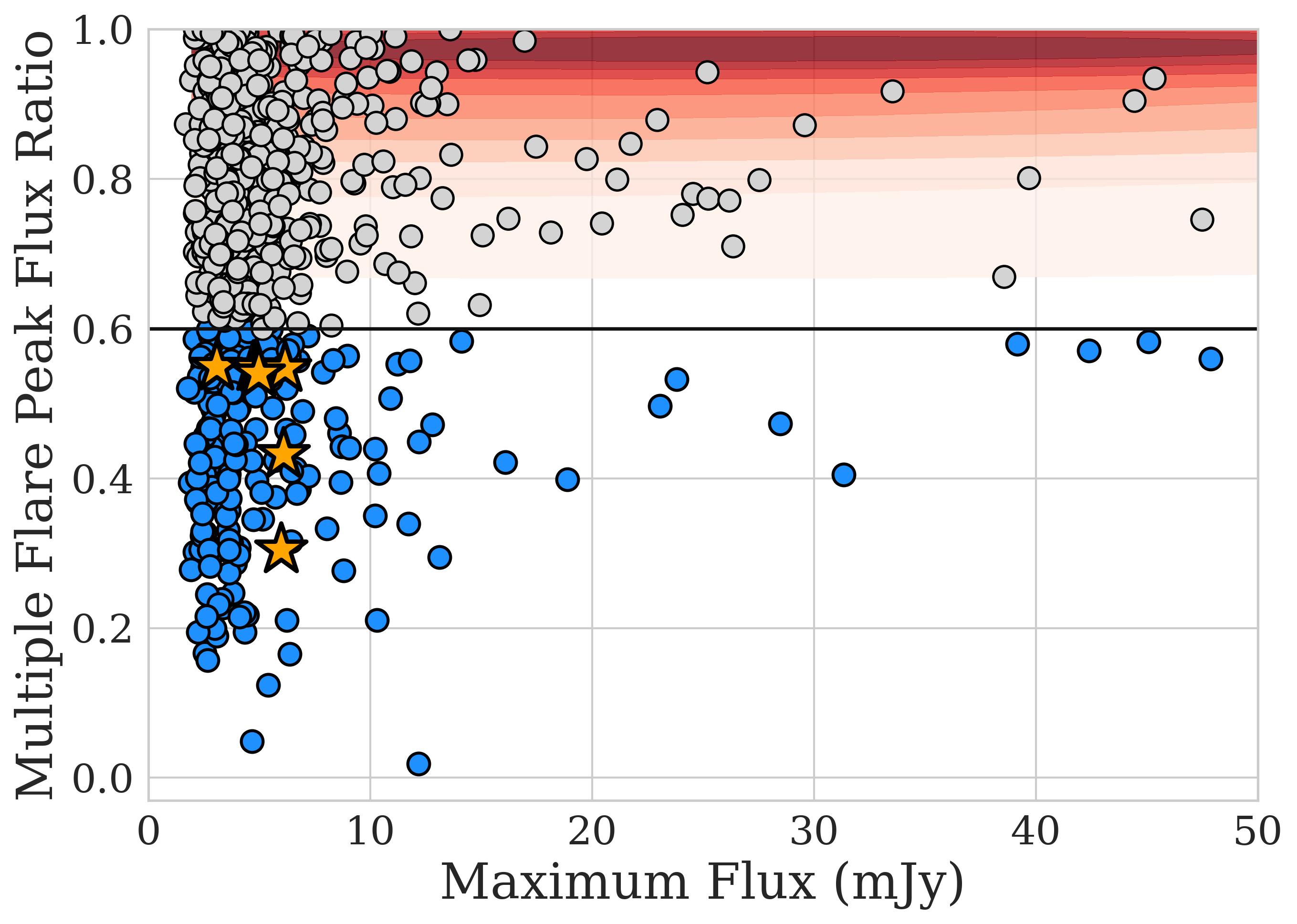}
\caption{Plot of Peak Flux Ratio, defined as the ratio of flux densities of the second brightest to brightest peak, as a function of maximum flux. Sources that pass the radio variability criteria described in Section~\ref{sec:radio_var} are shown as grey points. Blue points indicate sources that additionally pass the lightcurve morphology criterion of a peak flux ratio of less than 0.6, described in Section~\ref{sec:lc_morph}, and shown as a horizontal black line. Distribution of the peak flux ratio and maximum flux for AGN in the VAST Pilot catalogue identified in the V\'eron catalogue are shown in red, with colour scale representing the kernel density estimate. Orange stars indicate sources that are in our final sample of candidates. Note that not every source in our final sample has a secondary peak with which to calculate a peak flare ratio and would thus not appear in this plot.}\label{fig:peak_ratio}
\end{figure}

Finally, to eliminate the candidates that could be selected based on a single spurious observation, we perform a test where we one-by-one remove each epoch from the light curve and recalculate the maximum flux, fractional flux variation, and variability statistic. We require that each source pass the aforementioned criteria regardless of which singular epoch is removed. 

After both of the criteria described in this subsection are implemented, 1078 sources remain in our sample, only one of which is identified as an AGN in the V\'eron catalogue.

\begin{figure*}[t] 
  \centering
  \begin{minipage}{0.9\textwidth}
    \includegraphics[width=1\textwidth]{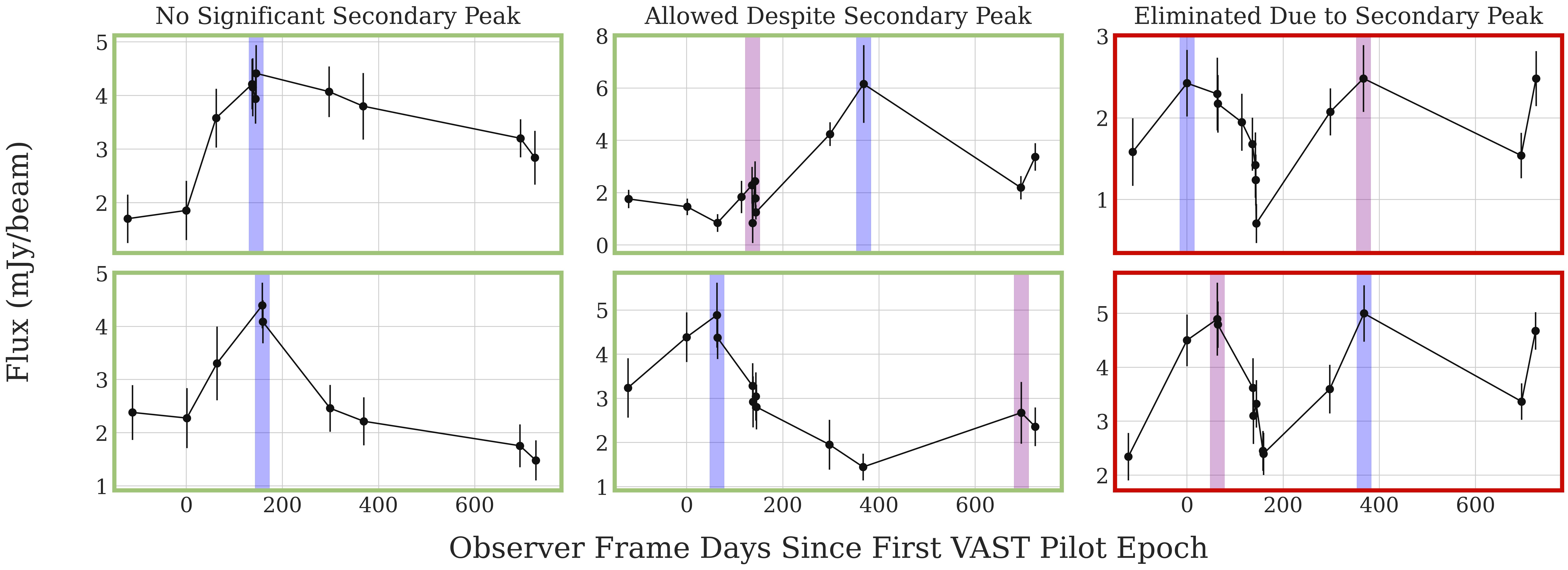}
  \end{minipage}
  \caption{Example lightcurves of VAST Pilot sources. The vertical blue and purple bars show the highest and second-highest peaks of the lightcurves, respectively. The four plots outlined in green pass all of the selection criteria described in Sections~\ref{sec:radio_var} and \ref{sec:lc_morph}. The plots outlined in red are eliminated by the lightcurve morphology criterion (Section~\ref{sec:lc_morph}). Top panel sources from left to right: VAST J230053.0$-$020732, VAST J104315.9$+$005059, VAST J042357.3$-$530255. Bottom panel sources from left to right: J213437.8$-$620433, VAST J221936.0$+$004724 VAST J213711.1$-$573146.}\label{fig:data_lc}
\end{figure*}

\subsection{Lightcurve Morphology}\label{sec:lc_morph}
The criteria from Section~\ref{sec:radio_var} eliminated all but one AGN from the sample of \cite{veron}.
 
However, Figure~\ref{fig:variability_param} shows that some AGN can exhibit fraction flux variations greater than 2. We therefore implement an additional criterion based on lightcurve morphology, to select sources whose variability resembles a single dominant flare (as expected for the models described in Section~\ref{sec:appearance}) rather than the ongoing radio variability typical of AGN. We note that this criterion may eliminate true TDEs with multiple distinct flaring episodes of similar luminosities---recently observed in the TDE ASASSN-15oi \citep{TDE_repeater}---if they occur within the two year window of the VAST Pilot. This will be further discussed in Section~\ref{sec:repeaters}.

To quantify how flare-like a particular source's lightcurve is, we define phases when the lightcurve is increasing or decreasing. A lightcurve enters an increasing phase when the flux density increases by more than one $\sigma$ from the immediately preceding epoch, where $\sigma$ is defined as the error combined in quadrature of both flux density measurements. Similarly, it enters a decreasing phase when the flux density decreases by more than one $\sigma$. We then define a \emph{peak} as the maximum flux density during an increasing phase. 

We consider any given lightcurve to be flare-like (i) if its lightcurve shows only a single peak by the criteria outlined above, or (ii) if multiple peaks are detected, that one peak is significantly predominant. We allow the latter because it is possible for TDEs to occur within galaxies with low-level AGN activity \citep{TDEs_in_AGN}. Figure~\ref{fig:peak_ratio} shows the Peak Flux Ratio, defined as the ratio of the flux density of the second highest peak to that of the highest peak, as a function of maximum flux density. On this plot we show our sample that passes the selection criteria described in Sections~\ref{sec:initial_cuts} and ~\ref{sec:radio_var} as well as AGN from the \cite{veron} catalog. 
 
AGN from this sample with multiple flares observed in the VAST Pilot appear exclusively above a peak flux ratio of 0.6, suggesting that the candidates with peak flux ratios below this value are inconsistent with a vast majority of known AGN, whose highest and second highest peak are closer in flux.

In Figure~\ref{fig:data_lc} we show examples of lightcurves without any secondary peak, a secondary peak below a ratio of 0.6, and a secondary peak above this ratio, the latter of which eliminates the source from our sample. There are 114 sources that pass this criterion, 33 of which are single flares, and 81 of which have a secondary peak below a ratio of 0.6. The flux ratio of lightcurve peaks is the final criterion that relies only on data from the VAST Pilot.

\subsection{Coincidence with a Galaxy Nucleus}\label{sec:coincidence}
TDEs occur in the presence of SMBHs, so we limit our main sample to variable radio sources whose localization regions overlap with the nuclei of known galaxies. To make this identification, we use several optical surveys as outlined below. The sensitivities of these optical surveys limit sample completeness; see Section~\ref{subsec:uncertain_rates} for how this factors into our TDE rate estimates.

\subsubsection{Optical Surveys} \label{subsec:optical_surveys}
We use five optical surveys with coverage overlapping the VAST Pilot footprint: Pan-STARRS (DR1) \citep{PS1}, the Sloan Digital Sky Survey (DR12) \citep[SDSS;][]{SDSS}, the Dark Energy Survey (DR2) \citep[DES;][]{DES}, the Skymapper Survey (DR1) \citep{skymapper}, and the \texttt{Gaia} Survey (DR2) \citep{GAIA}. All of the sources that passed our radio variability criteria described in Section~\ref{sec:radio_var} have coverage in at least one of these optical surveys.

\subsubsection{Coincidence with an Optical Source}\label{sec:initial_concidence}
We perform an initial test for nearby galaxies by cross-matching our VAST sources with the optical catalogues listed in Section~\ref{subsec:optical_surveys}. At this stage, we choose a radius larger than the positional uncertainties in order to capture any source with a potentially coincident host galaxy. We find that 73 of our radio transients have a cataloged optical source within 2$\arcsec$. 

We then use multiple metrics to eliminate any optical sources that are likely stars or quasars. First, we eliminate sources that have a measured \texttt{Gaia} parallax or proper motion value above 3\,$\sigma$. Second, we eliminate any sources that were classified as stars or quasars in SDSS and/or Pan-STARRS. These surveys both use various combinations of photometric, color-based classification and spectral energy distribution templates, as well as the difference between point-spread function and \cite{Kron} photometry \citep[e.g.,][]{star_search} to classify sources as stars, galaxies, or quasars. After removing these objects we are left with 60 VAST targets.

\subsubsection{Coincidence with Galaxy Nucleus}\label{sec:nuclear_class}
We next examine which VAST sources have positions that overlap with the nuclei of their host galaxies. The synthesized beam of VAST has a full width at half maximum (FWHM) of 12 – 20$\arcsec$. For isolated point sources with a signal to noise ratio (SNR) $>$10, this results in an average positional uncertainty of approximately $0.5\arcsec$.

However, there are additional systematic uncertainties related to astrometric offsets between the VAST Pilot and optical surveys. While the exact level of the offset can be dependent on both the field and location within a given image, we do not calculate this on an object-by-object basis, but rather include it as an overall systematic error. To quantify the magnitude of this error, we compare the positions in the VAST Pilot of AGN identified in the V\'eron catalogue to their closest optical sources in SDSS. We find that the offsets are randomly distributed with standard deviations of 0.41$\arcsec$ and 0.34$\arcsec$ in RA and Dec, respectively. To calculate a final error on the VAST Pilot positions, we combine this astrometric offset with the weighted average of the statistical uncertainties from the \emph{Selavy} detections.

\begin{deluxetable}{lc}[h]
\tablecolumns{2} 
\tablecaption{Summary of Selection Criteria\label{tab:filter_steps}}
\tablehead{\thead{Criteria} & \thead{\texttt{\#} of Sources \\Remaining}}
\startdata 
VAST Pilot Point Source Catalogue & 1\,068\,985\\
Initial Quality Cuts (Section~\ref{sec:initial_cuts}) & 263\,393\\
Radio Variability (Section~\ref{sec:radio_var}) & 723\\
Lightcurve Morphology (Section~\ref{sec:lc_morph})& 114\\
Optical Coverage (Section~\ref{sec:coincidence})& 114\\
Optical Source within $2\arcsec$ (Section~\ref{sec:initial_concidence})& 73\\
Removing Stars and Quasars (Section~\ref{sec:initial_concidence})& 60\\
Coincidence After Centroiding (Section~\ref{sec:nuclear_class}) & 12\\
\enddata 
\end{deluxetable}

We find the centroid of the optical galaxies in cutouts of the PanSTARRS, SDSS, and DES images using the python package \texttt{photutils} \citep{photutils}. We then eliminate any sources where the offset between the VAST position and optical centroid is more than two times their combined error. This leaves 12 sources in our sample, nine of which have an offset $\leq$1$\sigma$; the remaining three have offsets between 1 and 2$\sigma$. These offsets range from 0.42 to 1.09\,arcsec.

\subsection{Summary of Filtering Process} \label{sec:criteria_summary}
The number of sources that pass each individual step of our filtering criteria can be viewed in Table~\ref{tab:filter_steps}. Our final sample of radio TDE candidates consists of 12 sources which are listed along with their key properties in Table~\ref{tab:flare_properties}.

\section{Properties of TDE Candidates}\label{sec:properties}
In Section~\ref{sec:candidate_selection}, we applied a set of criteria to select highly variable radio sources that have positions that overlap with the nucleus of their host galaxies. In addition to TDEs, this population of transients may include high amplitude flares from AGN, as well as supernovae and GRB afterglows. Here we describe the multiwavelength properties of these potential nuclear transients and their host galaxies. This information will be used to select a ``gold'' sample of TDE candidates, and to inform our discussion of the nature of the entire sample in Section~\ref{sec:discussion}.

\subsection{Radio Lightcurves}
We calculate the luminosities and timescales of the transient radio flares seen in each lightcurve in order to place them in the context of TDEs and other transient sources. The lightcurve of each source in the final sample can be viewed in Appendix \ref{appendix:lc_and_hosts}. We restrict our data to the VAST Pilot rather than cross-matching with archival radio surveys as to not introduce additional uncertainty from observations at different frequencies and angular resolutions.

\subsubsection{Procedure for Calculating Light Curve Parameters}\label{sec:calculate_param}
To quantify the flare timescale and maximum luminosity, we begin by estimating the level of any underlying persistent (i.e. non-flaring) flux density that is present within the VAST light curve. For two of our twelve sources (VAST J230053.0$-$020732 and VAST J015856.8$-$012404; see Table~\ref{tab:flare_properties}) we determine by visual inspection that the flare encompasses the entire observed lightcurve and that there is no direct evidence for an underlying persistent radio flux density. For the other ten sources, the flares either appear to brighten after a series of relatively flat VAST detections, or fade to a roughly constant flux density before the end of the VAST Pilot. In these cases, we attempt to quantify the level of persistent flux density observed. Specifically, we identify the constant flux density that is consistent (within errors) of the highest number of measured flux density values in the lightcurve, while strictly requiring that it is within 1 sigma of the lowest measured flux density value. 

We then linearly interpolate the lightcurve between the observed fluxes. We chose to report timescales when the flare is above 50\% of flare's maximum flux density. The rising timescale, $t_{\rm 1/2, rise}$, and decline timescale, $t_{\rm 1/2, decline}$, are defined as the time elapsed between when the interpolated lightcurve crosses 50\% of the maximum flux density and the time of the maximum flux density, and then adjusted based on redshift to be in the rest-frame. The flare's maximum flux density is defined as the maximum flux density of the lightcurve with the estimated persistent flux density subtracted off (see Figure~\ref{fig:measure_param_example}, as well as other examples in Appendix~\ref{appendix:lc_and_hosts}).

\begin{figure}[t] 
  \centering
    \includegraphics[width=\linewidth]{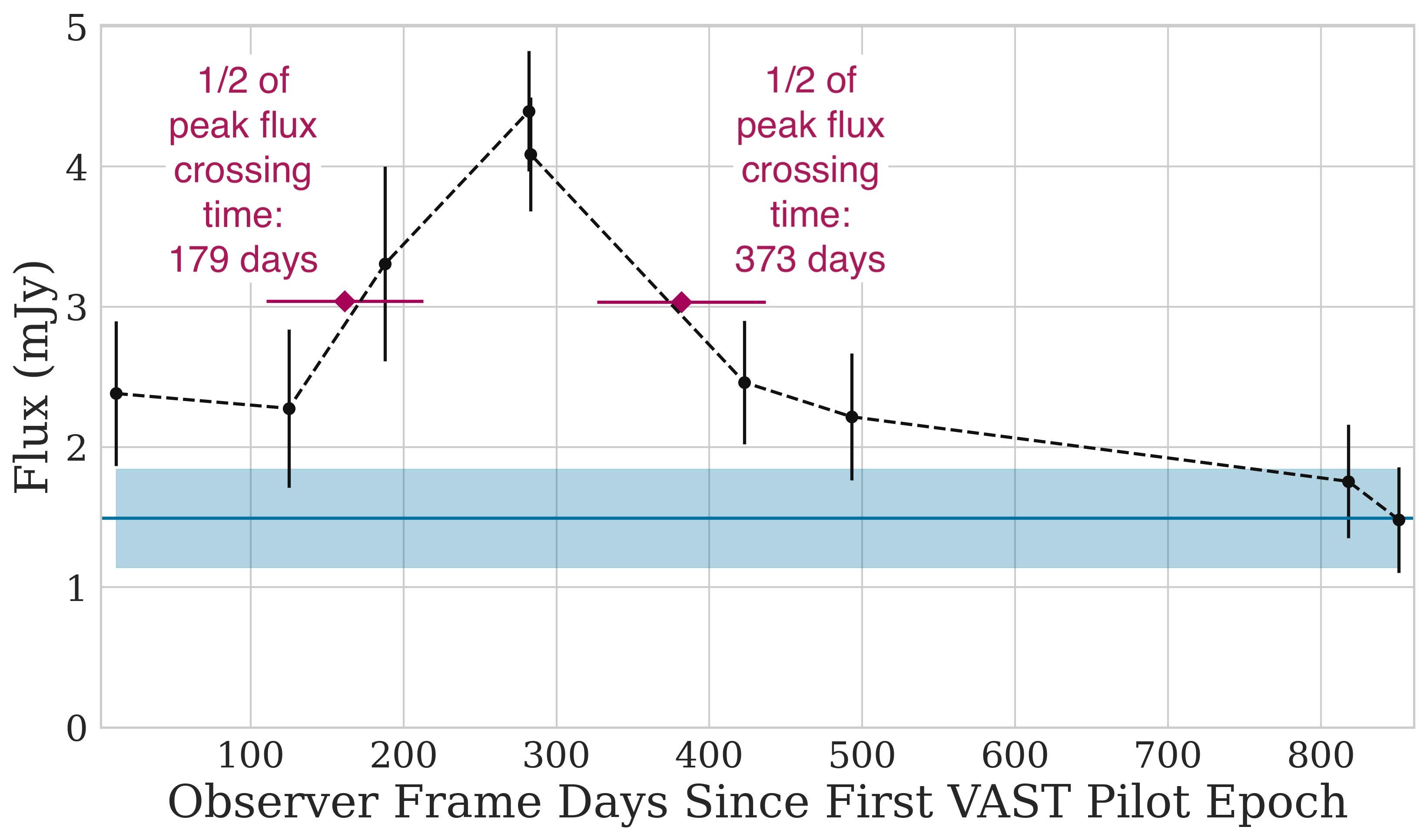}
  \caption{An example lightcurve (VAST J213437.8$-$620433) to show how peak luminosity and timescale are estimated. The inferred persistent flux density and its error are shown in blue. The linearly interpolated lightcurve is shown as a dotted line. The calculated start and end of the primary flare, defined as the times when the flux density passes 50\% of the peak flux density as measured from the persistent flux density, are shown in purple.}\label{fig:measure_param_example}
\end{figure}

We calculate the timescales in the source's rest frame according to its photometric redshift. 
For all twelve of the sources in our sample, photometric redshifts were taken from the the catalogues of SDSS, Pan-STARRS, or DES. These catalogues calculated photometric redshifts using training sets that included photometric and spectroscopic observations of galaxies as a reference, and then estimating the redshift using a local linear regression model \citep{Beck2016,PS1_redshifts,photozs_DES}. The photometric redshifts of our sample range from 0.06 to 0.8 (see Table~\ref{tab:host_properties} for values and uncertainties).

\begin{figure*}[t] 
  \centering
  \begin{minipage}{0.4\textwidth}
    \includegraphics[height=6.8cm,keepaspectratio]{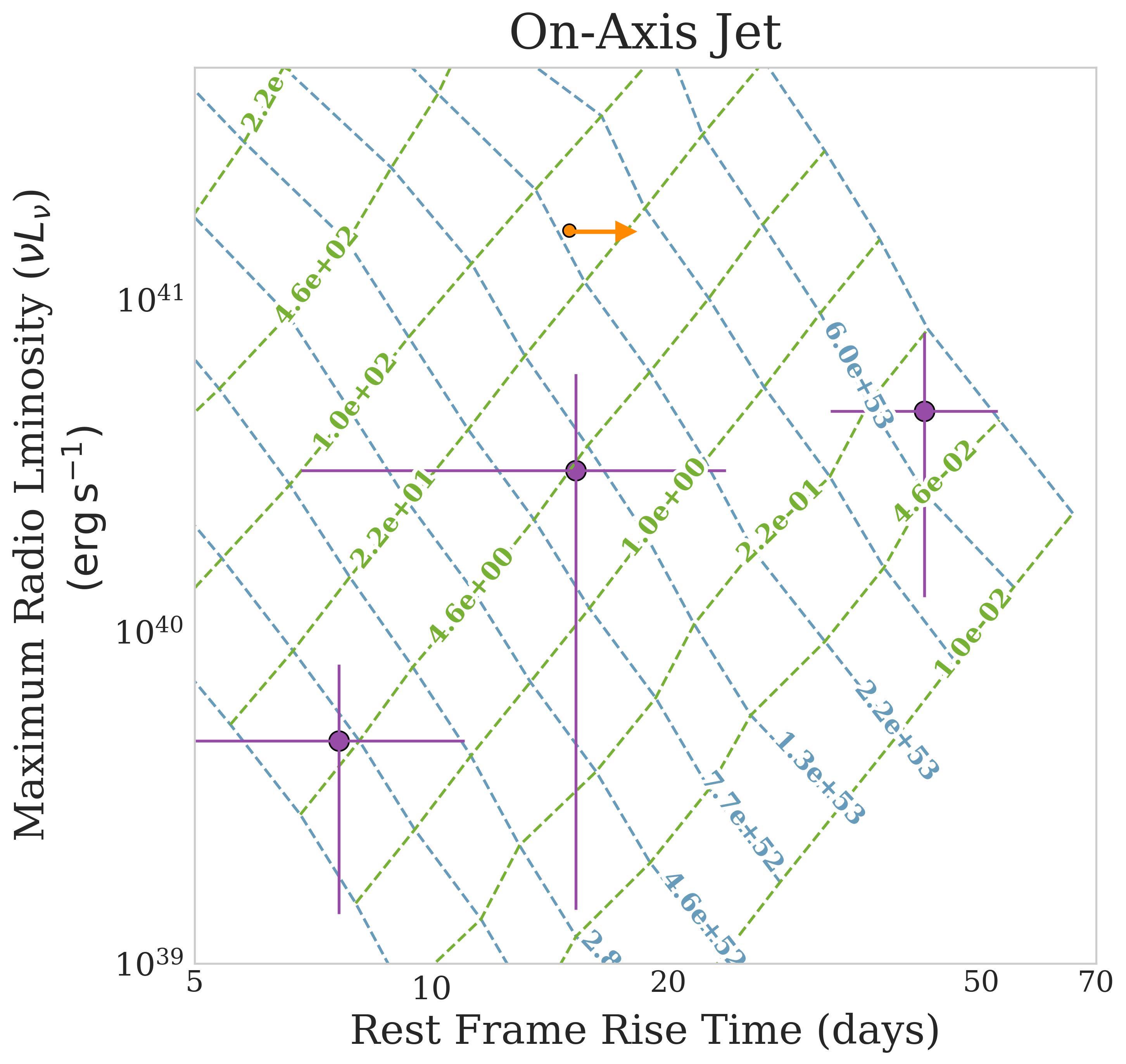}
  \end{minipage}
  \begin{minipage}{0.57\textwidth}
    \includegraphics[height=6.8cm,keepaspectratio]{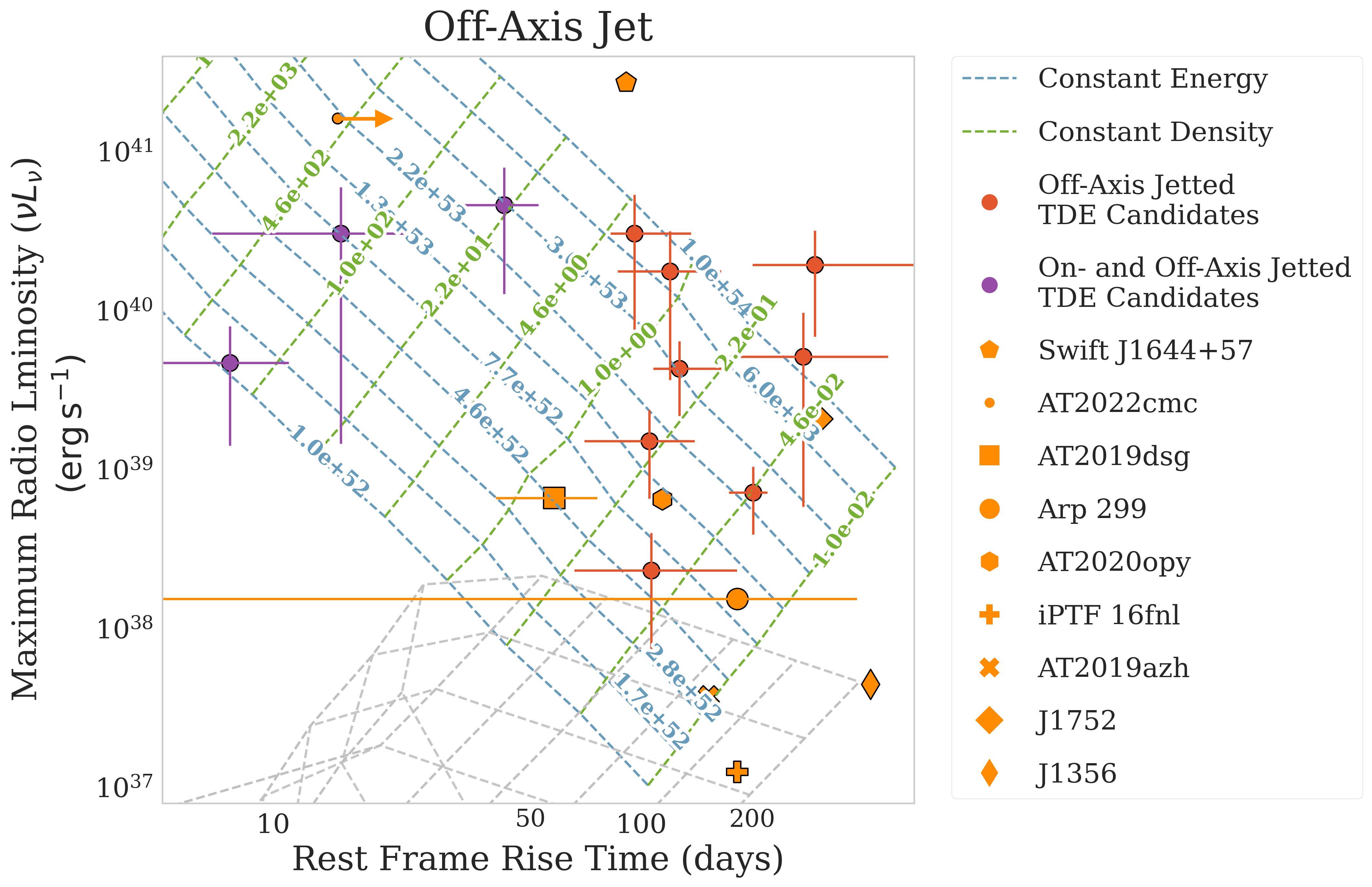}
  \end{minipage}
  \caption{Plot of maximum luminosity, $\nu {\rm L}_{\nu}$, and rest-frame rise times, $t_{\rm 1/2, rise}$, for the models described in Section~\ref{subsec:models} given specific energies and densities. The grid lines show the energies and densities used in our Monte Carlo simulation based on reasonable estimations for previously observed sources, with blue and green indicating lines of constant energy and density, respectively. Results for on- and off- axis jets are shown in the left and right plots, respectively. Results for the non-jetted sources are also shown as grey lines in the right plot. $\nu {\rm L}_{\nu}$ and $t_{\rm 1/2, rise}$ along with their errors for our sample, as calculated using the methods described in Section~\ref{sec:calculate_param}, are plotted in purple and red; purple indicates the sources consistent with both the on- and off-axis grids (and are thus visible in both panels) and red indicates the sources only consistent with the off-axis grid. We note that the brightest source in our sample (VAST J011148.1$-$025539, $\nu {\rm L}_{\nu}$=1.7$\times$ 10$^{41}$erg\,s$^{-1}$) does not appear in this figure as $t_{\rm 1/2, rise}$ is not observed. Seven radio TDEs with observed $t_{\rm 1/2, rise}$ are shown: Swift J1644+57 \citep{1644_z}, AT 2022cmc \citep{AT2022_cmc_zhou} AT 2019dsg \citep{at2019dsg}, Arp 299 \citep{arp299}, AT 2020opy \citep{AT2020opy}, iPTF 16fnl \citep{delayed_flares}, AT2019azh \citep{azh_sfaradi, azh_goodwin, somalwar_pop} VLASS J1752, J0813, and J1356 \citep{somalwar_pop}. Note that $t_{\rm 1/2, rise}$ of AT 2022cmc is depicted as an arrow as the flare is still rising. $t_{\rm 1/2, rise}$ and $\nu {\rm L}_{\nu}$ are both calculated in the rest-frame.} \label{fig:lum_vs_rise_time_models}
\end{figure*}

Uncertainties on each of these parameters (persistent flux, $t_{\rm 1/2, rise}$, $t_{\rm 1/2, decline}$, and peak flux) were all calculated using a Monte Carlo approach to produce 10\,000 versions of each lightcurve, based on the flux density uncertainty at each epoch. In the rest-frame, $t_{\rm 1/2, rise}$ ranges from $\sim$6 to 280 days, $t_{\rm 1/2, decline}$ ranges from $\sim$9 to 482 days, the maximum flux densities range from 3.4 to 6.2 mJy/beam and the persistent flux densities (where present) range from 0.7 to 2.3 mJy/beam, as listed in Table~\ref{tab:flare_properties}. The total time that the flares are above 50\% flux density ranges from 19 to 590 days. In 9 of the 12 cases $t_{\rm 1/2, decline}$ is longer than $t_{\rm 1/2, rise}$.

\subsubsection{Inferred Radio Luminosity}\label{sec:radio_lum}
In order to compare rest-frame luminosities at a consistent frequency across our sample, we need to apply a k-correction. We assume that the radio fluxes of our sources are dominated by synchrotron emission and can be well described by $S_\nu \propto \nu^{\alpha}$, where ${\alpha}$ is the spectral index and $S_{\nu}$ is the flux density at frequency $\nu$. In this case, the rest-frame luminosity $L_{\nu}$ is given by
\begin{equation}\label{equn:lum}
L_{\nu} = \frac{4\pi{D_{L}^2}S_{\nu}} {(1+z)^{1+\alpha}},
\end{equation}
where $z$ is the redshift and $D_{L}$ is the corresponding luminosity distance. \referee{We assume a spectral index of $\alpha=-0.75$ \citep{spec_index} for our sources, which assumes an electron energy distribution index of 2.5 and that \mbox{$\nu_{sa}$<$\nu_m$<$\nu$<$\nu_c$}, where $\nu_{sa}$ is the synchrotron self-absorption frequency, $\nu_m$ is the typical synchrotron frequency of the minimal electron in the power law, and $\nu_c$ is the synchrotron cooling frequency \citep{granot_sari}. This is an approximation as we do not have spectra, the p-value will vary between sources as well as over time \citep{14li_a,Goodwin_2022,repeaters}}. For $S_{\nu}$ of each source we use the peak flux density with the inferred persistent flux density subtracted. When coupled with the photometric redshifts of our sources, the inferred rest frame 888 MHz radio luminosities of our flares range from 2.7$\times10^{29}$ to 1.6$\times10^{32}$\,erg\,s$^{-1}$\,Hz$^{-1}$ (see Table~\ref{tab:flare_properties}). These correspond to $\nu {\rm L}_{\nu}$ values that range from 4$\times 10^{38}$ to 1.7$\times 10^{41}$\,erg\,s$^{-1}$ for our twelve sources.

\subsubsection{Broad Implications of Inferred Luminosities and Timescales in the Context of TDEs}\label{sec:timescale}
To understand if our estimated lightcurve parameters are broadly consistent with expectations of TDEs, we compare both to models and to previous TDE observations. We calculate the rest-frame 888 MHz peak luminosities ($\nu {\rm L}_{\nu}$) and rest-frame $t_{\rm 1/2, rise}$ of our simulated TDE lightcurves (see Section~\ref{sim_pop}) using the same methodology described in section~\ref{sec:calculate_param}. 
Figure~\ref{fig:lum_vs_rise_time_models} shows $t_{\rm 1/2, rise}$ and $\nu {\rm L}_{\nu}$ for a range of models as green and blue dotted lines. The dotted blue lines depict models with constant input energies spanning 10$^{52}$ to 10$^{54}$\,erg, while the dotted blue lines show models of a blastwave expanding into constant density medium ranging from 10$^{-2}$ to 10$^{4}$\,cm$^{-3}$. As described above, these ranges broadly span those that have been observed in TDEs previously \citep[e.g.,][]{1644_z, 14li_a}. However, we emphasize that these models are approximations. We do not use them to make definite claims about the outflow energy and circumburst density of individual events, but rather to assess whether the candidates are broadly consistent with TDEs of different classes.

The left and right plots show on- and off-axis models respectively. The viewing angle changes how various energies and densities will map onto timescale and luminosity. In particular, off-axis jets generally display longer rise times and lower peak luminosities for similar physical parameters. Also shown in the right hand plot are the models of non-relativistic outflows, which peak at luminosities $\lesssim$10$^{38}$ erg/s. Figure~\ref{fig:lum_vs_rise_time_models} shows that our sources have luminosities and rise times broadly consistent with our models of jetted and particularly off-axis TDEs. 
This first conclusion is based mainly on the high luminosities and the second relies on the relatively long $t_{\rm 1/2, rise}$ of our objects compared to on-axis models. Both of these are broadly robust predictions from a variety of models, including those with non-constant ambient media---although we note that \cite{new_jetted_model} recently showed viewing angle and Lorentz factor can be degenerate (allowing some off-axis jets to masquerade as Newtonian outflows and vice versa). 

\referee{We additionally note that our choice of free parameters in the models (see Section~\ref{subsec:models}) affects the resulting luminosities and timescales. In particular, changing our original choice of $\epsilon_e$ $=$ 0.2 and $\epsilon_b$ $=$ 0.01 to both equal 0.1 increases the radio luminosity of our on-axis jetted models by a factor of 5.6 and our off-axis models by a factor of 3.9. Perhaps most relevant is that it increases the maximum luminosity of the non-jetted models by a factor of 1.4, increasing the maximum peak luminosity of a non-jetted model in our sample from 2.1$\times$10$^{38}$ to 3.0$\times$10$^{38}$\,erg\,s$^{-1}$. In this case the least luminous sources in our sample (namely VAST J213437.8$-$620433 and VAST J015856.8$-$012404) could be interpreted as a non-jetted outflow. Multiwavelength modeling of GRB afterglows have found a large range of possible values for $\epsilon_b$, typically in the range of 10$^{-5}$ $-$ 10$^{-1}$ \citep[e.g.,][]{wijers_grb, santana_grb}. If we instead lower our choice of $\epsilon_b$ to equal 10$^{-3}$, while keeping $\epsilon_e$ $=$ 0.1, the maximum luminosities of our models decrease by a factor of 4.6, 4.2, and 3.6 for the on-axis, off-axis, and non-jetted models respectively. Additionally, the timescales in this case are significantly affected, decreasing by a factor of 2.5, 1.2, and 6.3 for the on-axis, off-axis, and non-jetted models respectively. For these parameters, the significantly shorter timescales of the non-jetted models do not reproduce the timescales of our faintest sources, and the more luminous and longer timescale sources in our sample are not consistent with either of the jetted models. Finally, we rerun our models with both with both $\epsilon_e$ and $\epsilon_b$ equal to 0.01, and find that the luminosities changed by a factor of 0.7, 2.6, and 8.2 for the on-axis, off-axis, and non-jetted models respectively. The decrease in luminosities for the on-axis case, and the increase in luminosities for the off-axis case would make these different scenarios harder to distinguish.}

\referee{Similarly, our choice of the initial Lorentz factor of the non-relativistic outflow, $\Gamma$$\approx$2 affects the final luminosity of the non-relativistic models. Varying $\Gamma$ to the minimal value of $\approx$1, but keeping the energy of the outflow the same, results in a maximum peak luminosity for the non-jetted models of $\approx$10$^{39}$, implying that our faintest sources may be consistent with a non-relativistic outflow.}

\referee{Also shown in Figure~\ref{fig:lum_vs_rise_time_models} are the rest-frame timescales and luminosities of previously observed TDEs \citep{1644_z, arp299, at2019dsg, AT2022cmc,AT2020opy,azh_sfaradi,somalwar_pop} as orange markers, calculated following the methodology outlined in Section~\ref{sec:calculate_param}. Observed radio flares from TDEs that do not have a sufficiently observed $t_{\rm 1/2, rise}$ are discussed in Section~\ref{sec:other_tdes}, including: ASASSN$-$14li \citep{14li_a}, AT 2020vwl \citep{AT2020vwl}, ASASSN-15oi \citep{TDE_repeater}, CNSS~J0019$+$00 \citep{Anderson_2019}, and XMSSL~J0740$-$85 \citep{XMMSL1}.} Many of the previously observed radio TDEs interpreted as jets are consistent with our models of jetted TDEs \citep[e.g. AT2022cmc;][]{AT2022cmc}, and sources that are thought to be non-relativistic are less luminous than our simulated models \citep[e.g., AT2019azh][]{azh_goodwin}. However Figure~\ref{fig:lum_vs_rise_time_models} also highlights how our models alone would not have predicted that Sw 1644 is an on-axis jetted TDE, as its specific combination of a long rise time and high luminosity fall outside of our model grid.

\begin{figure}[t]
\includegraphics[width=\linewidth]{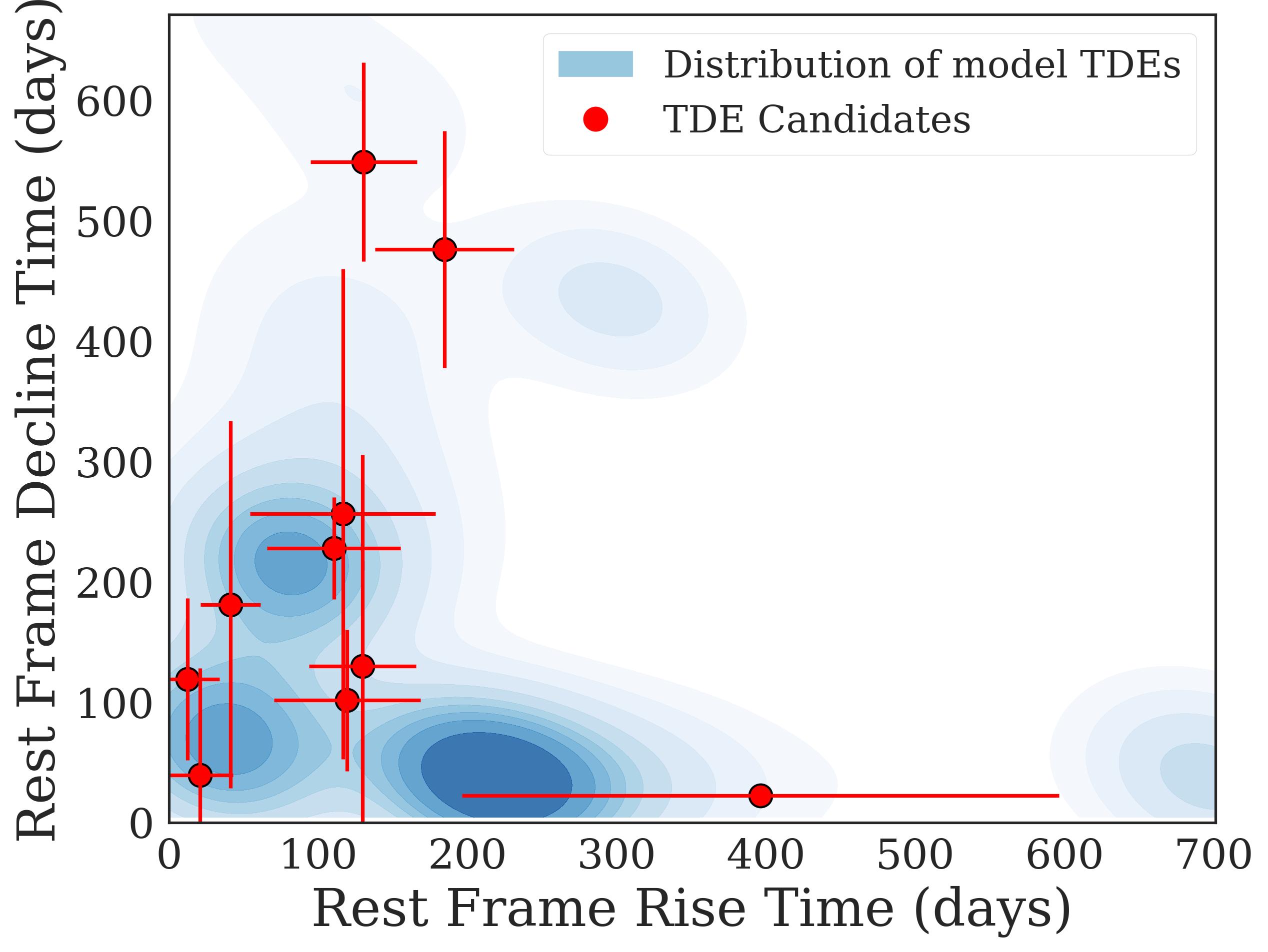}
\caption{Rise ($t_{\rm 1/2, rise}$) and decline ($t_{\rm 1/2, rise}$) lightcurve timescales, defined as the time elapsed between when the interpolated lightcurve crosses 50\% of the maximum flux density and the time of the maximum flux, as described in Section~\ref{sec:calculate_param}. The blue gradient shows the distribution of model TDEs and the red markers show TDE candidates in our final sample. Timescales of model TDEs are calculated after being projected onto the VAST Pilot cadence, as described in Section~\ref{sim_pop}.} \label{fig:timescales}
\end{figure}

Finally we note that some of our sources have timescales significantly shorter than the majority of TDEs. In particular, VAST J093634.7$-$054755, has measured $t_{\rm 1/2, rise}$ and $t_{\rm 1/2, decline}$ timescales of 10$^{+30}_{-7}$ and 9$^{+23}_{-4}$ days, respectively. However, the measured rise and decline timescales of our simulated model TDEs projected onto the VAST cadence (see Section~\ref{sim_pop}), can also have very fast timescales; consistent with the observed lightcurves. Figure~\ref{fig:timescales} shows the density of simulated sources with a particular measured rise and decline timescale alongside the rise and decline timescales of our observed sources, which overlap even for our fastest measured timescales. \referee{Additionally, \cite{goodwin_2023} detected an already fading radio outflow from the TDE AT2020vwl, 118 days after optical detection. This could imply that our faster $t_{\rm 1/2, rise}$ timescales are indeed plausible.} We therefore do not eliminate these sources a priori as plausible TDE candidates. However, we discuss other possible interpretations for these events in Section~\ref{sec:nature_of_sources}

\subsection{A Search for Multiwavelength Transient Counterparts}
TDEs produce flares across the electromagnetic spectrum. Detections or constraints from upper limits on each source's multiwavelength emission can help determine if a flare consistent with a TDE or other astronomical transient was emitted. We searched for existing archival data for transient emission at other wavelengths associated with our flares.

\subsubsection{Gamma Ray Bursts}\label{sec:grb}
We cross-matched the coordinates of the sources in our sample to two collections of observed GRBs. The first collection is compiled by the IceCube team and is updated on a weekly basis\footnote{\url{https://user-web.icecube.wisc.edu/~grbweb_public/Summary_table.html}, accessed February 28, 2024}. The second is compiled by Jochen Greiner and aims to encompass any GRB localized to within $\sim$1 square degree\footnote{\url{https://www.mpe.mpg.de/~jcg/grbgen.html}, accessed February 28, 2024}. These catalogues include measurements from BeppoSAX \citep{BeppoSAX}, the Burst and Transient Survey Experiment \citep[BATSE;][]{BATSE} onboard the Compton Gamma Ray Observatory \citep[CGRO;][]{CGRO}, the All-Sky Monitor \citep[ASM;][]{ASM} onboard the Rossi X-ray Timing Explorer \citep[RXTE;][]{rxte}, the Interplanetary Network \citep[IPN;][]{IPN}, the High Energy Transient Explorer \citep[HETE;][]{hete}, the International Gamma-Ray Astrophysics Laboratory \citep[INTEGRAL;][]{INTEGRAL}, AGILE \citep{agile}, Fermi's Gamma-Ray Burst Monitor \citep[GBM;][]{gbm} \& Large Area Telescope \citep[LAT;][]{lat}, Monitoring All-Sky X-Ray Images \citep[MAXI;][]{maxi}, \& the Burst Alert Telescope \citep[BAT;][]{swift-bat}, X-ray telescope \citep[XRT;][]{xrt} \& Ultra-Violet/Optical Telescope \citep[UVOT;][]{uvot} onboard the SWIFT Mission \citep{swift}.

In order to identify possible GRB associations, we require both that (i) the localization region of the GRB overlaps with the VAST TDE candidate and (ii) the GRB occurred within a two year window prior to the observed peak of the radio flare. The two year coincidence was chosen as a conservative limit given most TDEs with observed radio lightcurves have emission within approximately $\sim$one year of the start of the event \citep[e.g.,][]{1644_z,arp299,Ravi_2021}. With these criteria, we found that all 12 candidates had potential associated GRBs. However, in many cases the GRB localization regions were very large ($>$1000 deg$^2$). In fact, for all 12 candidates, multiple GRBs were detected within the allowed two year window whose localization regions formally overlapped with those of the VAST TDE candidate.

We tested the significance of these coincidences by performing the same cross-correlation using 1000 randomized versions of our sample's coordinates. Each time all 12 sources were still coincident. We therefore cannot conclude that any of our sources have a significant association with a detected GRB.

We therefore also performed more restrictive searches. If we limit our search to GRBs that occurred within 30 days prior to our events, we find events coincident with two of our radio flares (VAST J093634.7$-$054755 and VAST J104315.9$+$005059). In both cases these localization regions are large (GRB201207A with 195.0 deg$^2$ and GRB200422A with 61.6 deg$^2$). 

Rerunning cross-correlation with randomized coordinates we still find two associations, indicating that we cannot rule out random coincidences.

\begin{ThreePartTable}
\begin{deluxetable*}{lcccccccc}[t]
\tablecolumns{9} 
\tabletypesize{\footnotesize}
\renewcommand{\arraystretch}{1.2}

\tablehead{\thead{Source Name} & \thead{VLASS Epoch 1\\Flux Density\\(mJy\,beam$^{-1}$)}  & \thead{VLASS Epoch 1\\Date\\(MJD)} &\thead{VLASS Epoch 2\\Flux Density\\(mJy\,beam$^{-1}$)} &  \thead{VLASS Epoch 2\\Date\\(MJD)} & \thead{VLASS Epoch 2\\Implied\\Spectral Index} & \thead{RACS\\Flux Density\\(mJy\,beam$^{-1}$)} & \thead{RACS\\Date\\(MJD)} & \thead{RACS\\Implied\\Spectral Index}}
\startdata 
J011148.1-025539 & -- & -- & -- & -- & -- & 3.4$\pm$0.3 & 59211 & 2.5$\pm$0.3 \\
J015856.8-012404 & -- & -- & -- & -- & -- & -- & -- & -- \\
J093634.7-054755 & 4.2$\pm$0.3 & 58619 & 3.6$\pm$0.3 & 59489 & 0.25$\pm$0.07 & 3.7$\pm$0.3 & 59232 & 0.8$\pm$0.1 \\
J104315.9+005059 & 3.1$\pm$0.3 & 58118 & 2.7$\pm$0.4 & 59072 & -0.2$\pm$0.2\tnote{a} & 3.0$\pm$0.3 & 59223 & -0.3$\pm$0.2\tnote{b} \\
J144848.2+030235 & -- & -- & -- & -- & -- & 1.9$\pm$0.4 & 59239 & -0.7$\pm$0.2\tnote{b} \\
J210626.2-020055 & 1.6$\pm$0.2 & 58042 & 2.2$\pm$0.4 & 59066 & -0.65$\pm$0.10 & 3.0$\pm$0.3 & 59235 & -1.14$\pm$0.08\tnote{b} \\
J212618.5+022400 & -- & -- & -- & -- & -- & -- & -- & -- \\
J213437.8-620433 & -- & -- & -- & -- & -- & 2.3$\pm$0.4 & 59238 & 1.1$\pm$0.2 \\
J215418.2+002442 & 2.3$\pm$0.2 & 58023 & 2.1$\pm$0.3 & 59049 & 0.1$\pm$0.1 & -- & -- & -- \\
J221936.0+004724 & 4.8$\pm$0.3 & 58024 & 5.0$\pm$0.3 & 59068 & 0.6$\pm$0.1 & 1.9$\pm$0.3 & 59230 & -0.5$\pm$0.2 \\
J230053.0-020732 & 7.9$\pm$0.3 & 58071 & 5.0$\pm$0.3 & 59066 & 0.5$\pm$0.1\tnote{a} & 7.3$\pm$0.9 & 59211 & 2.2$\pm$0.1\tnote{b} \\
J234449.6+015434 & 4.2$\pm$0.2 & 58020 & 4.0$\pm$0.3 & 59096 & 1.2$\pm$0.4\tnote{a} & -- & -- & -- \\
\enddata 
{\noindent$^{a}$The VLASS observation for this spectral index calculation occurs during the flare.\\
\noindent$^{b}$The RACS observation for this spectral index calculation occurs during the flare.}

\caption{Detections of our sample in Epochs 1 and 2 of VLASS (2-4 GHz) and RACS-mid (1.367\,GHz). Calculation of spectral indices described in Section~\ref{subsec:vlass_compare}. In some cases, the VLASS Epoch 2 and RACS-mid observations occurred many days apart from any of the VAST Pilot epochs, see Figures~\ref{fig:J011148.1} to \ref{fig:J234449.6} in Appendix~\ref{appendix:lc_and_hosts}. The errors are derived from the flux values alone and do not account for the possible flux variations in time of the source. \label{tab:vlass_racs}}
\end{deluxetable*}
\end{ThreePartTable}

\subsubsection{Optical Flares}\label{sec:optical}
To check for coincident optical transients, we first queried the Transient Name Server\footnote{\url{https://www.wis-tns.org/}, accessed February 28, 2024} at the coordinates of each source in our sample. None had a cataloged optical transient coincident within 2$\arcsec$.

We cross-matched with the near-Earth object WISE survey \citep[NEOWISE;][]{neowise} and found that nine of our twelve sources had coincident data. We analysed the single exposure source catalogues and found that none of these sources had any flaring activity present in the data. 

We also cross-matched with g, i, and r-band data from the ZTF (DR19)\citep{ztf_data}, and found that eleven of our twelve candidates were within ZTF's observational footprint. Of those, nine had data coincident with our host galaxies from the ZTF forced-photometry service. \referee{We also cross-match our sample with the Asteroid Terrestrial-impact Last Alert
System \citep[ATLAS;][]{atlas}, which has observations for all twelve of our host galaxies.}

The ZTF and ATLAS light curves are shown in Appendix~\ref{appendix_ztf}, with the duration and peak time of the identified radio flare, as well as the duration of the entire VAST pilot, labeled. 
No clear optical flare is present in any of the ZTF or ATLAS lightcurves for our sources. \referee{We note that the source VAST J213437.8$-$620433 is not detected in ZTF and is observed in ATLAS only after the observation period of the VAST Pilot.}

To determine if a flare could have been present but not discernible in the lightcurve, we inject mock TDE flares into the \referee{difference image forced photometry} ZTF lightcurves. In particular, we add the g-band flux densities of three example TDE flares from the ZTF population of optical TDEs presented by \cite{TDE_demographics} after correcting for the relative distances to those events and our host galaxies. We choose AT2020yue, AT2021qth, and AT2020wey, with peak absolute magnitudes of $-$17.4, $-$19.2, and $-$21.5 mag, respectively, in order to represent the broad luminosity range of the observed ZTF population. \referee{We estimate which of the injected ZTF flares would have been detectable for each of our nine  candidates with ZTF coverage. We consider an injected flare to be detectable if the peak of the injected flare is greater than 2$\sigma$ higher than the mean flux of the host galaxy, estimated from the difference image forced photometry. For each host galaxy, the faintest detectable flare, of the three flares that we tested, is shown in Table~\ref{tab:host_properties}.} Overall, we find that while bright events like AT2020wey would have likely been identifiable in the ZTF light curves for all but two of our targets, fainter flares such as AT2020yue (and in some cases AT2021qth) could easily have been missed given the distances to these galaxies and the moderate flux density variations observed.

In this vein, we note that \cite{somalwar_pop} found, from their population of TDEs that were identified with both the VLA and ZTF, that radio-bright TDEs tended to have fainter and cooler optical flares, compared to the sample of ZTF TDEs in its entirety. This could suggest that if our radio-bright sources \emph{are} TDEs, the optical flares may not be sufficiently bright to be detectable given the distance to our sources. It is also possible that even if the flares launched were on the more luminous end of observed optical TDE flares, the flare could have occurred either during a gap in ZTF's observations (e.g.\ the $\sim$100 days between observing seasons) or prior to when ZTF began observations of a given field. In the latter case, we note that while the VAST Pilot overlaps temporally with ZTF, radio flares of TDEs have been observed up to four years after the primary optical emission \cite{repeaters}.

\subsubsection{Multiwavelength Radio Flare} \label{subsec:vlass_compare}
\referee{In order to probe the behavior of our identified radio flares at higher frequencies, we cross-matched our sample with epochs 1 and 2 of VLASS \citep{VLASS}, a radio survey conducted with the VLA at 2-4\,GHz, and the mid-band observation of RACS which was observed at 1.367 GHz. The first VLASS epoch was observed between September 2017 and July 2019---before the beginning of the VAST Pilot Survey (in August 2019)---while the second epoch was observed between June 2020 and March 2022 and has partial overlap with the VAST Pilot. Of the twelve sources in our sample, seven were detected in VLASS and 8 in RACS, see Table~\ref{tab:vlass_racs}. We use these observations to constrain both the high-frequency variability and spectral index of of our identified radio flares.}

\referee{\emph{Variability:} For three of the seven sources with VLASS detections, the observation in the second epoch occurred while the flare in the VAST Pilot Survey was 'active'. For the source VAST J104315.9+005059, no significant variability was observed between the two VLASS epochs. However, the observation in the second epoch of VLASS occurs very near the beginning of the flare, as shown in Figure~\ref{fig:J104315.9}, and therefore may not actually be probing the flaring activity. Similarly, for VAST J234449.6+015434, no significant variability is detected. This VLASS observation occurs very near the end of the flare (Figure~\ref{fig:J234449.6}), which for this VAST lightcurve has a large uncertainty in timing. Finally, for the source VAST J230053.0-020732, the flux density actually decreased by a factor of 1.6 between the two epochs of VLASS (Figure~\ref{fig:J230053.0}). The decrease in flux may be indicative of multiple flares, however higher cadence data at this frequency would be required to interpret this behaviour.} 

\referee{\emph{Spectral Index:} For 5 of our 12 sources, see Table~\ref{tab:vlass_racs}, there is an observation in either the second epoch of VLASS or RACS that coincides with the period of time when the source is flaring in VAST, see Section~\ref{sec:calculate_param}. We can therefore use the observations from these surveys to constrain the spectral shape of the source during the flare. While the observations in these surveys overlap with the flaring period, they also range from 18 to 149 days away from the closest VAST observation. We therefore first linearly interpolate the lightcurve to estimate the flux density in the VAST Pilot at the time of the VLASS and RACS observations.} 

\referee{We assume, as in Section~\ref{sec:radio_lum}, that the spectrum follows a simple power law parameterized by the spectral index. We then use the flux density from the interpolated VAST lightcurve as well as the flux density measured in RACS or VLASS to infer a spectral index. These are listed in Table~\ref{tab:vlass_racs}. While useful to provide context, we highlight two important caveats about these spectral indices. First, as described above, there are offsets in time between the VLASS/RACS and VAST observations, and our linear interpolation might not fully encapsulate the time evolution of the flare. Second, these spectral indices are the result of the \emph{combined} flux of the flare and any persistent radio source from the host galaxy. While we could subtract the persistent flux inferred at VAST frequencies from the analysis in Section~\ref{sec:calculate_param}, the sparse temporal coverage in VLASS/RACS prohibit a analogous assessment at higher frequencies.}

\referee{Despite these caveats, we note that two sources, VAST J144848.2$+$030235 and VAST J210626.2$-$020055 show negative spectral indices with $\alpha \lesssim -0.7$, perhaps indicating optically-thin synchrotron emission. One source, J104315.9+005059, shows a flatter spectrum, indicating the spectral peak is near the probed frequencies of 0.888 to 1.367\,GHz when comparing to RACS or 0.888 to 3\,GHz when comparing to VLASS. Finally, two sources show a positive spectral slope with $\alpha \gtrsim 0.5$, indicating the peak of the spectrum may be at even higher frequencies. Positivie spectral indices could indicate the transient is still optically thick at these frequencies. They could also be consistent with the peaked-spectrum radio sources presented by \cite{peaked-spectrum}, whose radio spectra were shown to peak between 72\,MHz and 1.4\,GHz. These sources are thought to be the preliminary stage of massive radio-loud AGN where the observed spectral peak is the result of two radio lobes with steep-spectra surrounding a flat spectrum AGN core \citep[e.g.,]{young_AGN}. Finally, another possibility that could potentially explain the positive spectral shape of some of the sources, as well as the lack of flaring observed between the VLASS epochs is the presence of some underlying scintillation, which we discuss further in Section~\ref{subsec:scintillation}.}

\subsection{Host Galaxy Properties}\label{sec:host_gal_prop}
Due to our selection criteria, all transients in our sample have localization regions that overlap with the nuclei of optical galaxies. We now examine several properties of the host galaxies, that could provide insight into the nature of the sources, discussed further in Section~\ref{sec:nature_of_sources}. We examine both the bulk properties and classification of the host, as well as possible origins of the persistent radio flux densities observed in many of our candidates.

\begin{figure}[t]
  \includegraphics[width=\linewidth]{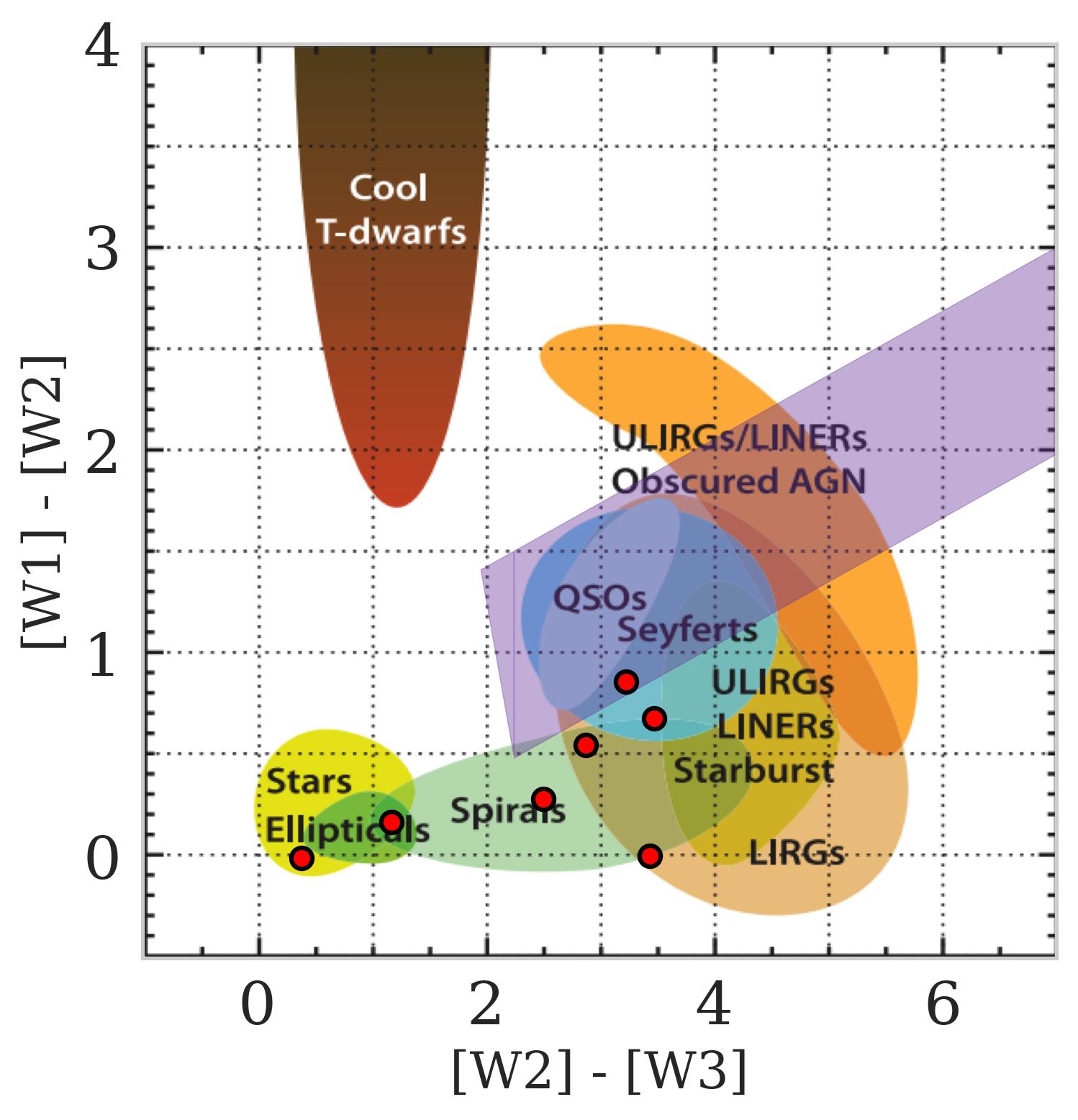}
  \caption{WISE colour-colour diagram showing different objects of classes adapted from Figure 12 of \cite{WISE}. W1, W2, and W3 represent the WISE infrared bands centered at 3.4, 4.6, and 12 $\mu$m. The purple region also shows the AGN criteria of \cite{agn_criteria}. Seven of our candidates were observed by ALLWISE \citep{ALLWISE} and are plotted in red.}
  \label{fig:wise}
\end{figure}

\subsubsection{IR Colours and Classification}\label{subsec:IR}
We examine the candidates' host galaxies using their infrared colours from the ALLWISE \citep{ALLWISE} catalogue, which are available for seven of the twelve sources in our sample. We add these host galaxies to the color-color diagram from \cite{WISE}, which shows the locations within this parameter space of various types of galaxies as shown in Figure~\ref{fig:wise}. Two of the host galaxies fall in the quiescent galaxy region, two are star-forming, and three fall in a region of overlap between luminous infrared galaxies (LIRGs) and Seyfert galaxies. 

\subsubsection{Prospector Modelling and Inferred Galaxy Mass}\label{subsec:prospector}
None of our twelve sources have associated spectra in SDSS or DES. We instead use the Bayesian fitting software \texttt{Prospector} \citep{prospector_l, prospector_j} to model the spectra of the host galaxies of our radio flares, as well as estimate the galaxies total masses. As input we provide the W1 (3.35\,$\mu{\rm m}$), W2 (4.6\,$\mu{\rm m}$), W3 (11.6\,$\mu{\rm m}$), and W4 (22.1\,$\mu{\rm m}$) magnitudes observed by ALLWISE and the u, g, r, i, z, and y AB magnitudes observed by SDSS, Pan-STARRS, or DES, where available. Additionally, one source (VAST J213437.8$-$620433) had a near-UV AB magnitude detection in the Galaxy Evolution Explorer \citep[GALEX;][]{galex}\footnote{The GALEX data \citep{10.17909/t9h59d} can be found in the Multimission Archive at Space Telescope (MAST).}. We also fix the photometric redshifts based on the values obtained from the optical catalogues; see Table~\ref{tab:host_properties}. 

We assume a $\tau$-model star formation history \citep{SFH_models}, a \cite{chabrier} initial mass function, and account for dust extinction assuming a \cite{calzetti} curve.
For two of the twelve sources, VAST J144848.2$+$030235 and VAST J215418.2$+$002442, \texttt{Prospector} was not able converge on a fit for the measured photometry. This may be due to an inaccurate photometric redshift. Neither source has infrared colours available, but the inability to properly fit the spectra may also be due to the presence of AGN activity. For the spectra successfully fit by \texttt{Prospector}, the inferred stellar mass formed for each host galaxy ranges from 4.5$\times$10$^{9}$ to 8.8$\times$10$^{11}$\,M$_{\odot}$ (see Table~\ref{tab:host_properties}). \referee{The masses of the host galaxies in our sample are consistent with those of optical TDE host galaxies \citep{TDE_demographics}, but concentrated on the higher mass end. Because our sample probes large redshifts, this may be an observational bias towards more massive and luminous galaxies. These more massive galaxies have lower TDE rates \citep{TDE_demographics}, potentially indicating that our sample of 114 TDE-like radio variables (see Table~\ref{sec:criteria_summary}) without an observed host galaxy, could be associated with a galaxy too faint to be observed in current optical surveys at these large redshifts.}

\subsubsection{Black Hole Mass Estimates}\label{subsec:BH_mass_estimate}

\referee{From our the calculated stellar masses of the host galaxies in our sample, we can estimate the mass of the central black hole using a parameterized scaling relation.} 

\referee{We use Equations 4 and 5 in \citep{BH_mass}}:
\begin{equation}
    \log({\rm M}_{\rm BH} / {\rm M}_{\odot}) = \alpha +\beta \log({\rm M}_{\rm stellar} / 10^{11}{\rm M}_{\odot})
\end{equation}
\begin{equation}
    \alpha=7.45\pm0.08;\beta=1.05\pm0.11
\end{equation}
\referee{Where ${\rm M}_{\rm BH}$ and ${\rm M}_{\rm stellar}$ are the mass of the central black hole and the stellar mass of the galaxy respectively, both in units of solar masses.}

\referee{For our sample, ${\rm M}_{\rm BH}$ ranges from $10^{6} - 10^{8}{\rm M}_{\odot}$. We note that this is only an approximation, as the above relation is based on the local universe whereas some of our sample is at considerably high redshifts. Our estimated black hole masses are again consistent with, but concentrated near the higher end of the optical sample found by \cite{TDE_demographics}. The majority of that sample of optically-discovered TDEs were between $10^{5} - 10^{7}{\rm M}_{\odot}$, with the maximum black hole mass being $10^{8.23}{\rm M}_{\odot}$.}

\subsubsection{Persistent Radio Flux}\label{subsec:persistent_flux}
As discussed in Section~\ref{sec:calculate_param}, ten of our twelve candidates have some level of persistent flux density in addition to the flaring component. The persistent luminosities implied by the sources' photometric redshifts range from 1.3$\times$10$^{29}$ to 2.9$\times$10$^{31}$\,erg\,s$^{-1}$. This additional component of flux density could originate from either star formation or an AGN present in the host galaxy, or a combination of both. Interpretation of this persistent flux density is important for understanding the possible origin of the variable component of the emission. 

\emph{Constraints on physical size:} One of our initial selection criteria was that the radio sources be unresolved in the VAST Pilot, which implies that the origin of the radio emission is less than 10$\arcsec$ in size.

We use \texttt{photutils}, to calculate the \cite{Kron} radius of each of our galaxies. The Kron radii, which contain $>$90\% of the galaxy flux, range from 0.8 to 1.8\,arcesc. Since these are all significantly smaller than 10\,$\arcsec$, we cannot rule out star formation as an explanation for the persistent flux density on the basis of galaxy radius.

\begin{figure}[t]
  \includegraphics[width=\linewidth]{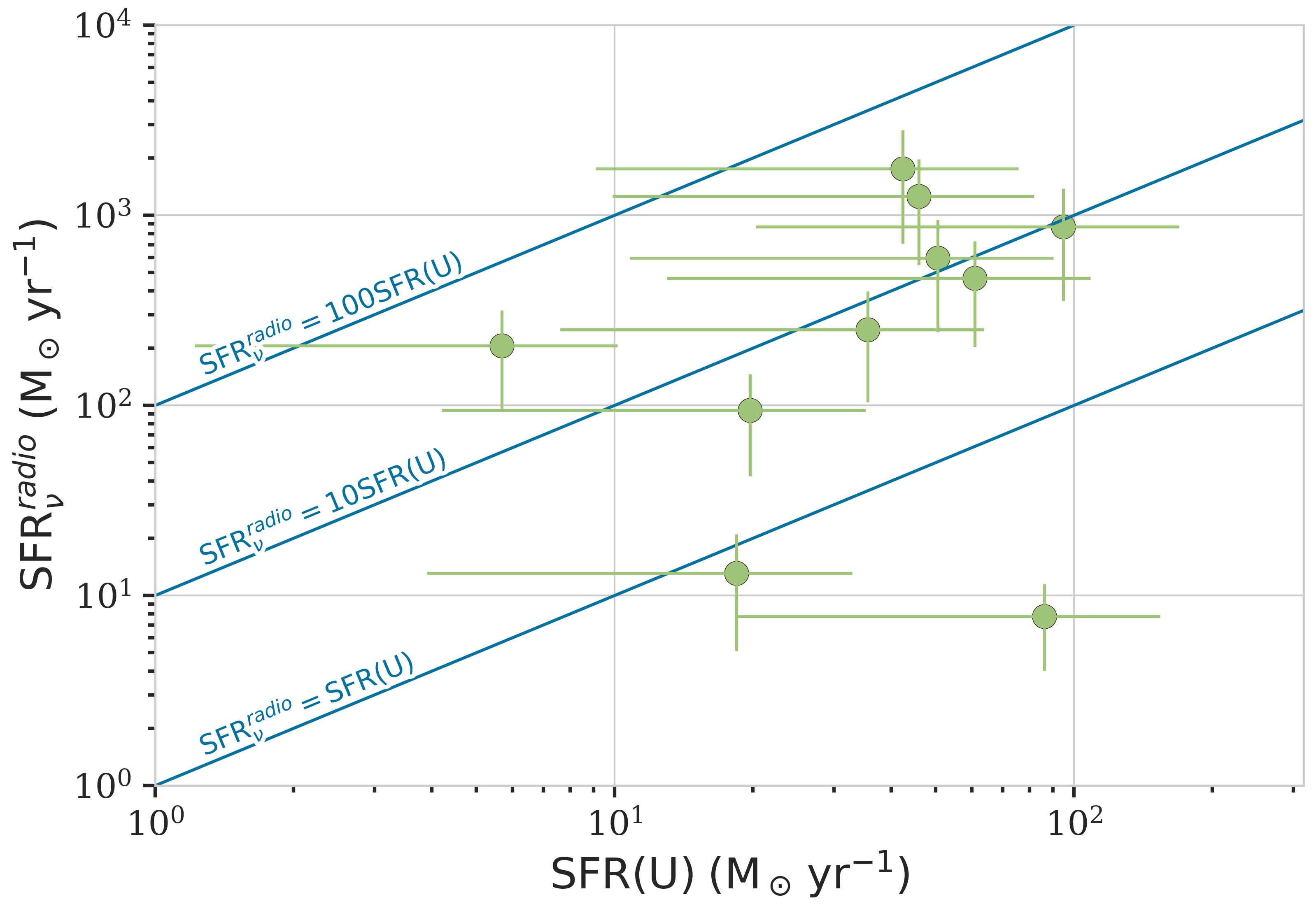}
  \caption{Star formation rate inferred from both U-band magnitude and persistent radio flux. Sources in our sample with adequate optical data and photometric redshifts are plotted. Blue lines indicate ratios of ${\rm SFR_{radio}} / {\rm SFR_{opt}}$ =1, 10, and 100.}
  \label{fig:sfr_comparison}
\end{figure}

\referee{\emph{Constraints on Spectral Index:} As mentioned in Section~\ref{subsec:vlass_compare}, we cross-matched our sample with epochs 1 and 2 of VLASS, as well as the mid-band observation of RACS. In this process, we found that six of our twelve sources had a detection in either VLASS or RACS after the observed flare (i.e. when we interpret the flux measured to be the persistent flux from the host galaxy). We use these points to estimate the spectral index of the persistent flux following the same process described in Section~\ref{subsec:vlass_compare}. In particular, similar to above, the VLASS/RACS observations occurred many days apart from any measurement in the VAST Pilot, ranging in offsets from 21 to 186 days, so interpolation was necessary. The resulting spectral indices are listed in Table~\ref{tab:vlass_racs}.}

\referee{One of the sources, VAST J210626.2$-$020055, has a spectral index of -0.65$\pm$0.10 at the time of the VLASS observation and -1.14$\pm$0.08 at the time of the RACS observation, broadly consistent with expectations for non-thermal emission. Three sources show shallow positive slopes, indicating a flatter spectrum, which may be evidence of thermal star formation \citep{SF_radio}, or a peaked spectrum source \citep{peaked-spectrum}, as discussed in Section~\ref{subsec:vlass_compare}. The remaining two sources show unusually steep positive spectral slopes which may be evidence of scintillation, we discuss this further in Section~\ref{subsec:scintillation}.}

\emph{Implications for radio star formation rate:} We can calculate the star formation rate (SFR) implied by the persistent radio flux density, assuming this flux density is entirely due to star formation. At the VAST frequency of 0.888\,GHz, the flux density is dominated by non-thermal emission \citep{SF_radio}. Assuming the flux density is entirely non-thermal, and adjusting for redshift, we use equation 12 from \cite{SF_radio}:

\begin{equation}
{\rm SFR}_{\nu}^{\rm radio} = 6.64\times 10^{-29} \left(\frac{\nu}{\rm GHz}\right) \times \left( \frac{L_{\nu}}{{\rm erg\,s^{-1}\,Hz^{-1}}}\right) {\rm M}_{\odot}{\rm yr}^{-1}
\end{equation}

\noindent the implied star formation rates range between $\sim$8 and 2000 M$_\odot$/yr.

\emph{Comparison to constraints on star formation from optical observations:} We next estimate how much star formation is expected from each galaxy based on its observed optical photometry to assess consistency with that derived from the persistent radio flux. 
As described above, we compute a best-fit galaxy spectrum for our twelve sources using \texttt{Prospector}. We extract the rest-frame U-band magnitudes from these spectra by performing synthetic photometry with the \cite{Bessel} U-band filter curve. 
We determine the inferred SFR using equation 11 from \cite{SF_optical}:
\begin{equation}
{\rm SFR(U)} = (1.4 \pm 1.1)\times 10^{-43} \left( \frac{{\rm L(U)}_{\rm obs}}{{\rm erg}\; {\rm s}^{-1}} \right) {\rm M}_{\odot}{\rm yr}^{-1},
\end{equation}

\noindent where L(U)$_{\rm obs}$ is the observer-frame U-band luminosity. Inferred SFR values range from $\sim$5 to 100 M$_\odot$/yr.

The resulting SFRs inferred from the persistent radio fluxes and the host galaxies' optical magnitudes are shown in Figure~\ref{fig:sfr_comparison} and Table~\ref{tab:host_properties}. All but three galaxies are consistent with having radio inferred star formation rates that are $\gtrsim$10 times higher than the optical inferred star formation rates. This implies that either $>$90\% of the star formation is obscured in the optical by dust (typical of luminous infrared galaxies; e.g. \citealt{lirg_obscure}) or something besides star formation (e.g. an AGN) contributes to the persistent radio emission observed. Of the three galaxies with ${\rm SFR_{radio}} / {\rm SFR_{opt}}$ $>$15, one does not have infrared colors from WISE, and the other two have IR colors consistent with LIRGs. Of the latter two, one is also consistent with the Seyfert region in Figure~\ref{fig:wise}.

One source, VAST J213437.8$-$620433, has a ratio below 1 (although this is only moderately significant given the large uncertainties on the U-band SFR calculations). This implies that the radio persistent flux density needs to be higher to account for the star formation inferred from the optical emission, or that the star formation is overestimated from the synthetic optical spectrum. The persistent radio flux density would need to be a factor of 6 higher, which is unlikely (see Figure~\ref{fig:lc_weird_SF}). In addition, the IR colours of VAST J213437.8$-$620433 overlap with the elliptical galaxy section of the WISE colour-colour diagram (see Figure~\ref{fig:wise}), implying that we do not expect significant star formation. Due to the redshift of the source, the photometry shifted to the observer's frame does not fully overlap with the U-band filter curve and we conclude the SFR implied by the optical colours of VAST J213437.8$-$620433 is likely overestimated. 

\subsubsection{Optical Variability}\label{subsec:optical_variability}
We first used ZTF \referee{and ATLAS observations} in Section~\ref{sec:optical} to determine if a TDE-like optical flare was observed. We now use these same ZTF observations, which can be viewed in Appendix~\ref{appendix_ztf}, to determine if there is evidence for significant optical variability that could indicate signs of an underlying AGN. We fit each optical lightcurve with a flat line and calculate the chi square statistic as $\chi^2 = \sum \frac{(m_i-\hat{m})^2}{\sigma_i^2}$, where $m_i$ are the observed magnitudes, $\hat{m}$ is the mean magnitude, and $\sigma_i$ is the error associated with each data point. One source, VAST J015856.8$-$012404 (Figure~\ref{fig:J015856.8}) had a statistically significant $\chi^2$, corresponding to a p-value of $<$0.01. This variability is likely due to an underlying AGN.

\begin{deluxetable}{lccc}[h]
\tablecolumns{4} 
\renewcommand{\arraystretch}{1.2}
\tablehead{\thead{Source Name} & \thead{IR Colours}& \thead{Excess Persistent\\Radio Flux}& \thead{Variable Optical\\lightcurve$^a$}}
\startdata 
J011148.1$-$025539&No Data&Yes&No\\  
J015856.8$-$012404&No Data&No&Yes\\  
J093634.7$-$054755&No&Yes&No\\  
J104315.9$+$005059&No&No&No\\  
J144848.2$+$030235&No Data&Yes&No\\  
J210626.2$-$020055&Yes&Yes&Yes\\  
J212618.5v022400&No Data&Yes&No\\  
J213437.8$-$620433&No&No &Yes\\  
J215418.2$+$002442&No Data&Yes&No\\  
J221936.0$+$004724&No&Yes&No\\  
J230053.0$-$020732&Yes&No&No\\  
J234449.6$+$015434&Yes&Yes&No\\  
\enddata 
{\raggedright $^a$ Inferred from the ZTF lightcurve where available. Remaining three sources (J213437.8$-$620433, J011148.1$-$025539, and J210626.2$-$020055) inferred from the ATLAS lightcurve.}
\caption{Host galaxy metrics that indicate presence of an AGN\label{tab:agn_Evidence}}
\end{deluxetable}

\subsubsection{Summary of Evidence for AGN Activity}\label{sec:summary_AGN}
In previous sections (\ref{subsec:IR}, \ref{subsec:prospector}, \ref{subsec:persistent_flux}, \ref{subsec:optical_variability}), we have examined how various host galaxy properties could imply an underlying AGN. We summarize this analysis in Table~\ref{tab:agn_Evidence}. Of the seven sources with IR colours, four did not overlap with the region of parameter space typically occupied by AGN. Of the ten sources with observed persistent flux, two can be explained by star formation according to the observed optical photometry and there were an additional two sources with no persistent flux density observed. The persistent flux density observed in the other 8 sources either indicated additional flux density from an AGN \emph{or} obscuration of star formation in the optical by dust. One source showed variability in the optical host galaxy emission likely due to an underlying AGN. 

\referee{One source, VAST J104315.9$+$005059}, shows (i) no signs of AGN activity from the IR colours in Figure~\ref{fig:wise},  (ii) no excess persistent flux, and (iii) no variability in the ATLAS lightcurve, see Table~\ref{tab:agn_Evidence}. We therefore consider this sources to be a particularly promising TDE candidate.

\section{Constraints on Volumetric Rates}\label{sec:rates}
We now use our simulated population of TDEs as well as the number of TDE candidates that we identified in the VAST Pilot to estimate the volumetric rate of TDEs. We note that this will be the rate of radio-bright jetted TDEs, as not all TDEs produce radio-emission that make them detectable in our survey. We discuss the implications of this in Section~\ref{sec:rate_implications}.

\subsection{Rate Calculation}\label{sec:rate_calculations}
We begin by using our simulated TDE population to determine the efficiency with which the VAST Pilot Survey is expected to detect TDEs. As a reminder, in Section~\ref{sim_pop} we created a set of $>$200\,000 theoretical jetted TDE lightcurves at redshifts between 0.05 and 2. These were calculated for a range of explosion energies and ambient densities for two viewing angles (to represent on and off-axis jets). We then ran a Monte Carlo simulation where we chose from this model grid and project the resultant lightcurve onto the VAST cadence and sensitivity. 

\begin{figure}[t]
  \includegraphics[width=\linewidth]{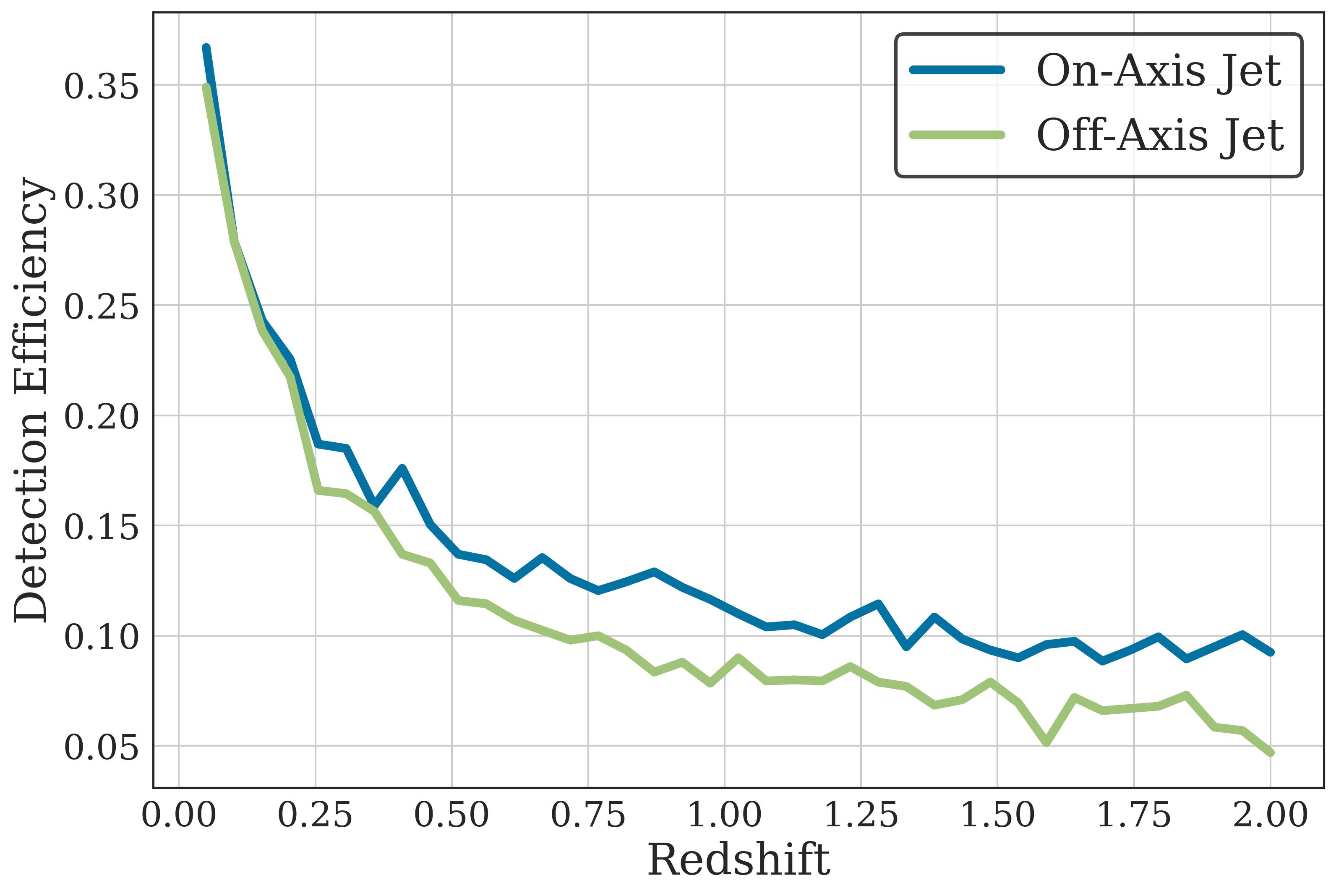}
  \caption{Detection Efficiency of model on-axis (blue) and off-axis (green) jetted TDEs in the VAST Pilot as a function of redshift. The relationship between efficiency and redshift is not perfectly smooth due to stochastic effects based on the number of models simulated.}
  \label{fig:detection}
\end{figure}

To calculate a detection efficiency, we take the simulated VAST lightcurves and apply the same selection criteria that we applied to the observed radio lightcurves, outlined in Section~\ref{sec:candidate_selection}. These criteria are: (i) be observed in at least three epochs, (ii) in two of those epochs have a measured flux density of $>$3$\sigma$, (iii) in one epoch have a measured flux density $\geq$\,10\,$\sigma$, (iv) have a fractional flux change of $\geq$2, (v) have a variability statistic, $V_s \geq$ 5. The detection efficiency is the fraction of simulated sources that pass these selection criteria. When calculating the volumetric rate with Equation~\ref{equn:rate}, the detection efficiency accounts for how many TDE are occurring that would not be detected using our data and methodology. In Figure~\ref{fig:detection}, we show our measured detection efficiencies as a function of redshift for both on and off-axis jetted TDEs. We focus on jetted TDEs, as all of our identified candidates have luminosities more consistent with jetted TDEs (see Section~\ref{sec:timescale}).

A volumetric rate can be calculated from a combination of the survey detection efficiency and number of TDE candidates identified as:

\begin{equation}\label{equn:rate}
R = \frac{N}{\sum{\epsilon_iV_it_i}}
\end{equation}

\noindent Where $\epsilon_i$, V$_i$, and t$_i$ are the detection efficiency, comoving volume, and proper time observed within each distance bin, i, respectively, and N is the number of TDE candidates in our sample. In our case, the observed volume is the comoving volume at that redshift multiplied by the fraction of the sky covered by the relevant fields of the VAST Pilot (see Figure~\ref{fig:coverage}). The observed time span starts at the earliest simulated TDE, 1000 days before the start of the VAST Pilot (see Section~\ref{sim_pop}), and ends at the last observation of the VAST Pilot epoch. \referee{The detection efficiency as a function of redshift is shown in Figure~\ref{fig:detection}.}

For the number of TDEs, N, as described in Section~\ref{sec:candidate_selection}, we detect a sample of twelve TDE candidates from the VAST Pilot. 
All twelve of these are consistent with an off-axis jet and three are consistent with both an on- and off-axis jet, depending on the inferred energies and densities; see Figure~\ref{fig:lum_vs_rise_time_models}. If we consider all twelve sources to be off-axis TDEs, this would imply \referee{$R_{\rm off-axis}$ = 0.80$^{+0.31}_{-0.23}$\,Gpc$^{-3}$\,yr$^{-1}$}. If we instead assume that the three TDEs consistent with both on- and off-axis jets are indeed on-axis jetted TDEs, we calculate $R_{\rm on-axis}$ = 0.15$^{+0.14}_{-0.08}$\,Gpc$^{-3}$\,yr$^{-1}$. The error bars on the rates are calculated using Tables 1 and 2 of \cite{Gehrels}, assuming a confidence level of 0.8413, corresponding to 1$\sigma$ Gaussian errors. Our rates are based on a range of energies and densities inferred from previously observed relativistic TDEs (see Section~\ref{subsec:models}). We discuss the implications of varying these, along with other parameters, in Section~\ref{subsec:uncertain_rates}. See Section~\ref{sec:rate_implications} for a comparison to other estimates of the volumetric rate of TDEs.

\begin{figure*}[ht] 
\centering
\begin{minipage}{0.45\textwidth}
\centering
\includegraphics[width=\textwidth]{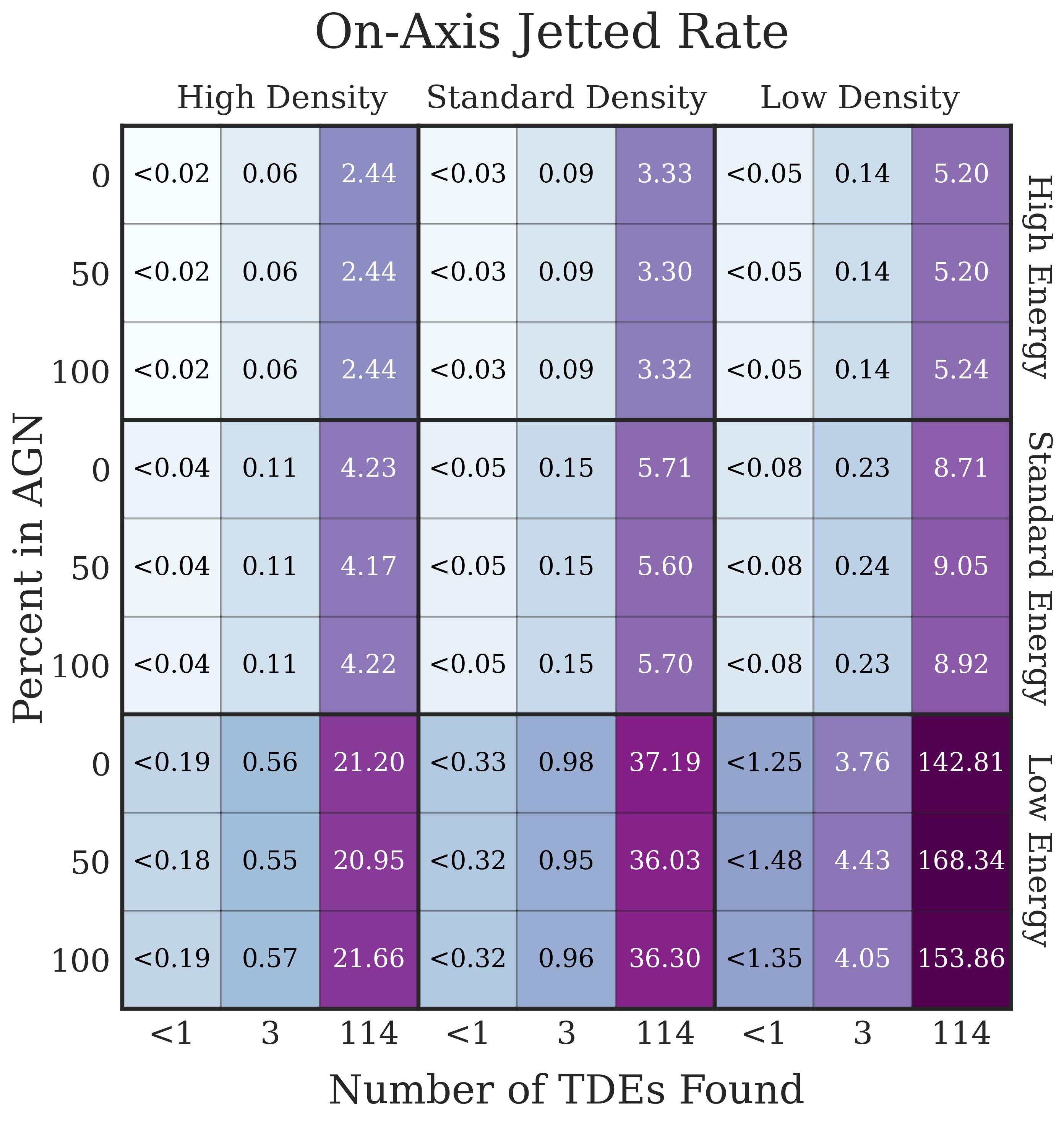}
\end{minipage}\hfil
\begin{minipage}{0.45\textwidth}
\centering
\includegraphics[width=\textwidth]{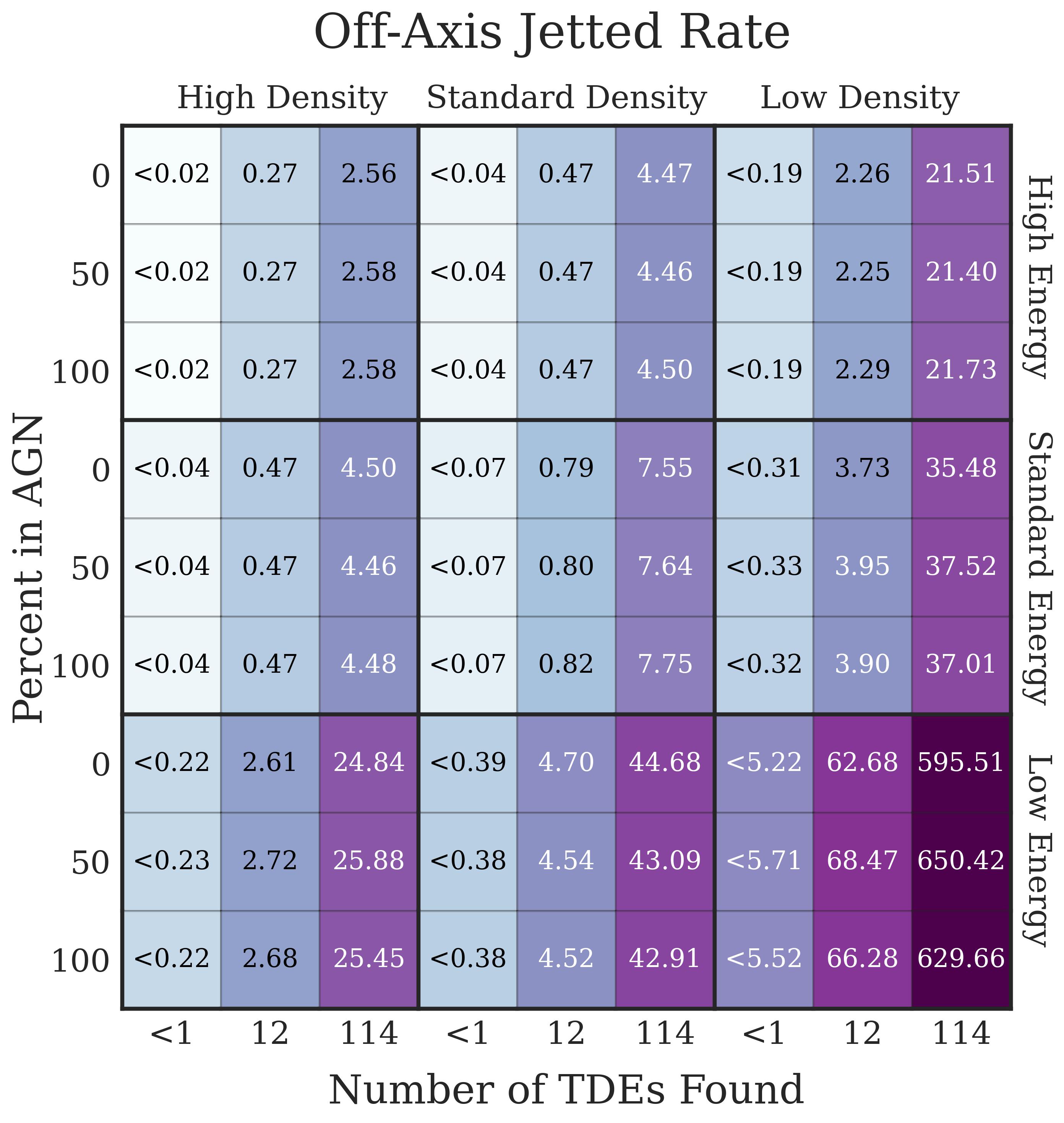}
\end{minipage}\hfil
\medskip
\caption{Estimated rates for a given set of parameters in units of Gpc$^{-3}$\,yr$^{-1}$ of on-axis (left) and off-axis (right) jetted TDEs. ``Percent in AGN'' is the percent of sources in the simulation with persistent flux densities typical of AGN added to their lightcurve. ``Number of TDEs Found'' is the assumed sample size when calculating the implied rate. ``High'', ``Standard'', and ``Low'' Density indicate the range of densities used for the simulation: $10^{1}$ to $10^{4}$, $10^{-2}$ to $10^{4}$, and $10^{-2}$ to $10^{1}$\,cm$^{-3}$, respectively. Similarly, ``High'', ``Standard'', and ``Low'' Energy indicate the range of energies used: $10^{53}$ to $10^{54}$, $10^{52}$ to $10^{54}$, and $10^{52}$ to $10^{53}$\,erg, respectively. Darker colours correspond to higher calculated rates.}
\label{fig:rate_grid}
\end{figure*}

\subsection{Uncertainties in the Rate Estimate}\label{subsec:uncertain_rates}
When calculating the final rates, there are several unknowns for which assumptions were made, namely, the luminosity function and distribution of timescales for TDEs (parameters in terms of outflow energies and circumnuclear densities in our simulations), and the fraction of TDEs occurring in AGN. Additionally, the final number of candidates that we classify as TDEs will clearly affect the calculated rate. Below we describe how varying each of these parameters affects the final rate estimation for on- and off-axis jetted TDEs. A summary of how each of these parameters affects implied rate can be seen in Figure~\ref{fig:rate_grid}.

\emph{Energy range}: The outflow energies for the simulated TDEs range from $10^{52}$ to $10^{54}$\,erg, based on observations of the relativistic TDE Sw J1644+57 \citep{1644_z,1644_b}. Rerunning the simulation and recalculating the rates using only the higher energies within this range ($10^{53}$ to $10^{54}$\,erg) causes the calculated rate to decrease by a factor of 0.42. Using only the lower energies ($10^{52}$ to $10^{53}$\,erg) causes the calculated rate to increase by a factor of 4.5.

\emph{Density range}: The circumnuclear densities for the simulated TDEs range from $10^{-2}$ to $10^{4}$\,cm$^{-3}$, based on the densities surrounding other SMBHs at various radii. Rerunning the simulation and recalculating the rates using only the higher densities within this range ($10^{1}$ to $10^{4}$\,cm$^{-3}$) causes the calculated rate to decrease by a factor of 0.42 .Using only the lower energies ($10^{-2}$ to $10^{1}$\,cm$^{-3}$) caused the calculated rate to increase by a factor of 3.8.

\emph{AGN Fraction}: One of the key properties of our sample was the persistent flux density present in ten of the twelve final candidates. Because the exact fraction of TDEs occurring in AGN is unknown, we reran the simulation with 0, 50, and 100\% of the simulated sources having an additional persistent flux density added to the lightcurve. We generated a sample of persistent luminosities to draw from, for each TDE simulated in an AGN, as follows. We calculated the level of persistent flux density in the sample of AGN identified in the VAST Pilot, following the methodology described in Section~\ref{sec:calculate_param}. We used the known redshifts for those sources to calculate the persistent luminosity of each AGN. We then drew randomly from this sample of AGN persistent luminosities and calculated the flux density that would be measured given the particular distance at which the TDE is simulated. We find that the detection efficiency and thus the calculated rate is largely unaffected by changing the fraction of TDEs occurring in AGN, only varying by $\pm$1\%.

\emph{Final Population Size}: There are 12 sources in our final sample of candidates, three of which are consistent with both on- and off- axis jets. These population sizes provide our best estimate to infer the volumetric rate of TDEs. We can also calculate a rate including every source that passes the radio variability criteria, regardless of if the source is classified as nuclear. This sample is then independent of the astrometric accuracy of the VAST Pilot, and of the completeness of the optical surveys used to identify host galaxies. The sample size in this case is 114 (see Table~\ref{tab:filter_steps}) and implies a volumetric rate of $<$14.75\,Gpc$^{-3}$\,yr$^{-1}$. If we instead assume that \emph{none} of the sources in this sample are TDEs, the implied rate is $<$0.04\,Gpc$^{-3}$\,yr$^{-1}$.

\section{Discussion} \label{sec:discussion}
We now discuss possible origins for the sources in our sample other than a TDE interpretation. We also compare our selected sample to previously observed TDEs with radio detections. Finally, we discuss the implications of our results for the rate of TDEs and prospects for future surveys.

\subsection{Nature of Sources} \label{sec:nature_of_sources}
In addition to the TDE scenario that were the main target of our search, other transient origins are also still possible for the sources in our final sample. Specifically we below investigate supernova, GRB, and AGN possibilities.

\subsubsection{Non-Nuclear Origin}
Each radio variable in our sample has a localization region that overlaps with the centroid of an optical source in one or more of the optical surveys SDSS, Pan-STARRs, DES, and Skymapper. However, it is possible that some of the sources are not truly associated with the nuclear regions of galaxies. This could be either because (i) the optical and radio sources are not truly associated but rather aligned by chance, (ii) the radio and optical are associated, but the transient originates somewhere else in the host galaxy, or (iii) the optical source is not a truly a galaxy, but a star. 

On the first point, we found angular offsets ranging between 0.42 and 1.09$\arcsec$, while the radii and magnitudes of our putative hosts range from 0.8 to 1.8$\arcsec$ and 14.2 and 22.5 mag, respectively. Using the method of \citet{Bloom2002} and \citet{Berger2010} we calculate a probability of chance coincidence of each individual source to be $\lesssim$0.003, indicating a chance coincidence is unlikely for any of the sources. However, given the typical precision of the VAST positions ($\sim$0.8$\arcsec$ when coupling centroiding errors with possible astrometric offsets between VAST and the optical surveys) and the inferred distances to our hosts, our constraints on the physical offsets of the radio sources are weak---often allowing for offsets of multiple kpc from the galaxy core. In this case, our discovered flares would be due to another type of astronomical transient such as supernovae or gamma-ray bursts, which will be discussed below.

We note that in Section~\ref{sec:initial_concidence} we explicitly excluded sources identified as stars. However, while many of our putative hosts show clearly elongated morphologies, a few (e.g. those associated with VAST J212618.5$+$022400 (Figure~\ref{fig:J212618.5}), VAST J093634.7$-$054755 (Figure~\ref{fig:J093634.7}), and VAST J210626.2$-$020055 (Figure~\ref{fig:J210626.2})) are faint sources near the detection threshold of the optical surveys and could not be confidently classified as either stars or galaxies. 

In these cases, the sources may be variable radio stars. For example, \cite{ASKAP_stars} conducted a search for M dwarf stars using ASKAP and found four known M dwarfs with variable radio emission at 0.888 GHz, with fractional flux changes greater than two. 

Stars can be variable in the radio with a large range of timescales; given the cadence of the VAST Pilot, it is possible for radio stars to be misconstrued as single flares and pass our radio variability and lightcurve morphology criteria. 
\subsubsection{Supernova Origin}
We next consider the case of a core-collapse supernovae\footnote{We specifically consider core-collapse supernovae, as deep searches for radio observations of Type Ia supernovae have all yielded non-detections \citep{no_radio_1a}}, in the case that our sources are not truly nuclear. As mentioned in Section~\ref{sec:radio_lum}, the spectral luminosities inferred for the radio flares in our sample range from 4$\times10^{29}$ to 2$\times10^{32}$\,erg\,s$^{-1}$\,Hz$^{-1}$ at a rest-frame frequency of 888\,MHz. These are all brighter than the \emph{most} luminous known radio supernova \citep[PTF 11qcj;][]{ptf11qcj}, which brightened to a 5 GHz radio luminosity of $1.7\times10^{29}$\,erg\,s$^{-1}$\,Hz$^{-1}$ approximately 2000 days post-explosion. If sources in our sample are supernovae, they would be uncharacteristically bright. However, our inferred luminosities rely on photometric redshifts, which have a large associated error. The dimmest source in our sample at 4$\pm$2$\times10^{29}$\,erg\,s$^{-1}$\,Hz$^{-1}$ may have a luminosity consistent with the brightest known supernova. However, this specific event (VAST J213437.8$-$620433, Figure~\ref{fig:lc_weird_SF}) exploded in an elliptical galaxy, making a core-collapse supernova origin unlikely \citep{SN_elliptical_rates}.  

\subsubsection{GRB Origin}
In Section~\ref{sec:timescale}, we demonstrated that our sources were broadly consistent with theoretical expectations for TDEs by comparing to models originally designed to model emission from GRBs. Using those same models, we also find that the ranges of implied isotropic equivalent energies and surrounding densities are broadly consistent not only with jetted TDEs but also with long-duration GRBs \cite[see, Figure 8 of][]{GRB_params}. 

The precision of the VAST astrometry does not allow us to assess the consistency of our population with the projected offsets of long-duration GRBs (approximately half of which have host-offsets $<$1\,kpc; \citep{grb_hosts}). However, we note that 11 of our 12 targets have host galaxy stellar masses inferred from \texttt{Prospector} modeling that are larger than 10$^{10}$ M$_{\odot}$, and hence larger than the stellar masses of a majority of long-duration GRB hosts \citep{Leloudas2015}. One of our events (VAST J213437.8$-$620433) also exploded in an elliptical galaxy. Thus, while the GRB interpretation is feasible for some individual events (and the VAST Pilot is also expected to detect GRB afterglows based on current afterglow rate estimates; \citep{GRB_radio_rates, vast_pilot_grbs}) GRBs are unlikely to explain our entire population.

We note that \cite{vast_pilot_grbs} conducted a search for GRB afterglows in the VAST Pilot. This search was designed to select sources whose lightcurves were well described by a power-law or a smoothly broken power-law function. Additional criteria were implemented to remove AGN from their sample. None of our sources were included in the sample of five GRB afterglow candidates that they identified. While the properties of our sources are generally consistent with GRBs, they were not selected as GRB candidates. This is likely due to the shape of their lightcurves having properties consistent with an AGN, such as persistent radio flux, or variability in the lightcurve that did not fit a smooth power-law function.

\subsubsection{AGN Origin}
As shown in Section~\ref{sec:summary_AGN}, two sources show no signs of AGN while the remaining ten indicate that the persistent flux densities of some of the candidates are likely due at least in part to radio emission from AGN. Our selection criteria were chosen to eliminate typical AGN flares. Thus if these sources are AGN, they are either TDEs occurring in AGN, or AGN flares that are particularly dominant. 

``Changing-look'' AGN are exceptions to normal AGN variability \citep[e.g.,][]{lamassa}. These are AGN that change from type-1 (shows both broad and narrow lines in their optical spectra) to type-2 (lacking broad lines). They also show extreme variability in their X-ray properties \citep[e.g.,][]{brandt}. \cite{merloni} investigated whether the first discovered changing-look quasar \citep{lamassa} was a flare produced by a TDE. One of the arguments that they use against this interpretation is that the gas mass implied by the broad line region of the spectrum is a few hundred solar masses, much more then would be expected by the disruption of a star. This suggests that changing-look AGN are a population of sources distinct from TDEs. \cite{clagn_Ruan} built a population of changing look quasars from an archival search of SDSS DR12 to investigate their emission mechanisms. Their results also disfavoured the TDE scenario in favour of an intrinsic dimming due to rapidly decreasing accretion rates.

Changing-look AGN are typically defined by their optical and X-ray properties. However, \cite{radio_changing_look_AGN} investigated the radio variability of Mrk 590, a changing-look AGN, and found that at 1.4 GHz, the AGN had a a 28\% flux density increase between the years 1983 and 1995 and a 46\% flux density decrease between 1995 and 2015, which were both correlated with the optical-UV and X-ray wavelengths. They show that this radio variability could be due to the increased accretion rate leading to a jet or wind that expanded before eventually fading as the accretion rate declined. While Mrk 590 showed extreme radio variability, this variability was over the course of decades, a much longer timescale than for the sources in our sample.

Due to the short timescale of the flares in our sample, we tentatively disfavor the changing-look AGN scenario. Optical spectra of candidate TDEs at the time of their flares would help to conclusively distinguish these possibilities by comparing to AGN and TDE spectra.

\subsubsection{Scintillation}\label{subsec:scintillation}
\cite{scintillation_in_ASKAP} present a population of six rapidly scintillating radio sources, variable on timescales of hours, detected using ASKAP. Our criteria on lightcurve morphology should select against sources that repeatedly vary on these short timescales. There is a possibility however that a scintillating source sampled at the cadence of the VAST Pilot could appear as a single flare and be included in the final sample, particularly for sources with short VAST $t_{\rm 1/2, rise}$ and $t_{\rm 1/2, decline}$ timescales such as VAST J093634.7$-$054755 (Figure~\ref{fig:J093634.7}).

\referee{The lack of flaring activity observed for three of our sources in VLASS, see Section~\ref{subsec:vlass_compare}, could indicate scintillation. For scintillation in the weak scattering regime, the RMS fractional flux variation has an inverse dependence on frequency. From Equation 6 in \citep{walker}, we can derive that the flux variations should be a factor of $\sim$5 times higher in the VAST Pilot, observed at a frequency of 0.888\,GHz than in VLASS, observed at a frequency of 3\,GHz. This may explain why variability is seen in the VAST Pilot but not in VLASS for some of our sources, however this could also be explained by the observations occurring near the beginning or end of the flare, see Section~\ref{subsec:vlass_compare}.}

\referee{In Section~\ref{subsec:persistent_flux}, we attempt to constrain the spectral shape of the host galaxy's persistent flux using observations from RACS and VLASS that overlap with the non-flaring period of the VAST observations. For two of the sources, VAST J011148.1$-$025539 and VAST J213437.8$-$620433, the measured flux densities in these surveys imply a somewhat steep positive spectral index, see Table~\ref{tab:vlass_racs}. However, this estimation ignores temporal differences as the RACS and VLASS observations in some cases were more than 100 days apart from a measurement in the VAST Pilot. In the case of VLASS, which is not an ASKAP survey, it also ignores instrumentation differences including angular resolution. If we take these spectral indices at face value, they likely indicate that the flares are at least in part due to scintillation. \cite{scintillation_example} present a sample of six rapidly scintillating AGN. The spectral indices implied by the flux densities measured at 4.9 and 8.4 GHz rapidly vary between positive and negative values, see Table 1 in \citep{scintillation_example}.}

\subsection{Comparison to Other Radio TDEs}\label{sec:other_tdes}

\subsubsection{Timescales and Luminosities}
In Section~\ref{sec:timescale}, we compared the luminosities and timescales of our identified radio flares to both theoretical models and previously observed radio TDEs. \referee{Figure~\ref{fig:lum_vs_rise_time_models} shows how our sample compares to several observed TDEs with measurable timescales.} Broadly, we find that the luminosities of our sample are consistent with those of TDEs classified as jetted, and are more luminous than those classified as non-relativistic. \referee{All of our sources are considerably brighter than most radio flares classified as a non-relativistic outflow, including: XMSSL~J0740$-$85 \citep{XMMSL1}, iPTF~16fnl, \citep{delayed_flares}, and AT~2019azh\citep{azh_sfaradi,azh_goodwin}. However, some TDEs classified as non-relativistic, in particular ASASSN$-$14li \citep{14li_a} and AT 2020vwl \citep{AT2020vwl} have luminosities only marginally below our least luminous source, VAST J213437.8$-$620433. We note that these were observed at a higher frequency than the VAST Pilot. Other observed flares, that are potentially the result of an off-axis jet including: ASASSN$-$15oi \citep{TDE_repeater}, CNSS J0019$+$00 \citep{Anderson_2019}, and ARP 299 \citep{arp299} have luminosities consistent with our sample.}

In particular, comparing to the recent population of TDEs that \cite{somalwar_pop} identified using both the VLA and ZTF, we find that five of their six sources appear at lower radio luminosities ($\lesssim$3$\times$10$^{38}$\,erg\,s$^{-1}$) than our twelve candidates. However, that search was sensitive to less luminous TDEs than the search presented in this paper, due to several properties of VLASS compared to the VAST Pilot. In particular, VLASS has a larger sky coverage (33\,885 square degrees), better sensitivity (typical image RMS of 0.12 mJy beam$^{-1}$), higher frequency (2-4 GHz), and larger time span ($\sim$three years). As a result, VLASS may be more sensitive to emission from sub-relativistic outflows. We will examine the implications of the fact that we did not identify any TDE candidates with luminosities below 10$^{38}$\,erg\,s$^{-1}$ in Section~\ref{subsubsec:non_jet_rate}, below.

By comparing their population of radio-selected and optically-detected TDEs to the general population of optically-detected TDEs, \cite{somalwar_pop} found that the radio-selected population was slightly more likely to occur in AGN host galaxies. They also found that they had fainter and cooler optical flares, compared to the population of optical TDE flares as a whole, potentially explaining why we did not detect any flares in the ZTF lightcurves for our sample's host galaxies (see Section~\ref{sec:optical}).

\subsubsection{Repeated Radio Flares From TDEs}\label{sec:repeaters}
In Section~\ref{sec:lc_morph}, we described how we select sources whose lightcurves have a single dominant flare. However, TDEs have been observed with multiple distinct flaring episodes, for example ASASSN-15oi \citep{TDE_repeater}. Recently, \cite{repeaters} have discovered 23 delayed radio flares from a population of optically selected TDEs, including two that exhibited re-brightening after a previously observed early flare. The nature of these delayed flares is currently debated (e.g. delayed jet-launching, off-axis jets, or sub-relativistic ejecta). However, we note that for the TDEs with multiple observed flares: (i) the initial flares had radio luminosities $\nu L_{\nu} < 10^{39}$\,erg\,s$^{-1}$, and (ii) the secondary flare occurred more than two years after the initial flare \citep[]{at2019dsg, delayed_flares}. Thus, given the sensitivity and duration of the VAST Pilot, for similar events it is reasonably likely that any bright delayed flare would be the \emph{only} flare present in the VAST lightcurve. However, as the diversity of radio TDEs is further explored and the time baseline of wide-field radio transient surveys increases, the need for methods to distinguish between AGN and TDE variability will become even more paramount.

\subsection{Implications of Rate Constraints} \label{sec:rate_implications}
In Section~\ref{sec:rates}, we showed how our sample (three candidate on-axis TDEs and twelve candidate off-axis TDEs) implies a physical volumetric rate for on-axis jetted TDEs of 0.15$^{+0.14}_{-0.08}$\,Gpc$^{-3}$\,yr$^{-1}$ and an off-axis rate of 0.80$^{+0.31}_{-0.23}$\,Gpc$^{-3}$\,yr$^{-1}$. Here we discuss how this value compares to previous estimates and implications for the rates of the jetted TDEs. 

\subsubsection{Comparison to Previous Estimates for Non-Jetted TDEs}\label{subsubsec:non_jet_rate}
We did not find any TDE candidates with luminosities consistent with non-jetted TDEs. While none were identified in this search, we know that it is possible for the VAST Pilot to detect radio emission from TDEs whose emission is not necessarily jetted: AT2018hyz was a mildly relativistic TDE at a redshift z$=0.0457$ that had an upper-limit followed by a single 1.3 mJy detection within the VAST Pilot \citep{AT2018hyz}. In addition, \cite{RACS_TDEs} recently identified radio emission in the RACS survey at the location of 4 optically identified TDEs that are consistent with non-relativistic outflows. However, our selection criteria (Section~\ref{sec:candidate_selection}), are stricter that a single detection, so it is not necessarily unexpected that no non-jetted TDEs, known or new, were found by the methodology of this paper. 

To test this, we use the same methodology described in Section~\ref{sec:rate_calculations}: we can calculate a detection efficiency and subsequently an upper limit on the volumetric rate implied by detecting $<$1 non-relativistic source in this search. This upper limit is $<$1.6$\times$\,10$^{3}$\,Gpc$^{-3}$\,yr$^{-1}$. We can compare this limit to previously estimated rates of TDEs. 

\cite{TDE_demographics} used a sample of optically-selected TDEs from ZTF over a three-year time period to infer the black hole mass range as well as to constrain the rates of TDEs, as a function of blackbody luminosity. Their inferred rates span several orders of magnitude, ranging from $\sim$10$^{-1}$ to 10$^{3}$\,Gpc\,$^{-3}$\,yr$^{-1}$ for luminosities ranging from $\sim$10$^{43}$ to 10$^{45}$\,erg\,s$^{-1}$. They integrate over their range of luminosities to find that the volumetric rate of optical TDEs with a blackbody luminosity $>$10$^{43}$\,erg\,s$^{-1}$ is 3.1$^{+0.6}_{-1.0}$\,$\times$\,10$^{2}$\,Gpc$^{-3}$\,yr$^{-1}$. 

Using an X-ray selected population of 13 TDEs from the eROSITA X-ray telescope on Spektrum-Roentgen-Gamma (SRG) \citep{erosita}, \cite{xray_pop} calculated the X-ray-loud TDE rate to be $\sim$2.3\,$\times$\,10$^{2}$\,Gpc$^{-3}$\,yr$^{-1}$ for sources with X-ray luminosities $>$10$^{43}$\,erg\,s$^{-1}$. According to these estimates, the rates of X-ray-loud and optical-loud TDE observations agree well. Our upper limit is consistent with this rate, but is also non-constraining, as it is approximately an order of magnitude higher than the estimates from optical and X-ray observations.

\subsubsection{Comparison to Previous Estimates for Jetted TDEs}
\cite{Metzger_2015} calculated empirically and theoretically expected rates of on-and off-axis jetted TDEs. The empirical rate was calculated based on the detection of Sw J1644+57. \emph{Swift} detected Sw J1644+57 in a comoving volume of $\sim$11\,Gpc$^{3}$ over a time period of ten years, implying that the on-axis jetted TDE rate is $R_{\rm on-axis}\approx$ 0.01\,Gpc$^{-3}$\,yr$^{-1}$. In addition, Sw J1644+57 had a Lorentz factor of $\sim$10 \citep[e.g.,][]{1644_b}, which corresponds to a beaming correction of $\sim$100 \citep{Metzger_2015}, implying $R_{\rm off-axis}\approx$ 1\,Gpc$^{-3}$\,yr$^{-1}$. 

For their theoretical rates, \cite{Metzger_2015} adopt a per galaxy rate of $\sim$10$^{-5}-10^{-4}$ yr$^{-1}$ from \cite{wang_merritt_TDE_rates} and \cite{stone_metzger}, a local galaxy density of $\sim$10$^{7}$\,Gpc$^{-3}$, a fraction of TDEs that launch jets of $\leq$10\% \citep{bower_2013,velzen_2013}, and a beaming factor of $\sim$100 \citep{1644_b,1644_z}. With these values, they calculate theoretical volumetric rates for on- and off-axis jetted TDEs of $R_{\rm on-axis}$ $\lesssim$0.1--1\,Gpc$^{-3}$\,yr$^{-1}$ and $R_{\rm off-axis}$ $\lesssim$10--100\,Gpc$^{-3}$\,yr$^{-1}$.

Our on-axis rate estimate, $R_{\rm on-axis}$ = 0.15\,Gpc$^{-3}$\,yr$^{-1}$, is significantly higher than the empirical estimate but consistent with the theoretically calculated rate. However, our on-axis rate is possibly an overestimation, as the three sources classified as on-axis jetted TDEs used to calculate this rate are all sources that were also consistent with off-axis jetted TDEs. Our upper limit if our survey detected no on-axis jetted TDEs ($<$0.05 \,Gpc$^{-3}$\,yr$^{-1}$) is consistent with the empirical estimates based on Sw J1644+57.

In contrast, our off-axis rate estimate, $R_{\rm off-axis}$ = 0.80 Gpc$^{-3}$ yr$^{-1}$, is consistent with the empirical rate estimated by the detection of Sw J1644+57 after a beaming correction has been applied. However, it is again possible that not all of our sources are true TDEs. If we assume that we did not detect any off-axis jetted TDEs, we would place an upper limit on the volumetric rates of on axis jetted TDEs of $<$0.07\,Gpc$^{-3}$\,yr$^{-1}$. This is over an order of magnitude lower than the empirical estimates based on Sw J1644+57.

It is also possible that our final candidate sample of 12 objects does not include every TDE within the VAST Pilot. To calculate an upper bound on the rate using our sample, we can assume that every source consistent with our radio variability criteria is a TDE. By ignoring the criteria on coincidence with the nucleus of a galaxy, we keep any potential candidate whose host galaxy may be sufficiently faint that it is not visible in one of the optical surveys. This results in a rate upper limit of $<$7.65\,Gpc$^{-3}$\,yr$^{-1}$ and $<$10.40\,Gpc$^{-3}$\,yr$^{-1}$ for on- and off-axis jets, respectively. Our upper bound on the off-axis jet rate agrees well with the theoretical estimate from \cite{Metzger_2015}. The upper-bound on on-axis jets is significantly higher than either the theoretical or empirical estimates from \cite{Metzger_2015}, although we note that on-axis upper limit assumes that \emph{every} source in the sample is an on-axis jet.

Finally, we note that \cite{somalwar_pop} used their population of TDEs discovered with VLASS to constrain the rate of radio-emitting TDEs to be $\gtrsim$ 10\,Gpc$^{-3}$\,yr$^{-1}$. While this estimate is more than an order of magnitude higher than ours, it is not inconsistent with our estimate, as their search was sensitive to less luminous sources (Section~\ref{sec:other_tdes}). Their sample could therefore include more sources with non-jetted emission. 

\subsubsection{Implications for Fraction of TDEs that Launch Jets}
Using our estimated rates, we can derive implications for the fraction of jetted TDEs by comparing to other TDE rate estimates. As stated in Section~\ref{subsubsec:non_jet_rate}, optical observations of TDEs imply a TDE rate of 3.1$^{+0.6}_{-1.0}$\,$\times$\,10$^{2}$\,Gpc$^{-3}$\,yr$^{-1}$. Our calculated rate of jetted TDEs, 0.80\,Gpc$^{-3}$\,yr$^{-1}$, is three orders of magnitude lower than this. Because our search specifically probes the population of \emph{jetted} TDEs, this implies that the fraction of TDEs that launch relativistic jets, f$_j$, is $\sim$0.26\%. The largest uncertainty in our sample comes from distinguishing between TDEs and other forms of radio transients such as AGN. However, other forms of transients being included in our sample would only lead to an overestimate in the rate calculation. \cite{fabio_jet_fraction} constrained the rate of TDEs that launch relativistic jets, f$_j$, to be 3$\times$10$^{-3}$ $<$ f$_j$ $<$ 1 using three jetted events detected by Swift \citep{burrows_2015, brown_2015} and a total TDE rate of 10$^{-3}$\,Gpc$^{-3}$\,yr$^{-1}$ \citep{velzen_2018}. Our estimate agrees with this result.

There is a discrepancy between the rate of TDEs implied by observations and the rate calculated from two-body relaxation, with the observational rates being consistently and significantly lower than the those inferred from models. \cite{stone_metzger} attempted to resolve this tension by adjusting several assumptions of the theoretical rate calculation. However, even their most conservative estimate, 3\,$\times$\,10$^{3}$\,Gpc$^{-3}$\,yr$^{-1}$ is significantly higher than the rates inferred from optical detections. If we instead assume that the theoretical estimate is correct, then our rate of jetted TDEs would imply f$_j$$\approx$0.02\%, also consistent with the estimates from \cite{fabio_jet_fraction}.

\subsubsection{Implications for Number of TDEs in the Full VAST Survey}
The full VAST survey will consist of 2174 hours across 4 years with $\sim$8\,000\,deg$^2$ of coverage. We have ran our simulation described in Section~\ref{sim_pop}, updating for the coverage, cadence, and sensitivity of the full VAST survey. We use this version of the simulation to calculate the detection efficiencies of on- and off-axis jetted TDEs. Combining our maximum calculated physical rates with these detection efficiencies, we can calculate the number of TDEs expected to be found in the extragalactic component of the full VAST Survey. Based on our calculation we expect to find 6 on-axis TDEs, assuming a rate of 0.15\,Gpc$^{-3}$\,yr$^{-1}$ or 26 off-axis TDEs assuming a rate of 0.80\,Gpc$^{-3}$\,yr$^{-1}$. 

Our search for TDEs in the VAST Pilot has shown that in order to conclusively classify the TDE candidates observed in the full VAST survey, we require astrometric precision so that we can compare positions with optical surveys and identify truly nuclear candidates. This will likely require followup observations to properly localize each source. We will also require deep optical imaging, potentially out to redshifts $\gtrsim$1, to identify host galaxies which may be faint. Definitively classifying any transients discovered will require multi-frequency follow-up, both in radio bands (to trace the evolution and energetics of the blastwave) and other wavebands (to identify potential counterparts).

\section{Summary}

We conducted a Monte Carlo simulation in which we generated millions of model radio TDEs with a range of energetics, distances, and viewing angles, and then projected them onto the VAST Pilot frequency and cadence. From this simulation we chose selection criteria to identify the largest number of TDEs while minimizing other forms of radio transients in the sample. This resulted in criteria based on radio variability, lightcurve morphology, and coincidence with the nucleus of an optical galaxy.

We present a sample of twelve radio TDE candidates identified in the VAST Pilot Survey. Eleven sources in our sample are consistent with off-axis jets based on their maximum luminosities and $t_{\rm 1/2, rise}$ timescales calculated from their lightcurves. The one source without a measurable $t_{\rm 1/2, rise}$ value has a luminosity also consistent with a jetted TDE. Three sources are consistent with both an on-axis and off-axis jet depending on the inferred energy of the outflow and circumnuclear density. In addition to the TDE interpretation, sources identified in this search may also be consistent with AGN, uncharacteristically bright supernovae, or gamma-ray bursts.

We apply the same selection criteria that we used to identify our sample to our simulated population to infer the efficiency with which we are able to detect TDEs at various redshifts. We combine this information with the number of candidates in our sample to estimate an implied volumetric rate of radio TDEs. We estimate a rate for on-axis jetted TDEs of 0.15 Gpc\,$^{-3}$\,yr$^{-1}$ and an off-axis jetted rate of 0.80\,Gpc$^{-3}$\,yr$^{-1}$. We found our rate estimates to be consistent with previous volumetric rate estimates and with estimates of the fraction of TDEs that launch jets. This search provides an additional independent constraint on the rate of TDEs.

Our search for TDEs in the VAST Pilot along with searches for radio-bright TDEs in other radio surveys like the one conducted using VLASS \citep{somalwar_pop} have begun to shed light on the properties and volumetric rates of TDEs. With the full VAST survey having now commenced, we will soon expand our sample size of radio TDEs. The fast cadence and longer time span of the full survey will increase our ability to find and classify these sources. The more detailed lightcurves will allow for more accurate measurements of the timescales, luminosities, and lightcurve morphologies. Upcoming instruments like the Square Kilometer Array \citep[SKA;][]{SKA},the Next Generation Very Large Array \citep[ngVLA;][]{ngvla} and the Deep Synoptic Array 2000 \citep[DSA-2000;][]{DSA2000} will expand our sample size even further, transforming our understanding of radio TDEs.

\begin{acknowledgments}
The authors thank Joshua Speagle, Yuyang Chen, James Leung, Benjamin Shappee, and Ashley Stock for useful conversations. 

H.D. acknowledges support from the Walter C. Sumner Memorial Fellowship and from the Natural Sciences and Engineering Research Council of Canada (NSERC) through a Postgraduate Scholarship. M.R.D. acknowledges support from the NSERC through grant RGPIN-2019-06186, the Canada Research Chairs (CRC) Program, and the Dunlap Institute at the University of Toronto. B.M.G. acknowledges support from the NSERC through grant RGPIN-2022-03163, and of the CRC Program. D.K. is supported by NSF grant AST-1816492. A.H. is grateful for the support by the the United States-Israel Binational Science Foundation (BSF grant 2020203) and by the Sir Zelman Cowen Universities Fund. This research was supported by the ISRAEL SCIENCE FOUNDATION (grant No. 1679/23). 

The Dunlap Institute is funded through an endowment established by the David Dunlap family and the University of Toronto.  Parts of this research were conducted by the Australian Research Council Centre of Excellence for Gravitational Wave Discovery (OzGrav), project number CE170100004. The ZTF forced-photometry service was funded under the Heising-Simons Foundation grant \#12540303 (PI: Graham).

\end{acknowledgments}
\vspace{5mm}

\software{\texttt{AFTERGLOWPY} \citep{afterglowpy}, \texttt{ASKAPSOFT} \citep{askapsoft}, \texttt{ASTROPY} \citep{astropy1,astropy2}, \texttt{EMCEE} \citep{emcee}, \texttt{MATPLOTLIB} \citep{matplotlib}, \texttt{NUMPY} \citep{numpy}, \texttt{PANSTAMPS}, \citep{panstamps}, \texttt{PHOTUTILS} \citep{photutils}, \texttt{PROSPECTOR} \citep{prospector_j,prospector_l}, \texttt{SCIPY} \citep{scipy}, \texttt{SELAVY} \citep{selavy}, and the \texttt{VAST PIPELINE} \citep{vast_pipeline}.}

\begin{ThreePartTable}
\begin{deluxetable*}{lcccccccccc}
\tabletypesize{\footnotesize}
\setlength\tabcolsep{3.8pt}

\tablehead{\thead{Name} &\thead{RA\\(J2000)} &\thead{RA err\tnote{a}\\(arcsec)} &\thead{DEC\\(J2000)} &\thead{DEC err\tnote{a}\\(arsec)} &\thead{Maximum\\Flux Density\\(mJy\,beam$^{-1}$)}&\thead{Time of\\Peak\\(MJD)}&\thead{Persistent\\Flux Density\\(mJy\,beam$^{-1}$)} &\thead{Rest Frame\\Peak Luminosity\\($10^{40}$\,erg\,s$^{-1}$)}&\thead{$t_{\rm 1/2, rise}$\\(days)}&\thead{$t_{\rm 1/2, decline}$\\(days)}}
\startdata 
J011148.1$-$025539&17.95&0.4&$-$2.93&0.4&6.2$\pm$1.4&58600&1.0$\pm$0.3&20$\pm$10&--&34$^{+653}_{-3}$\\
J015856.8$-$012404&29.74&0.4&$-$1.40&0.4&5.46$\pm$0.34&58859&$<$1.8&0.07$\pm$0.03&180$^{+20}_{-60}$&450$^{+80}_{-100}$\\
J093634.7$-$054755&144.14&0.4&$-$5.80&0.4&5.99$\pm$0.43&58859&2.3$\pm$0.4&5$\pm$3&10$^{+30}_{-7}$&9$^{+23}_{-4}$\\
J104315.9$+$005059&160.82&0.3&0.85&0.3&6.4$\pm$1.3&59090&0.7$\pm$0.6&0.18$\pm$0.09&90$^{+40}_{-40}$&200$^{+50}_{-30}$\\
J144848.2$+$030235&222.20&0.4&3.04&0.4&5.95$\pm$0.64&59417&1.1$\pm$0.3&2$\pm$1&300$^{+400}_{-100}$&17$^{+4}_{-4}$\\
J210626.2$-$020055&316.61&0.4&$-$2.02&0.4&5.05$\pm$0.42&59447&1.9$\pm$0.4&0.8$\pm$0.4&200$^{+200}_{-100}$&--\\
J212618.5$+$022400&321.58&0.4&2.40&0.4&5.14$\pm$0.42&58867&1.1$\pm$0.3&6$\pm$3&30$^{+10}_{-10}$&60$^{+290}_{-10}$\\
J213437.8$-$620433&323.66&0.4&$-$62.08&0.4&4.60$\pm$0.39&58880&1.5$\pm$0.3&0.04$\pm$0.02&110$^{+40}_{-40}$&80$^{+50}_{-30}$\\
J215418.2$+$002442&328.58&0.4&0.41&0.4&4.41$\pm$0.47&58867&1.0$\pm$0.3&0.6$\pm$0.3&6.2$^{+0.8}_{-4.2}$&80$^{+30}_{-20}$\\
J221936.0$+$004724&334.90&0.4&0.79&0.4&5.25$\pm$0.54&58785&1.5$\pm$0.3&2$\pm$1&100$^{+30}_{-40}$&50$^{+120}_{-40}$\\
J230053.0$-$020732&345.22&0.4&$-$2.13&0.4&4.95$\pm$0.39&58867&$<$1.7&0.4$\pm$0.2&110$^{+30}_{-20}$&482$^{+6}_{-42}$\\
J234449.6$+$015434&356.21&0.5&1.91&0.5&3.42$\pm$0.43&58859&0.8$\pm$0.2&4$\pm$2&60$^{+120}_{-20}$&100$^{+300}_{-200}$\\
\enddata 
{\raggedright $^{a}$Errors on RA and Dec are a statistical error from \emph{Selavy}. As noted in Section~\ref{sec:nuclear_class}, there is an additional uncertainty due to astrometric offset with optical surveys.}

\caption{Flare properties of the sources in our final sample. Methodology for determining persistent flux, $t_{\rm 1/2, rise}$, and $t_{\rm 1/2, decline}$ are described in Section~\ref{sec:calculate_param}. If no persistent flux density (see Section~\ref{sec:calculate_param}) is observed, the limit is determined from the minimum measured flux.\label{tab:flare_properties}} 

\end{deluxetable*}
\end{ThreePartTable}

\begin{ThreePartTable}
\begin{deluxetable*}{lcccccccccc}
\tabletypesize{\footnotesize}
\setlength\tabcolsep{2.3pt}

\tablehead{\thead{Name} &\thead{Photometric\\Redshift}&\thead{Host Galaxy\\z mag}& \thead{Offset\\(arcsec)} & \thead{Offset\\(kpc)} & \thead{Offset\\(Num $\sigma$)} & \thead{Host\\Stellar Mass\\log $\frac{M}{M_{\odot}}$} & \thead{Black Hole\\ Mass\\log $\frac{M}{M_{\odot}}$} & \thead{Radio\\SFR\\(${\rm M}_{\odot}{\rm yr}^{-1})$} & \thead{Optical\\SFR\\(${\rm M}_{\odot}{\rm yr}^{-1})$} &\thead{Faintest\\Detectable\\ZTF TDE}}
\startdata 
J011148.1$-$025539&0.8$\pm$0.4\tnote{a}&22.6\tnote{a}&0.79$\pm$0.80&21$\pm$21&1.0&11.49$^{+0.03}_{-0.40}$&8$^{+1}_{-1}$&2000$\pm$1000&40$\pm$30&No ZTF\\
J015856.8$-$012404&0.076$\pm$0.009\tnote{b}&14.9\tnote{b}&0.99$\pm$0.78&1.7$\pm$1.3&1.3&10.27$^{+0.06}_{-0.31}$&7$^{+1}_{-1}$&--&30$\pm$20&AT2020wey\\
J093634.7$-$054755&0.50$\pm$0.05\tnote{a}&19.9\tnote{a}&0.57$\pm$0.75&8$\pm$11&0.8&10.62$^{+0.05}_{-0.11}$&7$^{+1}_{-1}$&1300$\pm$700&50$\pm$40&AT2020yue\\
J104315.9$+$005059&0.107$\pm$0.005\tnote{a}&15.8\tnote{a}&0.58$\pm$0.72&1.5$\pm$1.8&0.8&11.477$^{+0.001}_{-0.004}$&8$^{+1}_{-1}$&14$\pm$8&20$\pm$10&AT2021qth\\
J144848.2$+$030235&0.35$\pm$0.02\tnote{b}&18.4\tnote{b}&1.09$\pm$0.76&10.3$\pm$7.2&1.4&10.74$^{+0.12}_{-0.05}$&7$^{+1}_{-1}$&300$\pm$100&40$\pm$30&AT2020yue\\
J210626.2$-$020055&0.2$\pm$0.2\tnote{b}&20.2\tnote{b}&0.54$\pm$0.76&3.3$\pm$4.7&0.7&9.65$^{+0.04}_{-0.05}$&6$^{+1}_{-1}$&200$\pm$100&6$\pm$4&No ZTF\\
J212618.5$+$022400&0.58$\pm$0.04\tnote{b}&18.9\tnote{b}&1.03$\pm$0.75&17$\pm$13&1.4&11.4$^{+0.6}_{-0.6}$&8$^{+1}_{-1}$&900$\pm$500&90$\pm$70&None\\
J213437.8$-$620433&0.059$\pm$0.005\tnote{c}&13.5\tnote{c}&0.49$\pm$0.83&0.7$\pm$1.1&0.6&10.3$^{+0.1}_{-0.3}$&7$^{+1}_{-1}$&8$\pm$4&90$\pm$70&No ZTF\\
J215418.2$+$002442&0.23$\pm$0.04\tnote{a}&17.8\tnote{a}&0.42$\pm$0.80&2.4$\pm$4.6&0.5&11.1$^{+0.2}_{-0.2}$&8$^{+1}_{-1}$&100$\pm$50&20$\pm$20&AT2020yue\\
J221936.0$+$004724&0.39$\pm$0.04\tnote{a}&18.0\tnote{a}&0.58$\pm$0.78&6.1$\pm$8.3&0.7&11.9430$^{+0.0003}_{-0.0003}$&8$^{+1}_{-1}$&500$\pm$300&60$\pm$50&AT2020yue\\
J230053.0$-$020732&0.18$\pm$0.03\tnote{b}&17.8\tnote{b}&0.63$\pm$0.75&2.8$\pm$3.4&0.8&10.9$^{+0.2}_{-0.1}$&7$^{+1}_{-1}$&--&12$\pm$9&AT2021qth\\
J234449.6$+$015434&0.59$\pm$0.09\tnote{b}&20.4\tnote{b}&0.55$\pm$0.83&9$\pm$14&0.7&10.3$^{+0.2}_{-0.8}$&7$^{+1}_{-1}$&600$\pm$400&50$\pm$40&None\\
\enddata 
{\raggedright $^a$ Data from Pan-STARRS $^b$ Data from SDSS $^c$ Data from DES}

\tablewidth{0pt}
\setlength\tabcolsep{2.5pt}
\renewcommand{\arraystretch}{2}

\caption{Host galaxy properties of the sources in our final sample. Calculation of the offset from the optical center of the host galaxy is described in Section~\ref{sec:coincidence}. Calculation of the radio and optical SFR, the host stellar mass, and the mass of the SMBH described in Section~\ref{sec:host_gal_prop}. The faintest detectable ZTF TDE is described in Section~\ref{sec:optical}. \label{tab:host_properties}}
\end{deluxetable*}
\end{ThreePartTable}
\clearpage
\newpage

\appendix

\setlength{\belowcaptionskip}{-10pt}
\setlength{\abovecaptionskip}{-3pt}

\section{Lightcurves and Optical Images}\label{appendix:lc_and_hosts}
For each source in our final sample, the lightcurve of integrated fluxes is shown in Figures~\ref{fig:J011148.1} to \ref{fig:J234449.6}, with blue points indicating fluxes measured by \emph{Selavy}. For sources not detected by \emph{Selavy} at certain epochs, the flux density at that position is estimated using a forced Gaussian fit. We visually inspected the sources at these epochs to distinguish between a true flux density estimate and an upper limit, i.e., where no source was detectable. We consider flux density estimates from forced photometry of less than 5\,$\sigma$ to be upper limits. For sources with only an upper limit at certain epochs, the variability criteria described in Section~\ref{sec:radio_var} and the lightcurve morphology criteria described in Section~\ref{sec:lc_morph} use upper limits as proxies for the flux density values. Yellow points and red arrows indicate forced photometry, where the flux density is estimated above or below our criteria of $<$5$\sigma$ respectively. The inferred persistent flux density and its error are shown in green. The linearly interpolated lightcurve is shown as a dotted line. The crossing times delimiting the duration of the primary flare, defined as the times when the flux density passes 50\% of the peak flux density as measured from the persistent flux, are shown in purple. Also shown on the right are the optical images of the host galaxies, with a blue point indicating the centroid of the optical host and a green cross indicating the VAST position with errors (see Section~\ref{sec:coincidence}).

\begin{figure*}[h] 
\centering
\includegraphics[width=\textwidth,keepaspectratio]{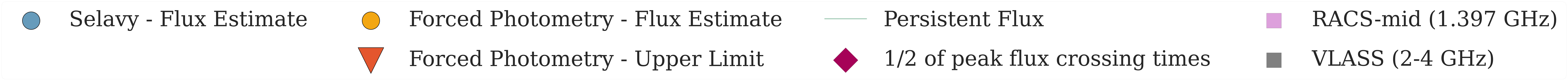}
\label{legend}
\end{figure*}

\begin{figure*}[h] 
\centering
\begin{minipage}{0.48\textwidth}
\centering
\includegraphics[height=5cm,keepaspectratio]{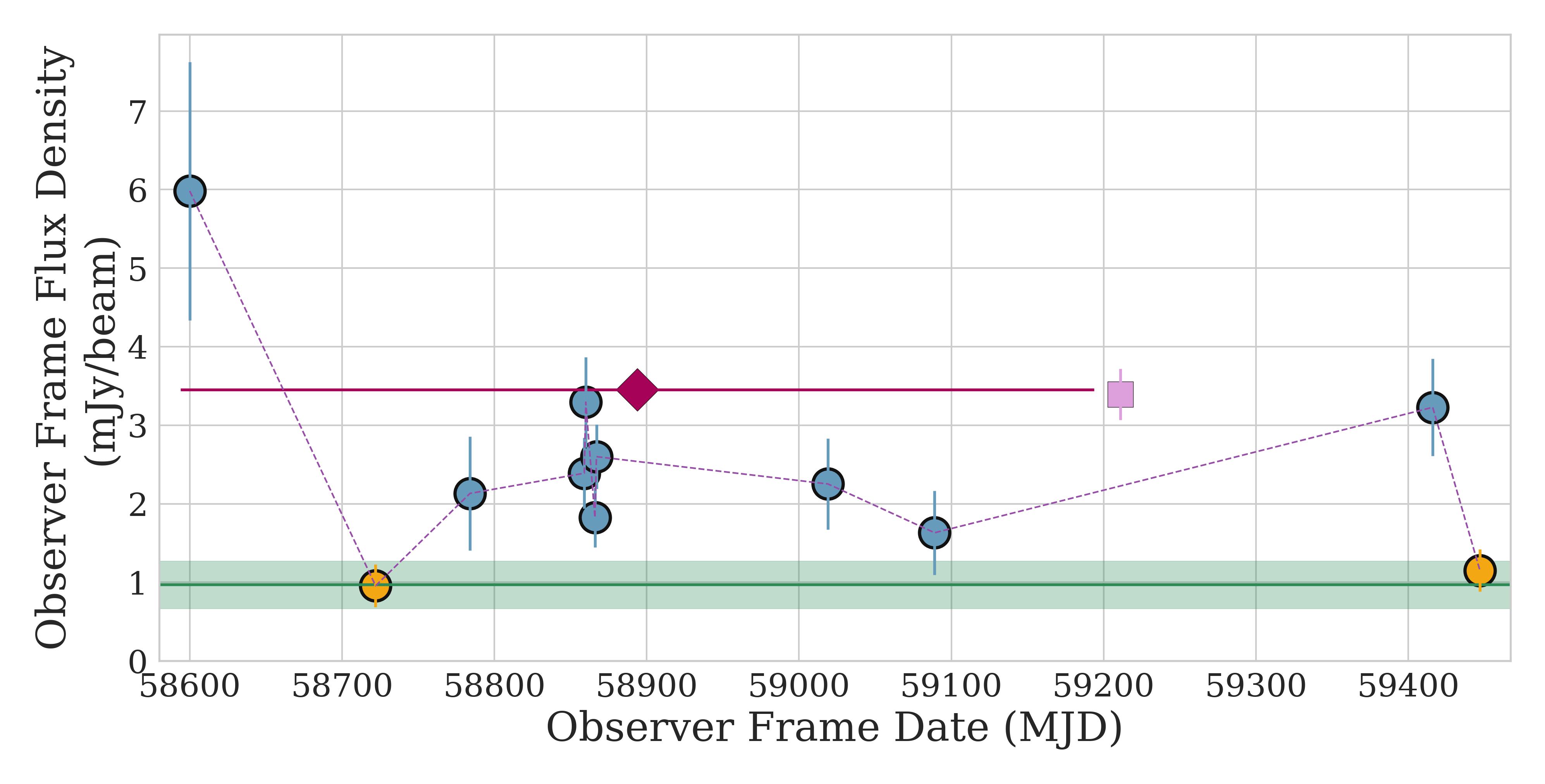}
\end{minipage}\hfil
\begin{minipage}{0.48\textwidth}
\centering
\includegraphics[height=5cm,keepaspectratio]{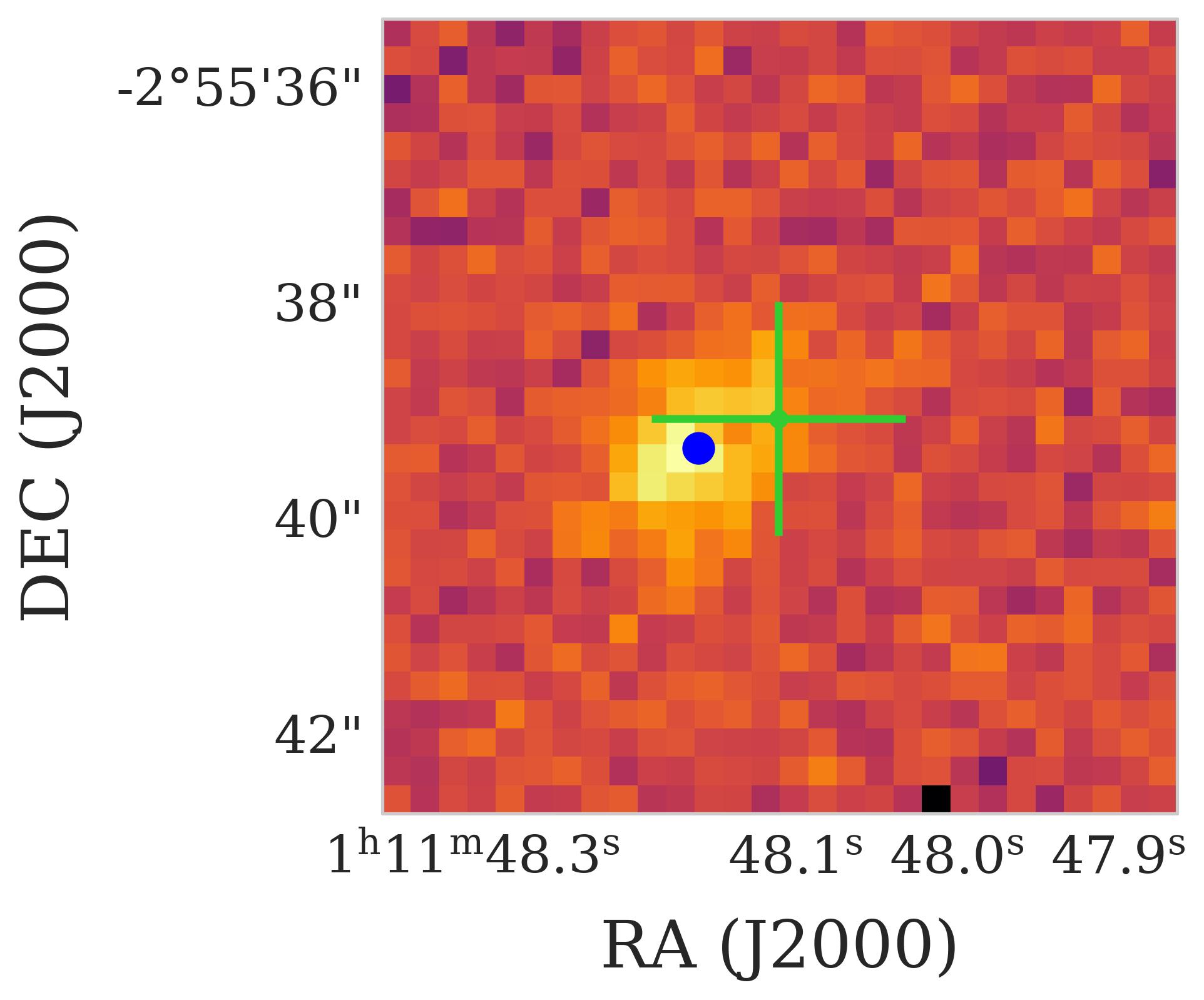}
\end{minipage}\hfil
\medskip
\caption{VAST J011148.1$-$025539}
\label{fig:J011148.1}
\end{figure*}

\begin{figure*}[h] 
\centering
\begin{minipage}{0.48\textwidth}
\centering
\includegraphics[height=5cm,keepaspectratio]{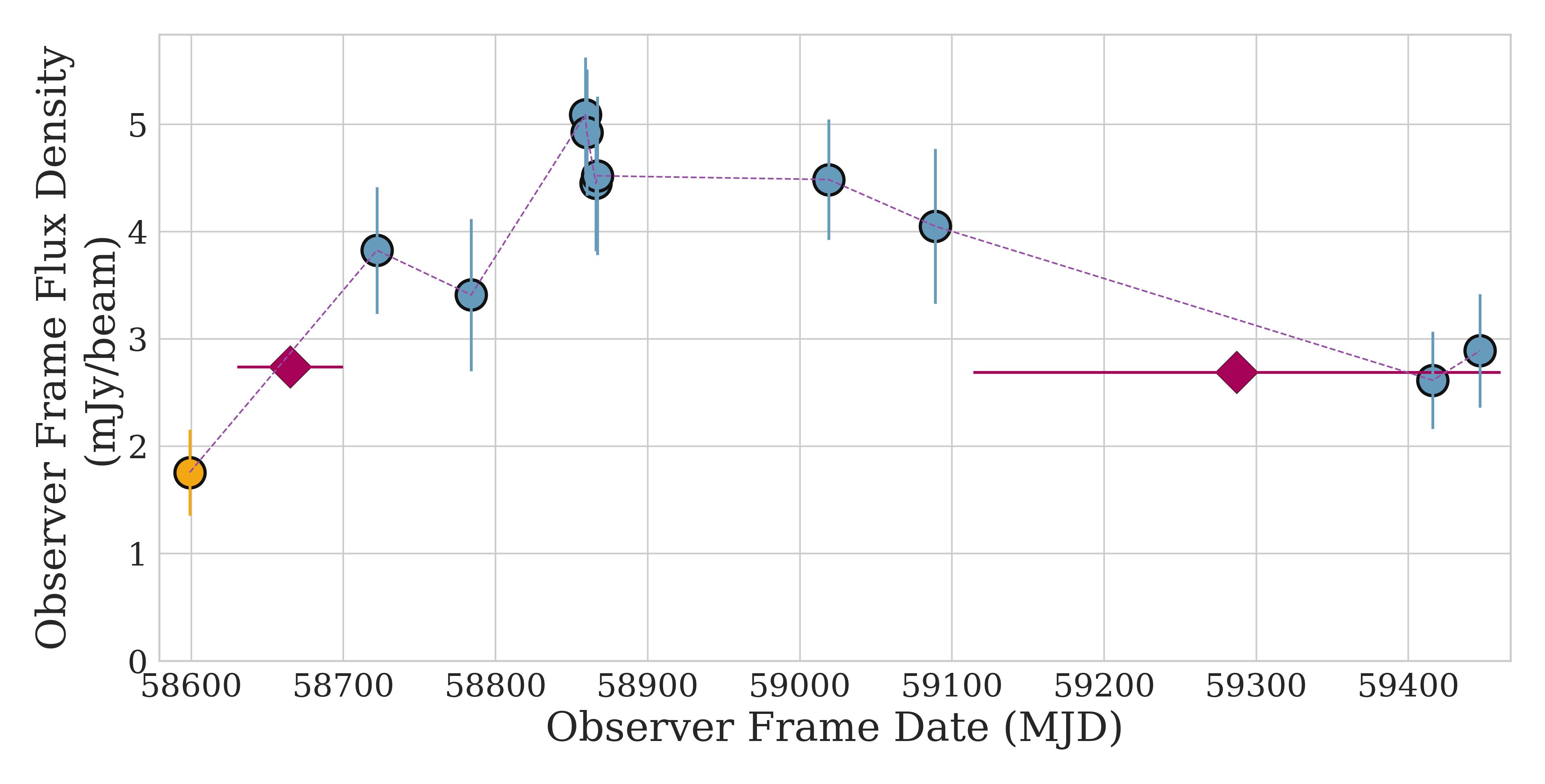}
\end{minipage}\hfil
\begin{minipage}{0.48\textwidth}
\centering
\includegraphics[height=5cm,keepaspectratio]{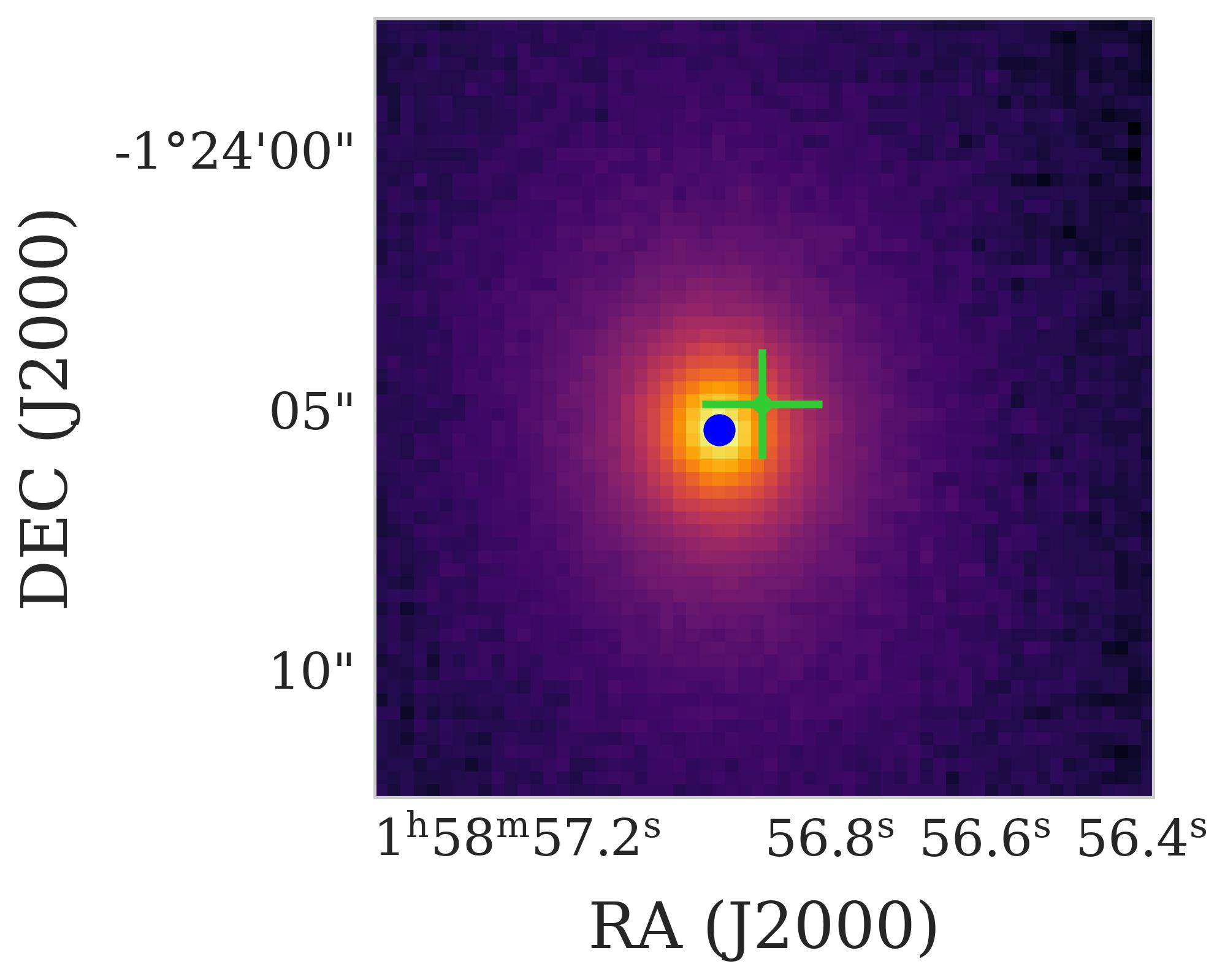}
\end{minipage}\hfil
\medskip
\caption{VAST J015856.8$-$012404}
\label{fig:J015856.8}
\end{figure*}

\begin{figure*}[h] 
\centering
\begin{minipage}{0.48\textwidth}
\centering
\includegraphics[height=5cm,keepaspectratio]{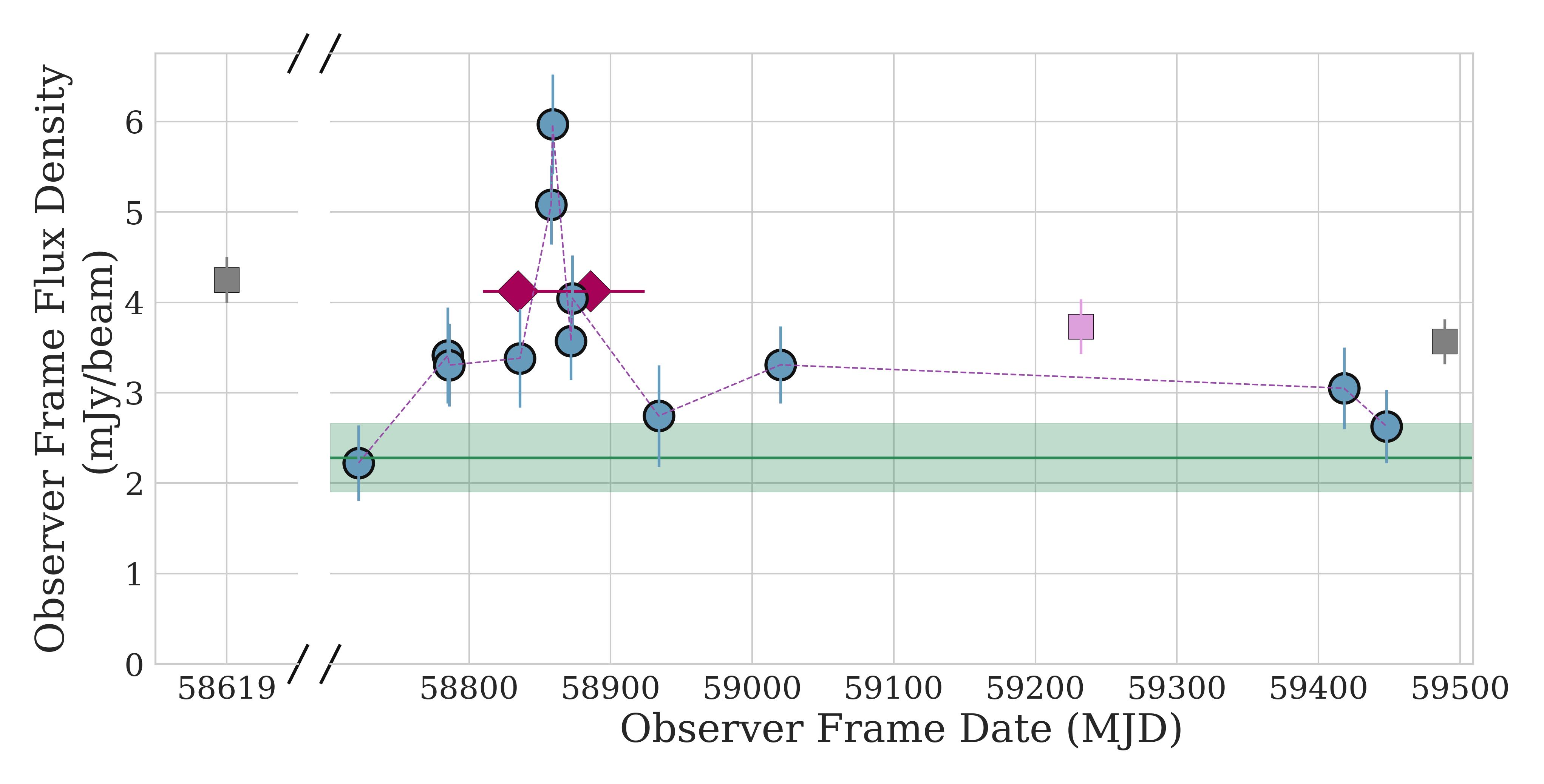}
\end{minipage}\hfil
\begin{minipage}{0.48\textwidth}
\centering
\includegraphics[height=5cm,keepaspectratio]{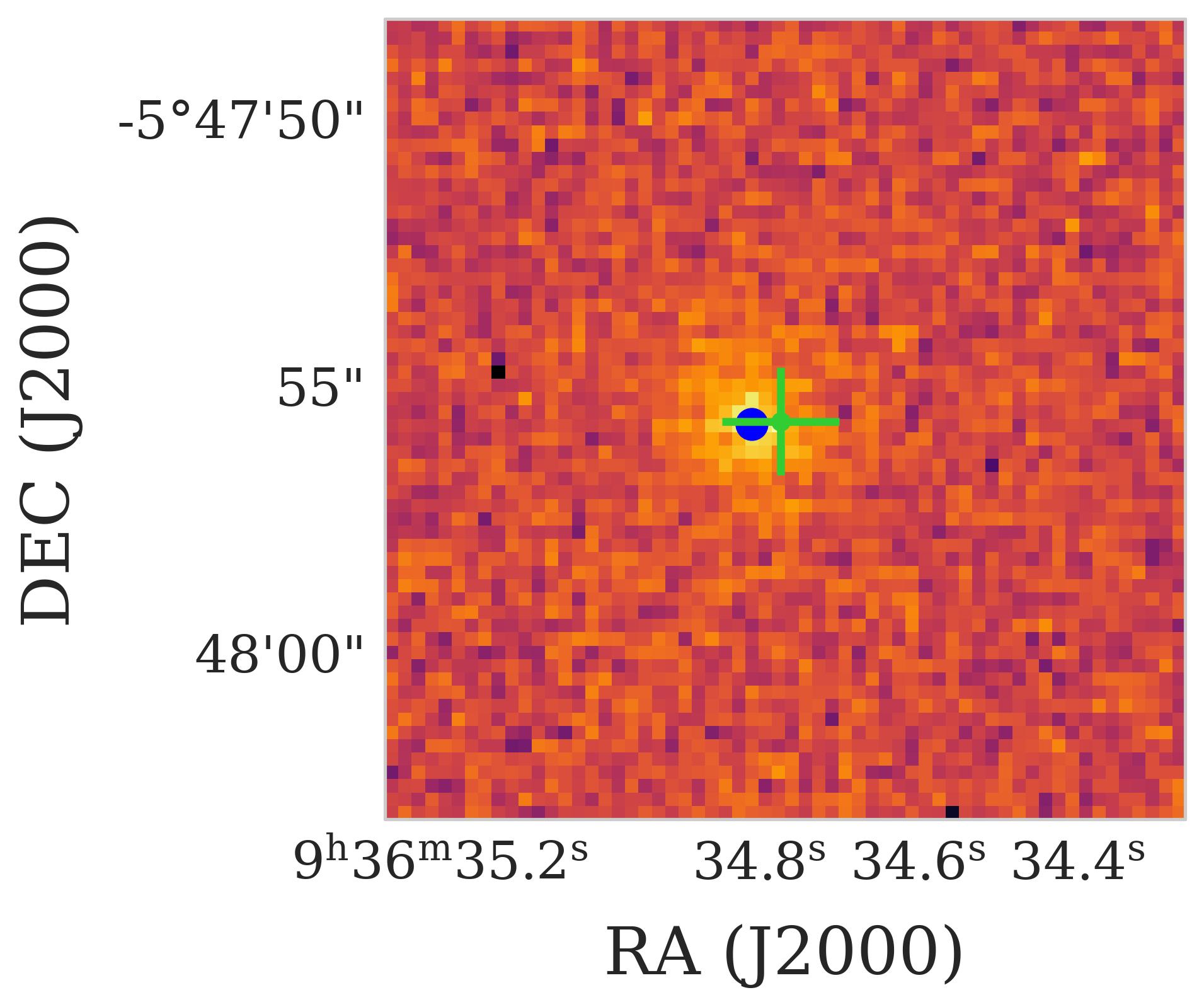}
\end{minipage}\hfil
\medskip
\caption{VAST J093634.7$-$054755}
\label{fig:J093634.7}
\end{figure*}

\begin{figure*}[h] 
\centering
\begin{minipage}{0.48\textwidth}
\centering
\includegraphics[height=5cm,keepaspectratio]{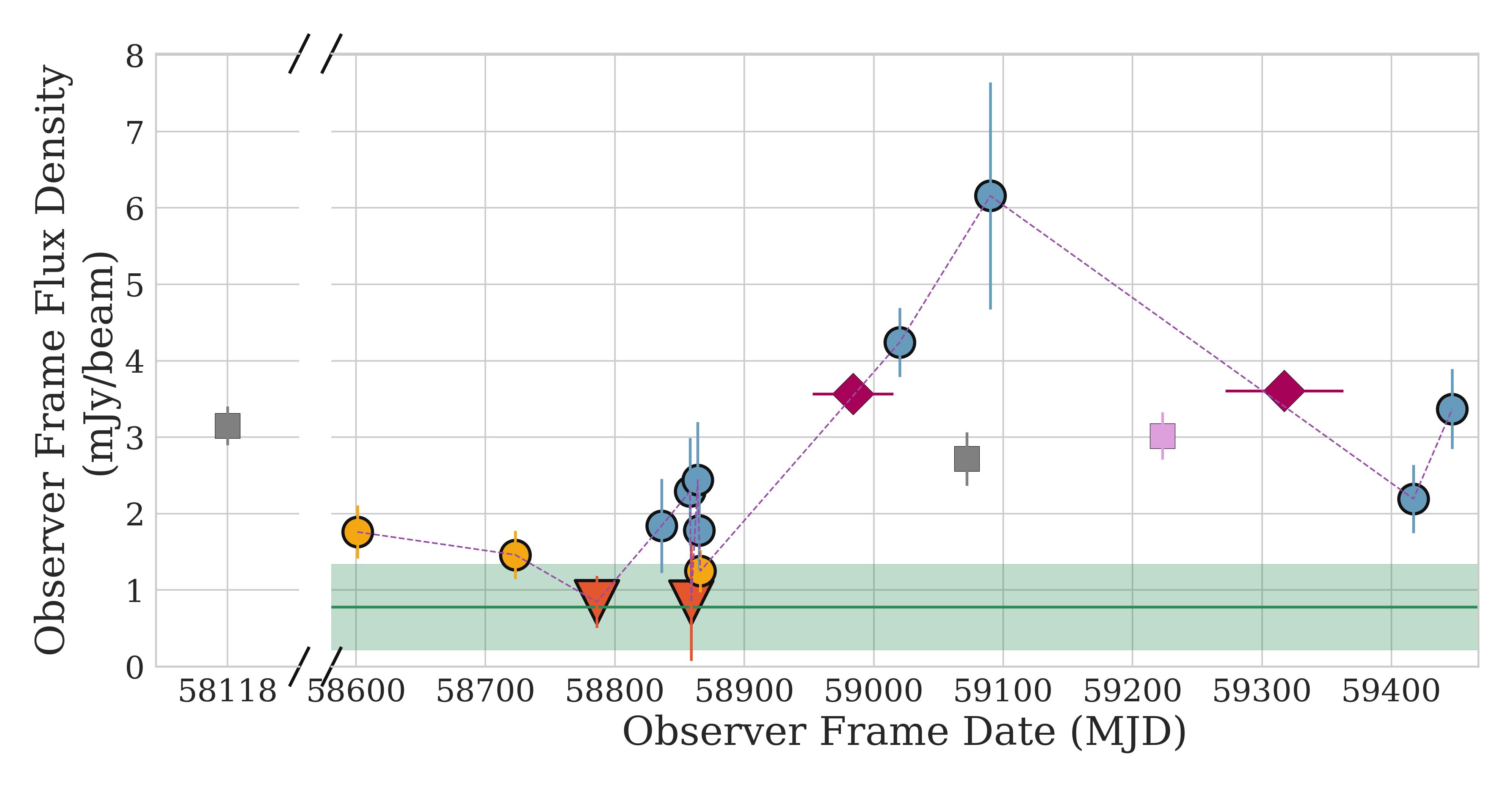}
\end{minipage}\hfil
\begin{minipage}{0.48\textwidth}
\centering
\includegraphics[height=5cm,keepaspectratio]{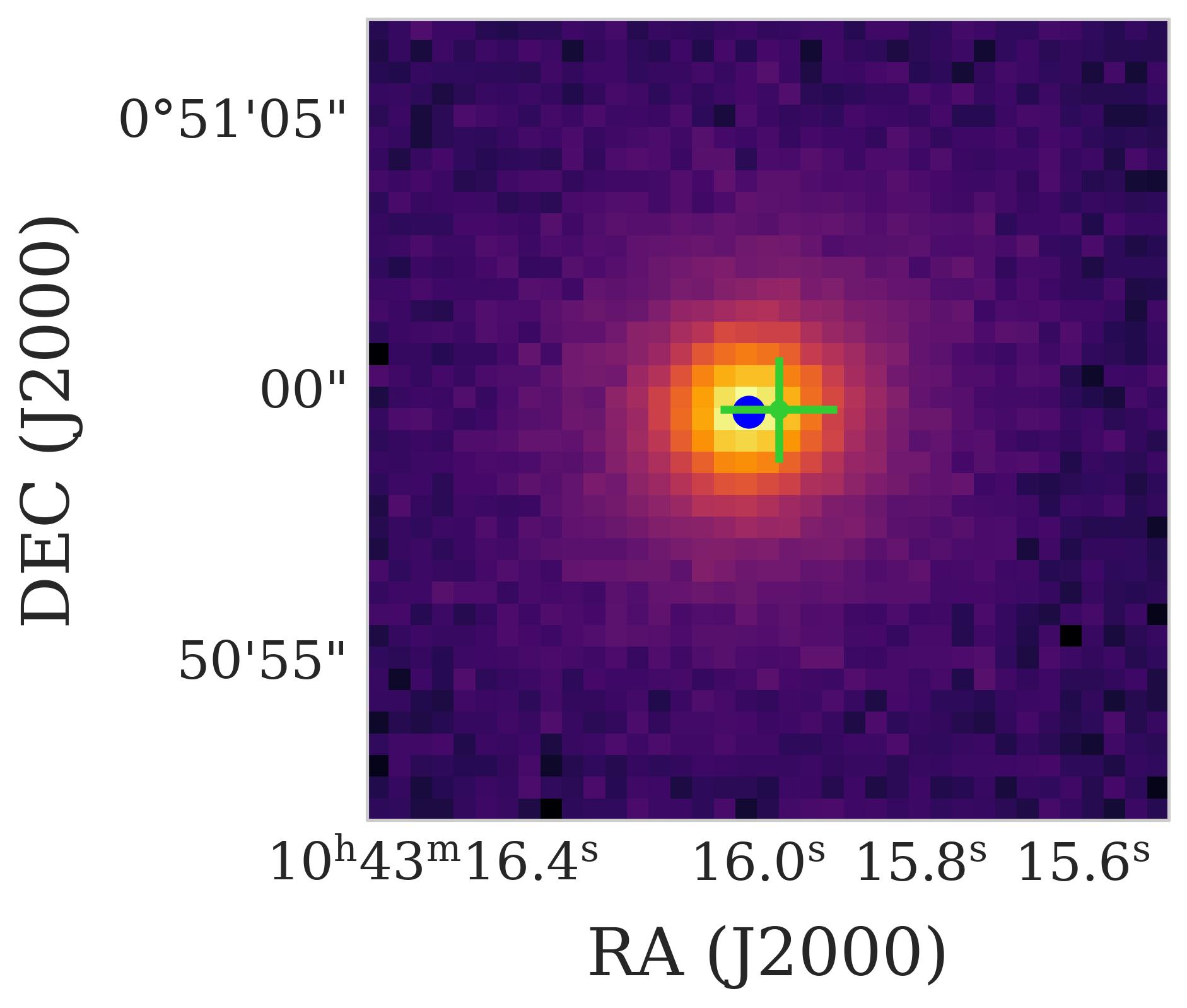}
\end{minipage}\hfil
\medskip
\caption{VAST J104315.9$+$005059}
\label{fig:J104315.9}
\end{figure*}

\begin{figure*}[h] 
\centering
\begin{minipage}{0.48\textwidth}
\centering
\includegraphics[height=5cm,keepaspectratio]{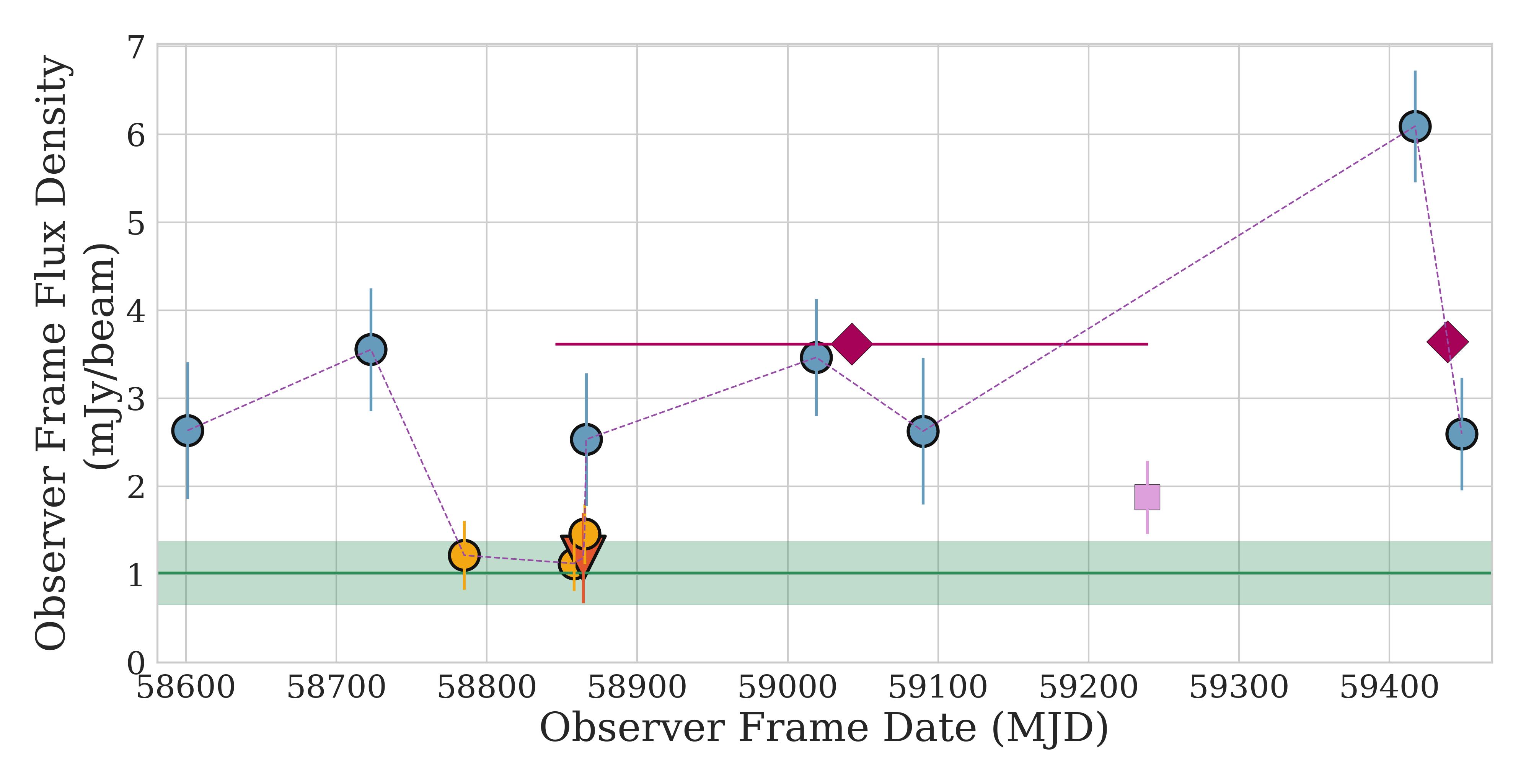}
\end{minipage}\hfil
\begin{minipage}{0.48\textwidth}
\centering
\includegraphics[height=5cm,keepaspectratio]{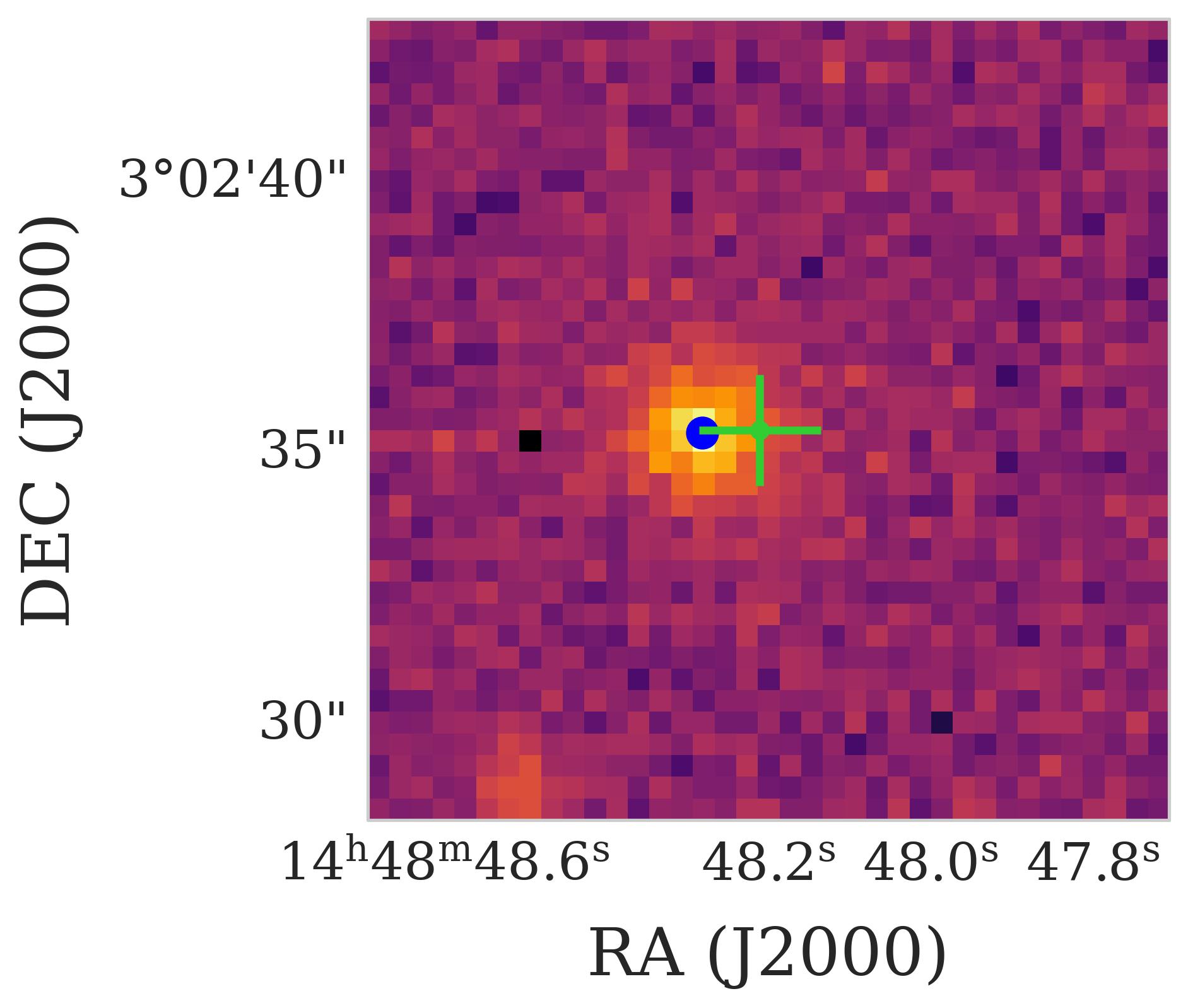}
\end{minipage}\hfil
\medskip
\caption{VAST J144848.2$+$030235}
\label{fig:J144848.2}
\end{figure*}

\begin{figure*}[h] 
\centering
\begin{minipage}{0.48\textwidth}
\centering
\includegraphics[height=5cm,keepaspectratio]{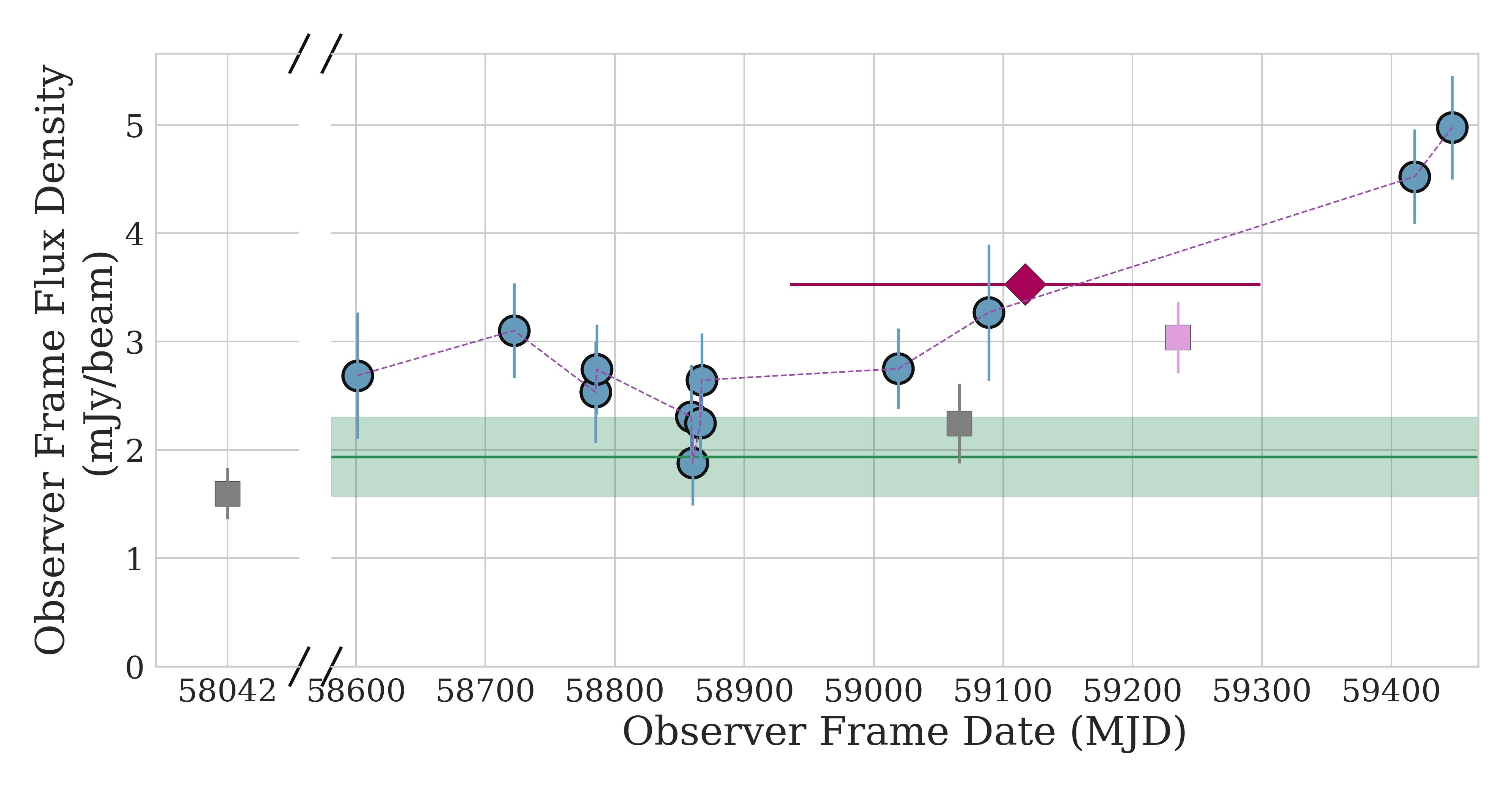}
\end{minipage}\hfil
\begin{minipage}{0.48\textwidth}
\centering
\includegraphics[height=5cm,keepaspectratio]{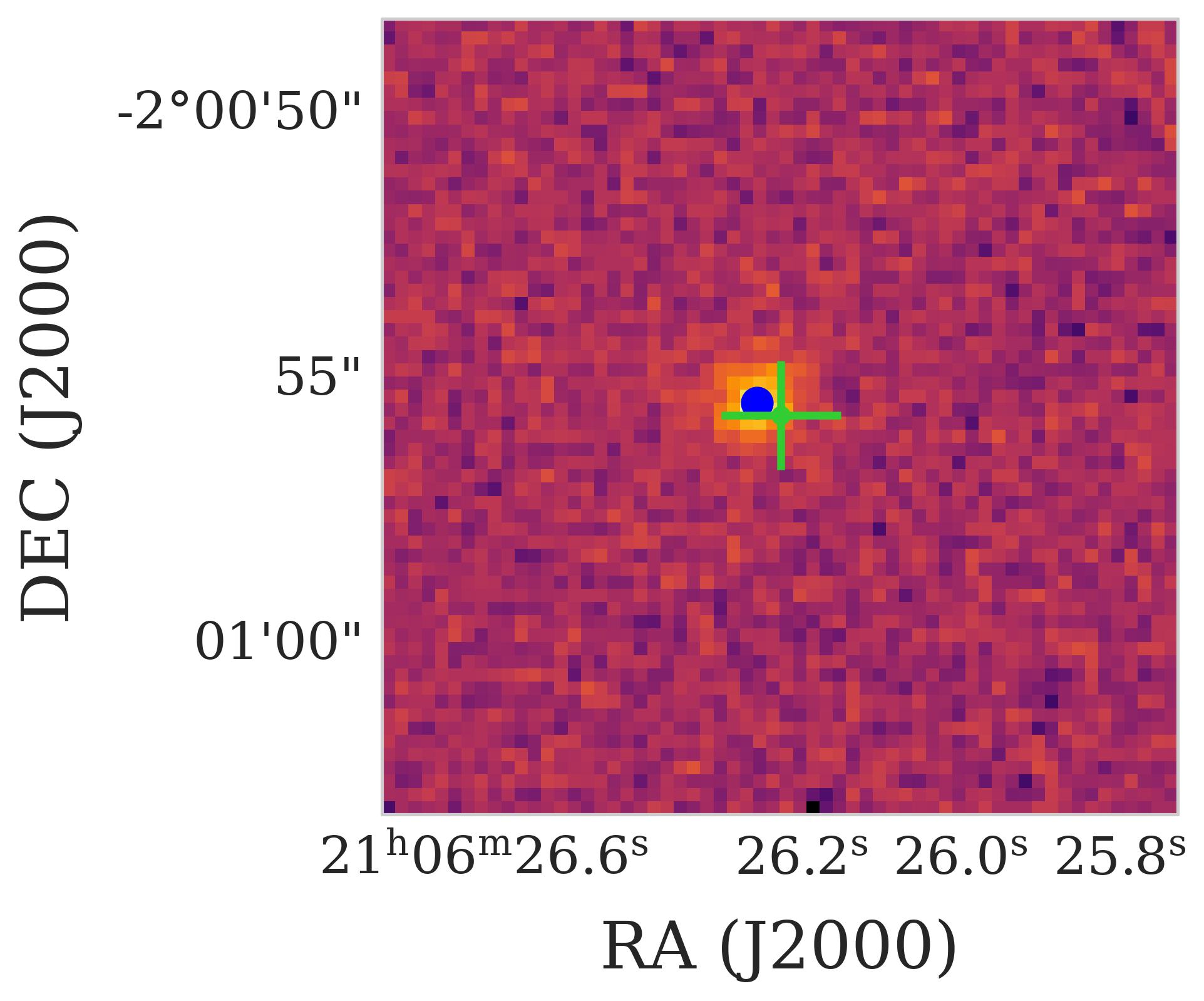}
\end{minipage}\hfil
\medskip
\caption{VAST J210626.2$-$020055}
\label{fig:J210626.2}
\end{figure*}

\begin{figure*}[h] 
\centering
\begin{minipage}{0.48\textwidth}
\centering
\includegraphics[height=5cm,keepaspectratio]{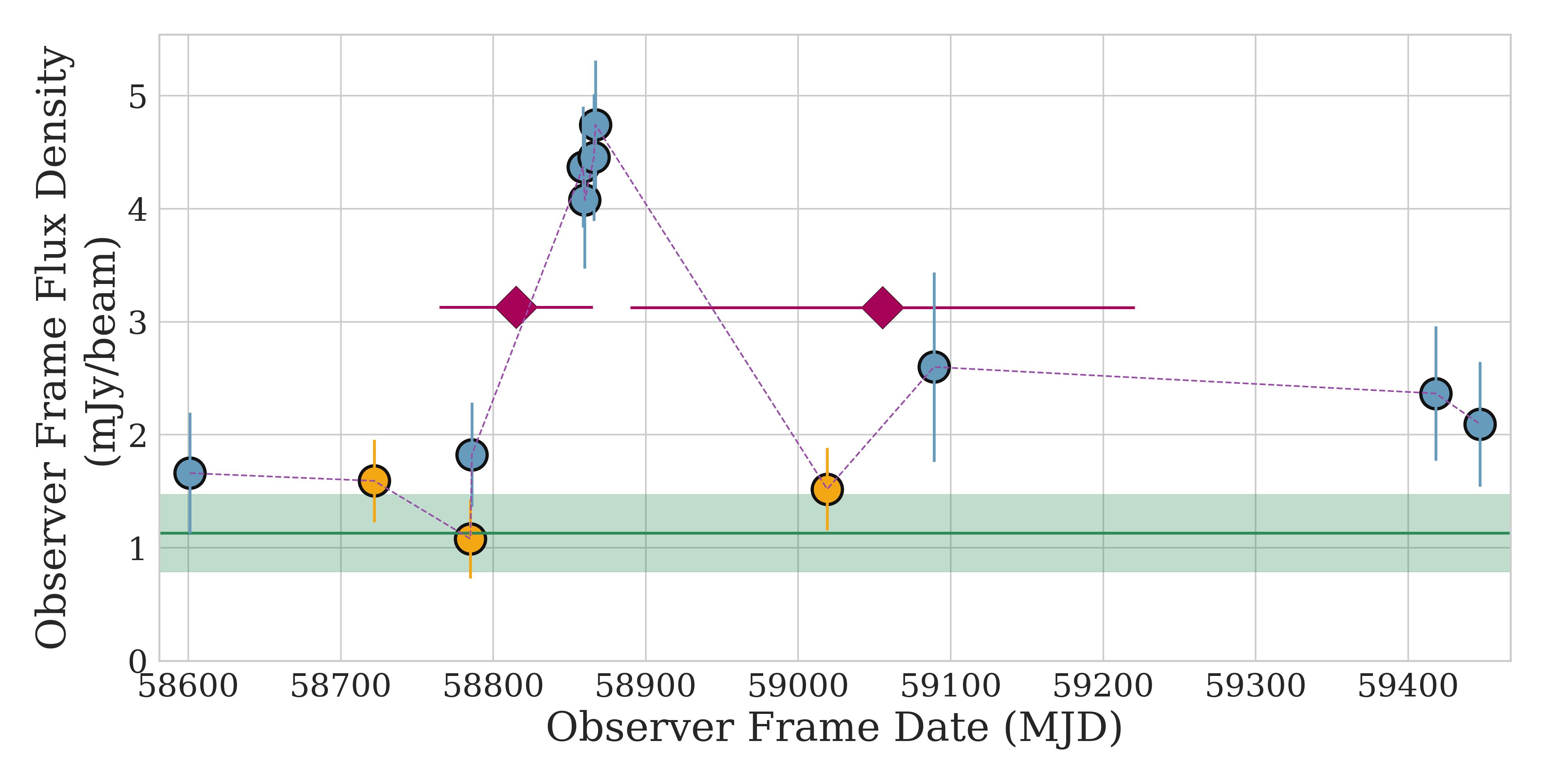}
\end{minipage}\hfil
\begin{minipage}{0.48\textwidth}
\centering
\includegraphics[height=5cm,keepaspectratio]{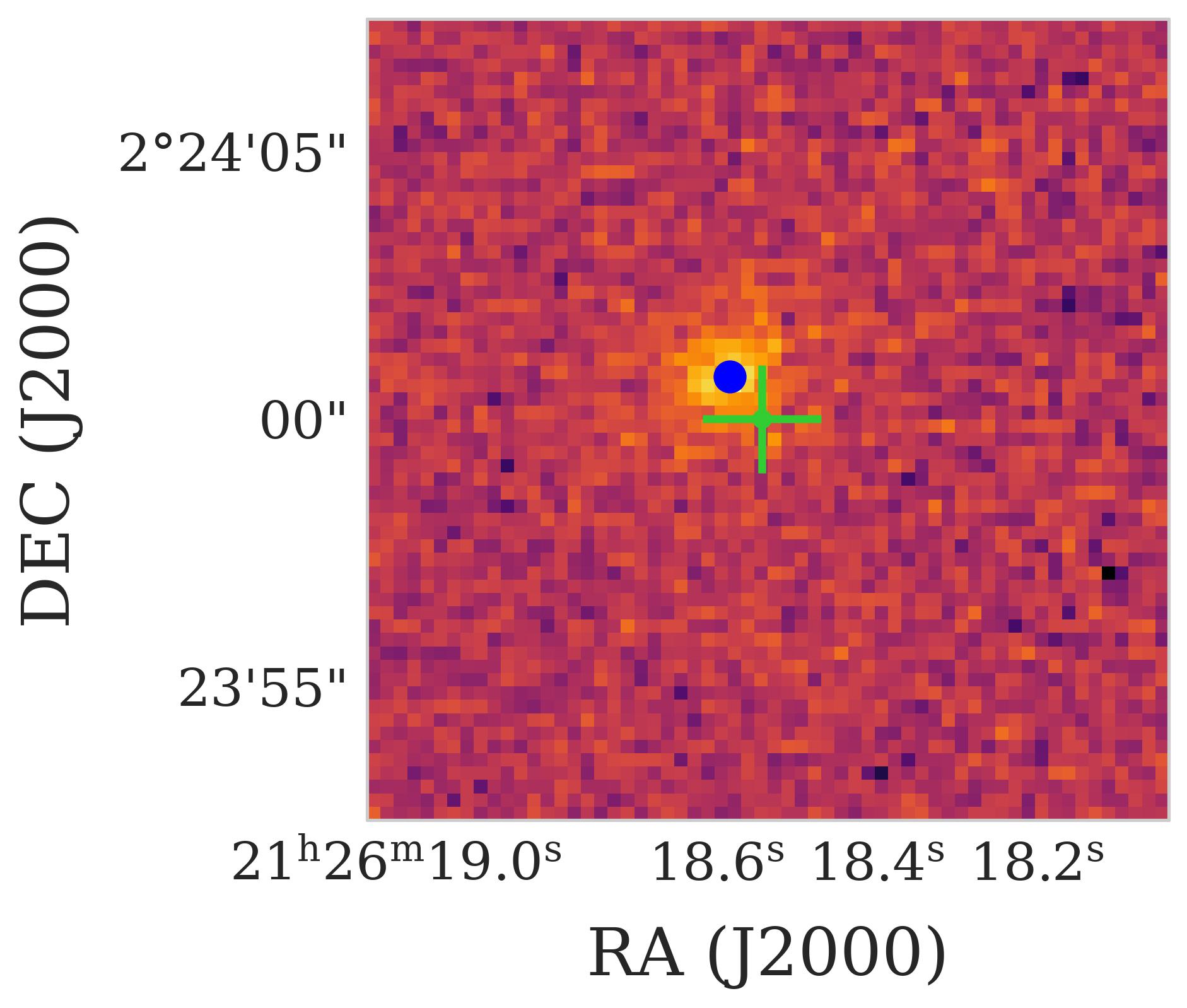}
\end{minipage}\hfil
\medskip
\caption{VAST J212618.5$+$022400}
\label{fig:J212618.5}
\end{figure*}

\begin{figure*}[h] 
\centering
\begin{minipage}{0.48\textwidth}
\centering
\includegraphics[height=5cm,keepaspectratio]{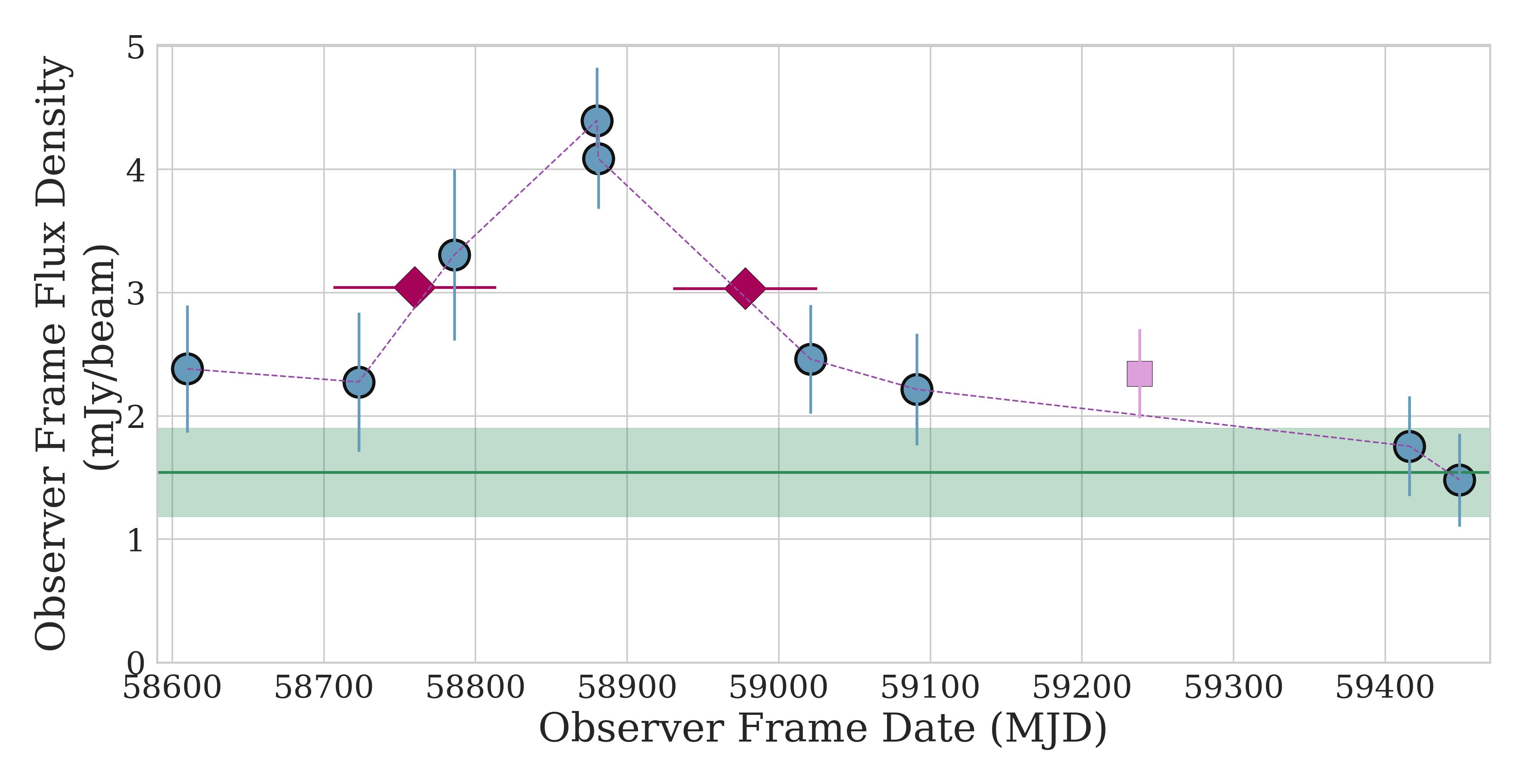}
\end{minipage}\hfil
\begin{minipage}{0.48\textwidth}
\centering
\includegraphics[height=5cm,keepaspectratio]{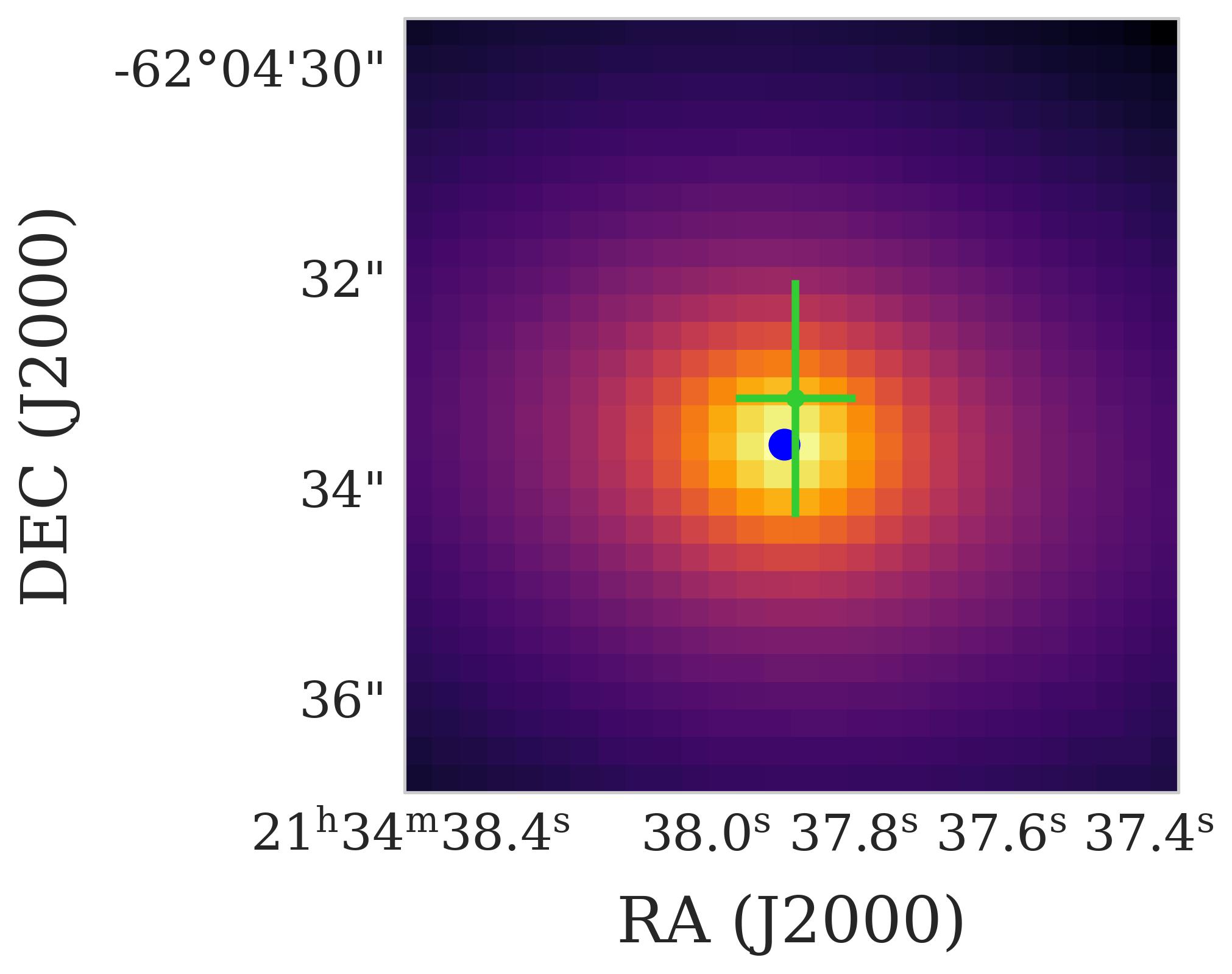}
\end{minipage}\hfil
\medskip
\caption{VAST J213437.8$-$620433}\label{fig:lc_weird_SF}
\end{figure*}

\begin{figure*}[h] 
\centering
\begin{minipage}{0.48\textwidth}
\centering
\includegraphics[height=5cm,keepaspectratio]{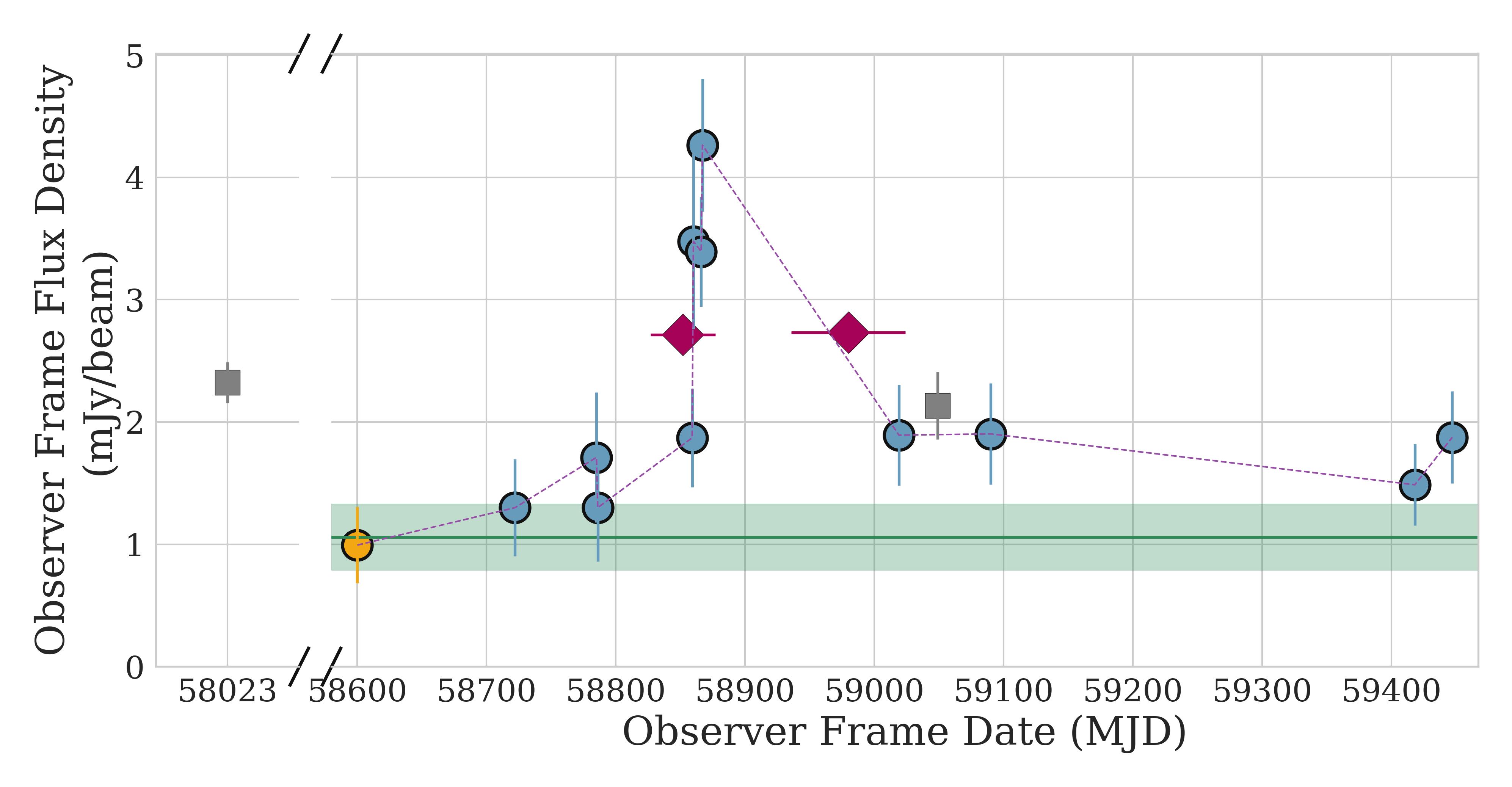}
\end{minipage}\hfil
\begin{minipage}{0.48\textwidth}
\centering
\includegraphics[height=5cm,keepaspectratio]{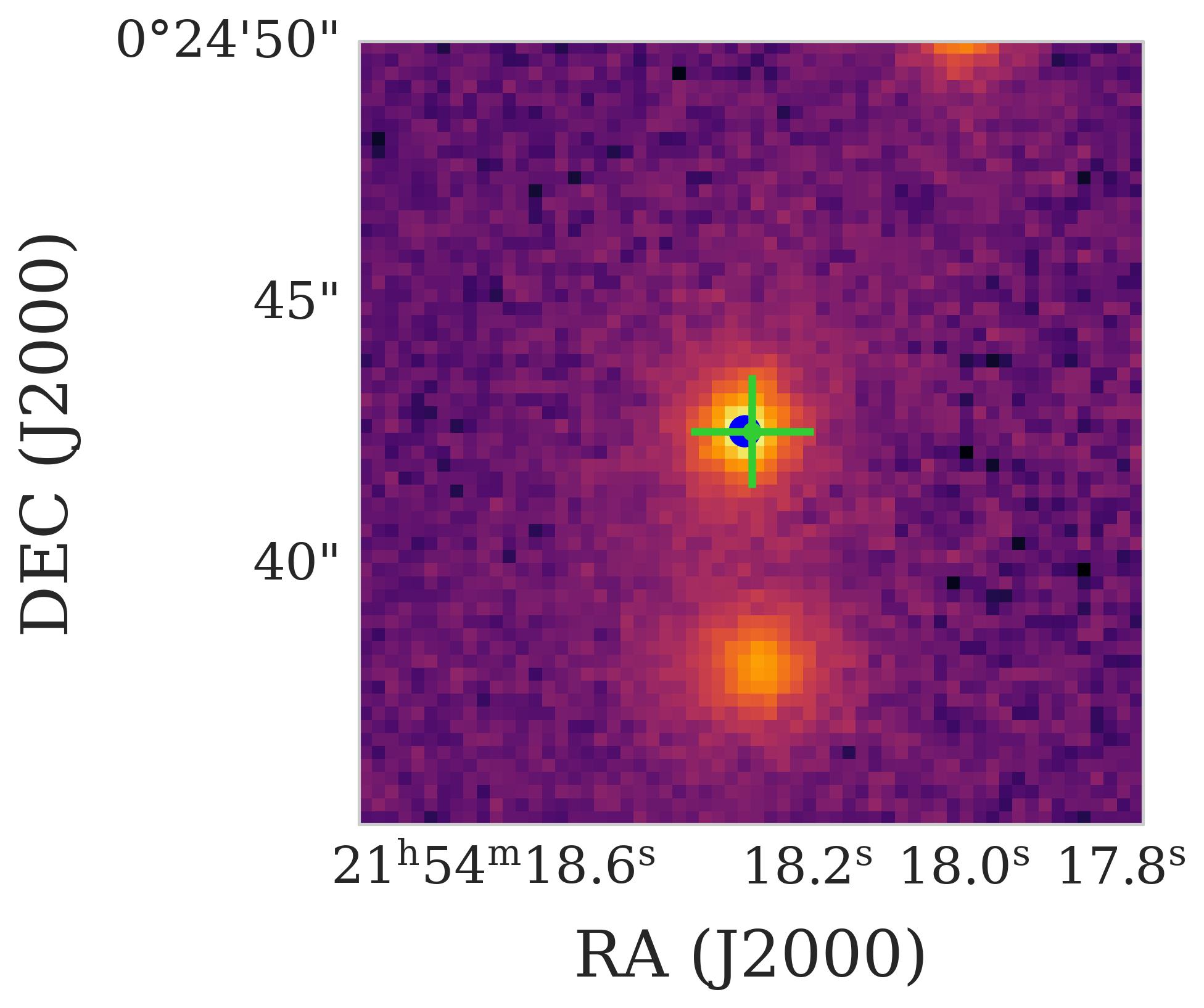}
\end{minipage}\hfil
\medskip
\caption{VAST J215418.2$+$002442}
\label{fig:J215418.2}
\end{figure*}

\begin{figure*}[h] 
\centering
\begin{minipage}{0.48\textwidth}
\centering
\includegraphics[height=5cm,keepaspectratio]{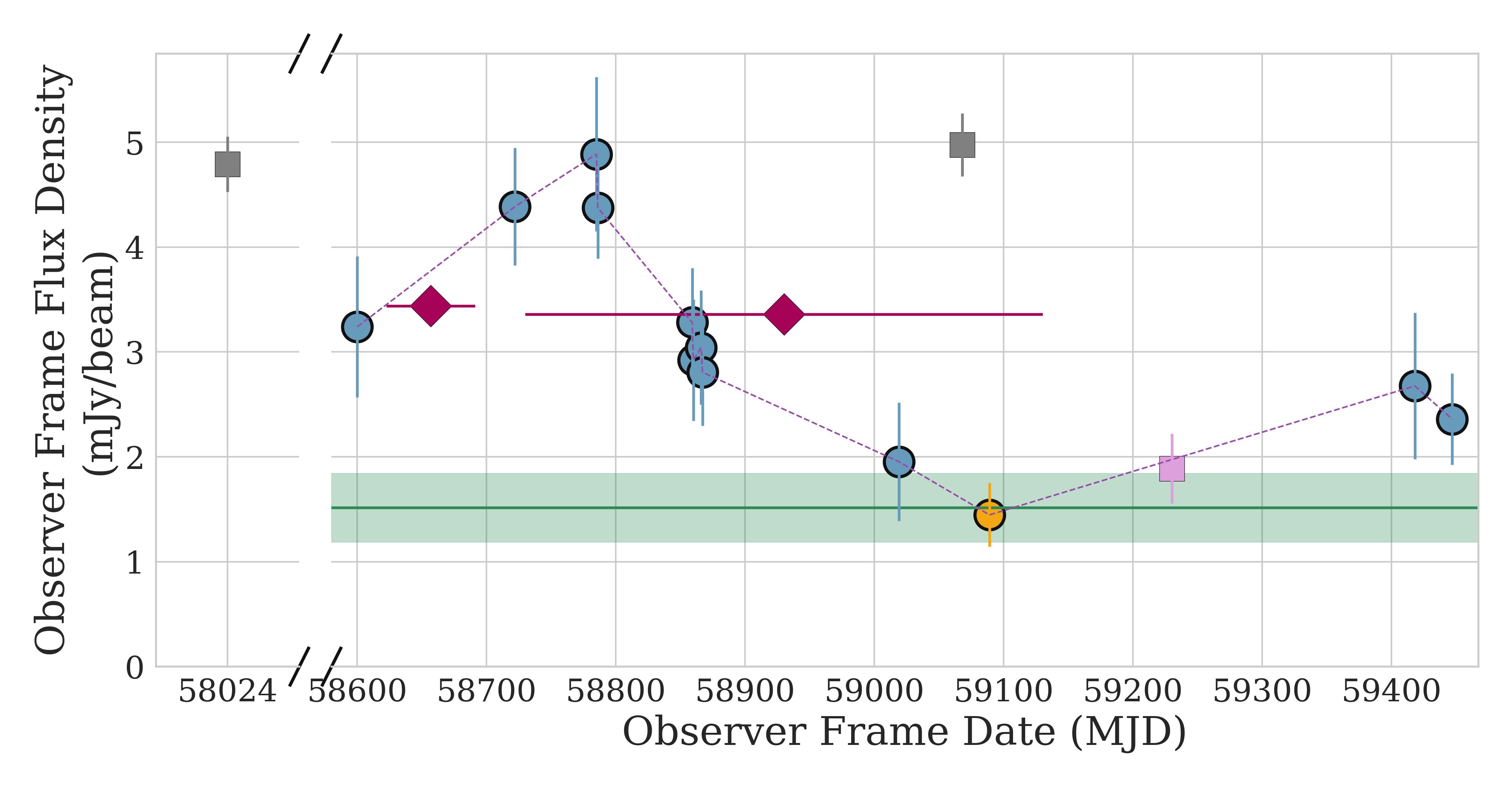}
\end{minipage}\hfil
\begin{minipage}{0.48\textwidth}
\centering
\includegraphics[height=5cm,keepaspectratio]{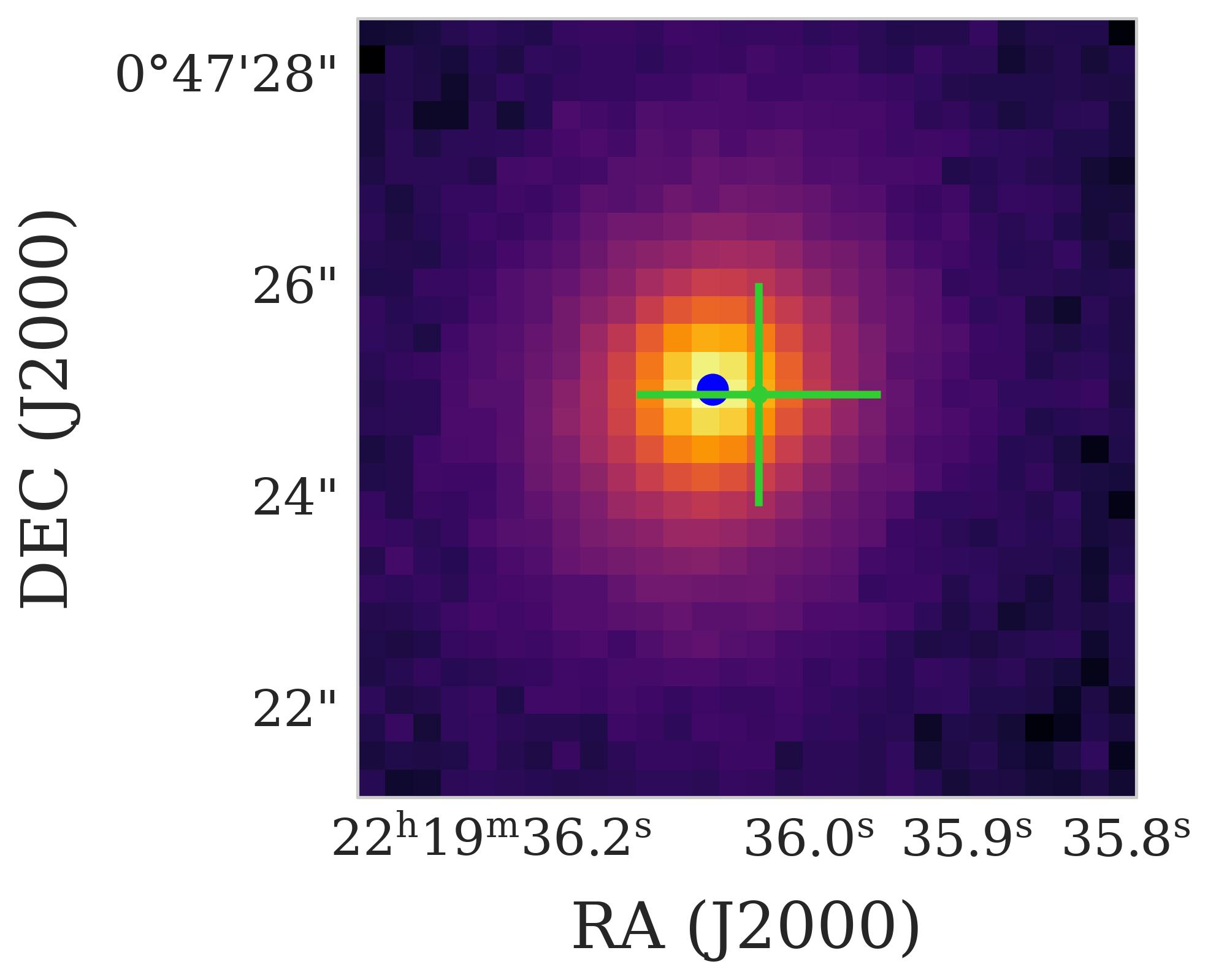}
\end{minipage}\hfil
\medskip
\caption{VAST J221936.0$+$004724}
\label{fig:J221936.0}
\end{figure*}

\begin{figure*}[h] 
\centering
\begin{minipage}{0.48\textwidth}
\centering
\includegraphics[height=5cm,keepaspectratio]{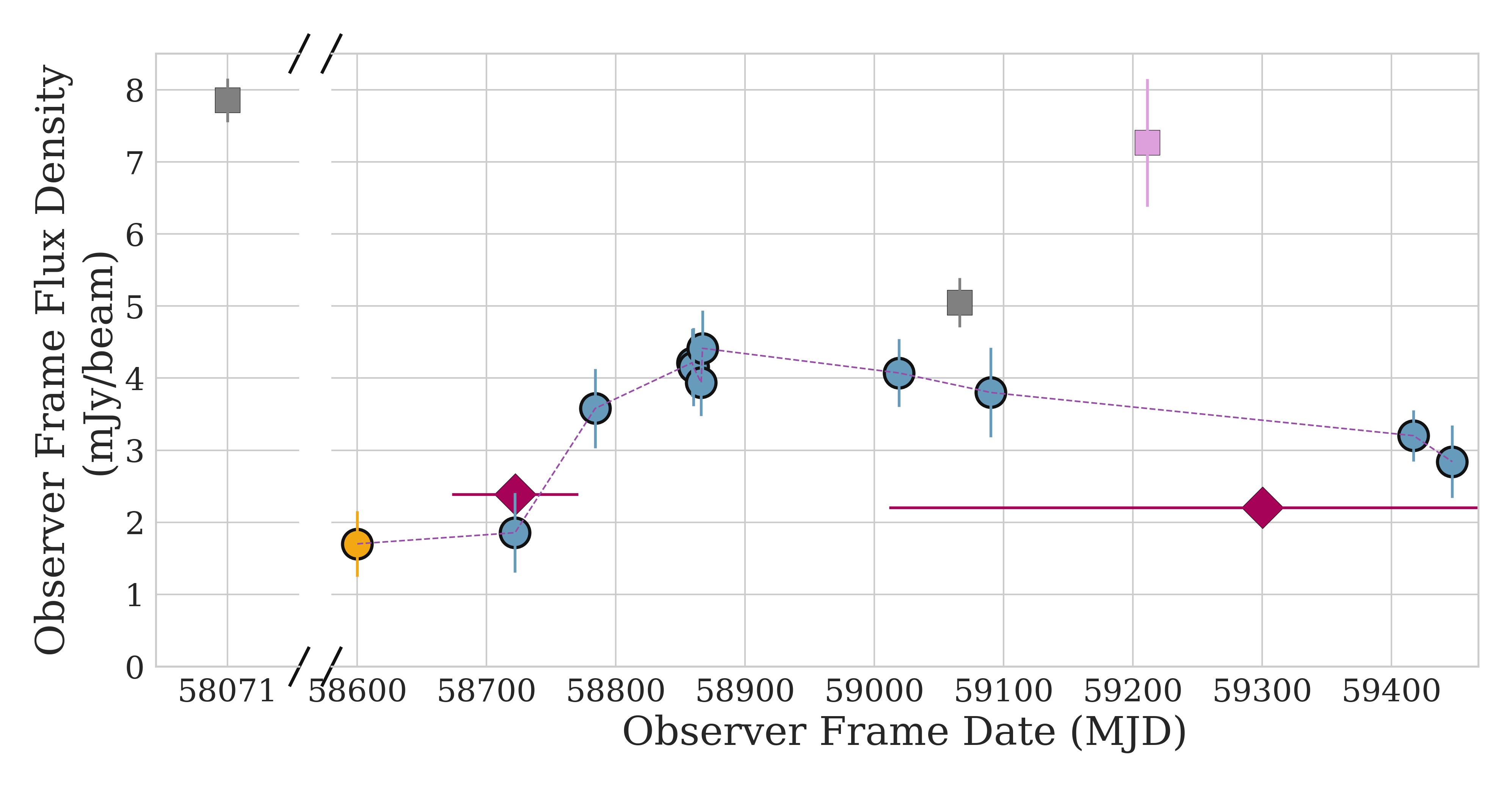}
\end{minipage}\hfil
\begin{minipage}{0.48\textwidth}
\centering
\includegraphics[height=5cm,keepaspectratio]{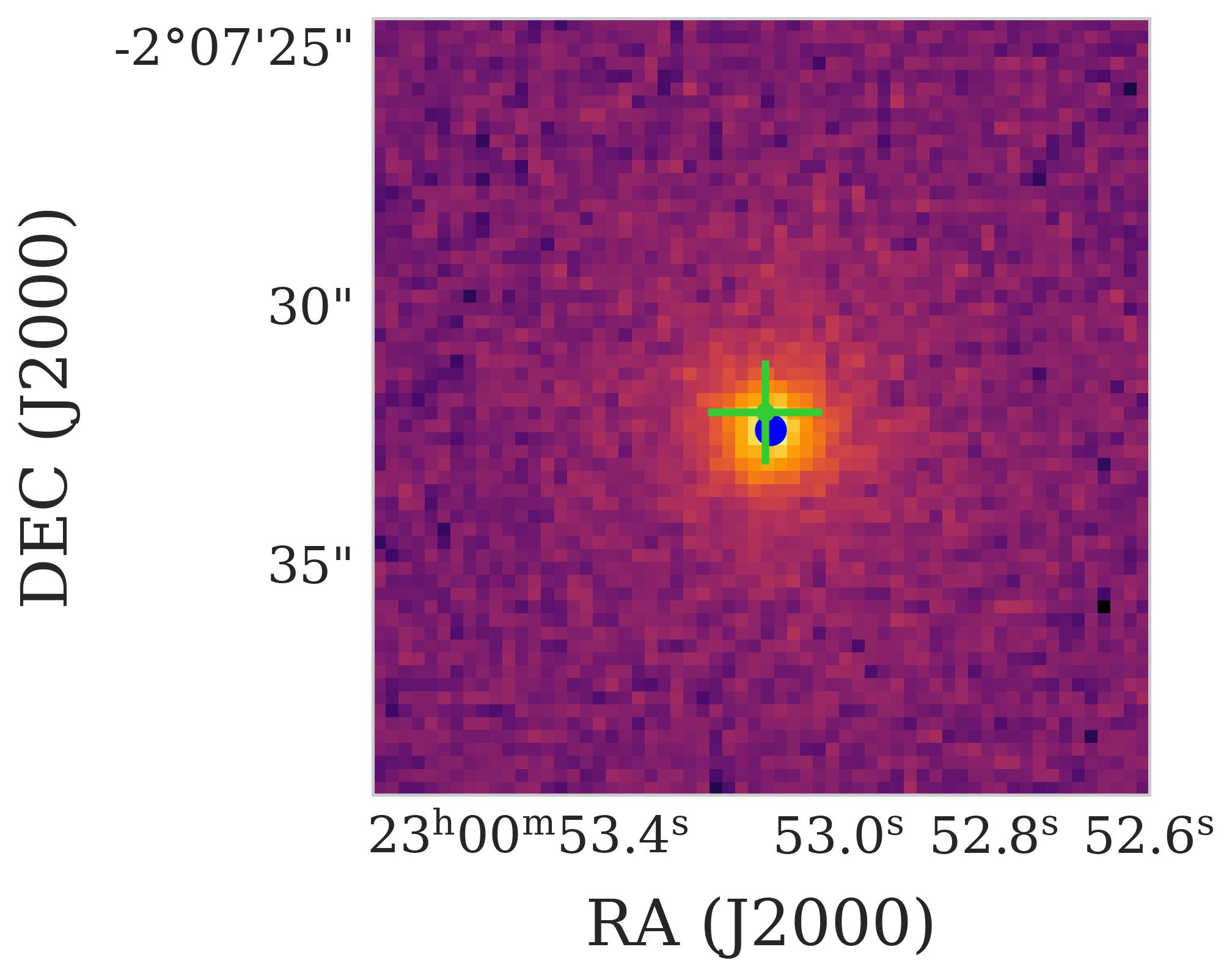}
\end{minipage}\hfil
\medskip
\caption{VAST J230053.0$-$020732}
\label{fig:J230053.0}
\end{figure*}

\begin{figure*}[h] 
\centering
\begin{minipage}{0.48\textwidth}
\centering
\includegraphics[height=5cm,keepaspectratio]{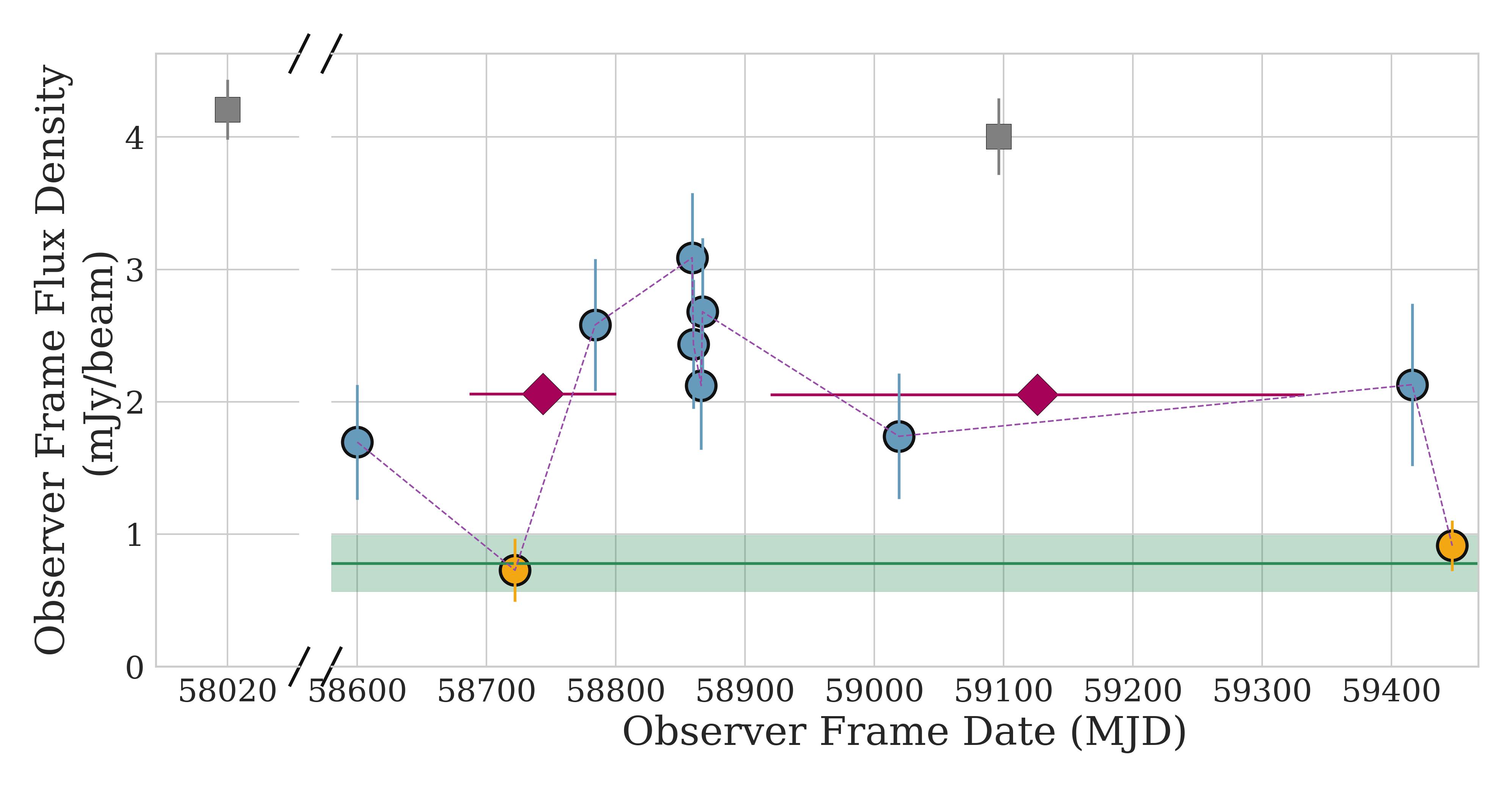}
\end{minipage}\hfil
\begin{minipage}{0.48\textwidth}
\centering
\includegraphics[height=5cm,keepaspectratio]{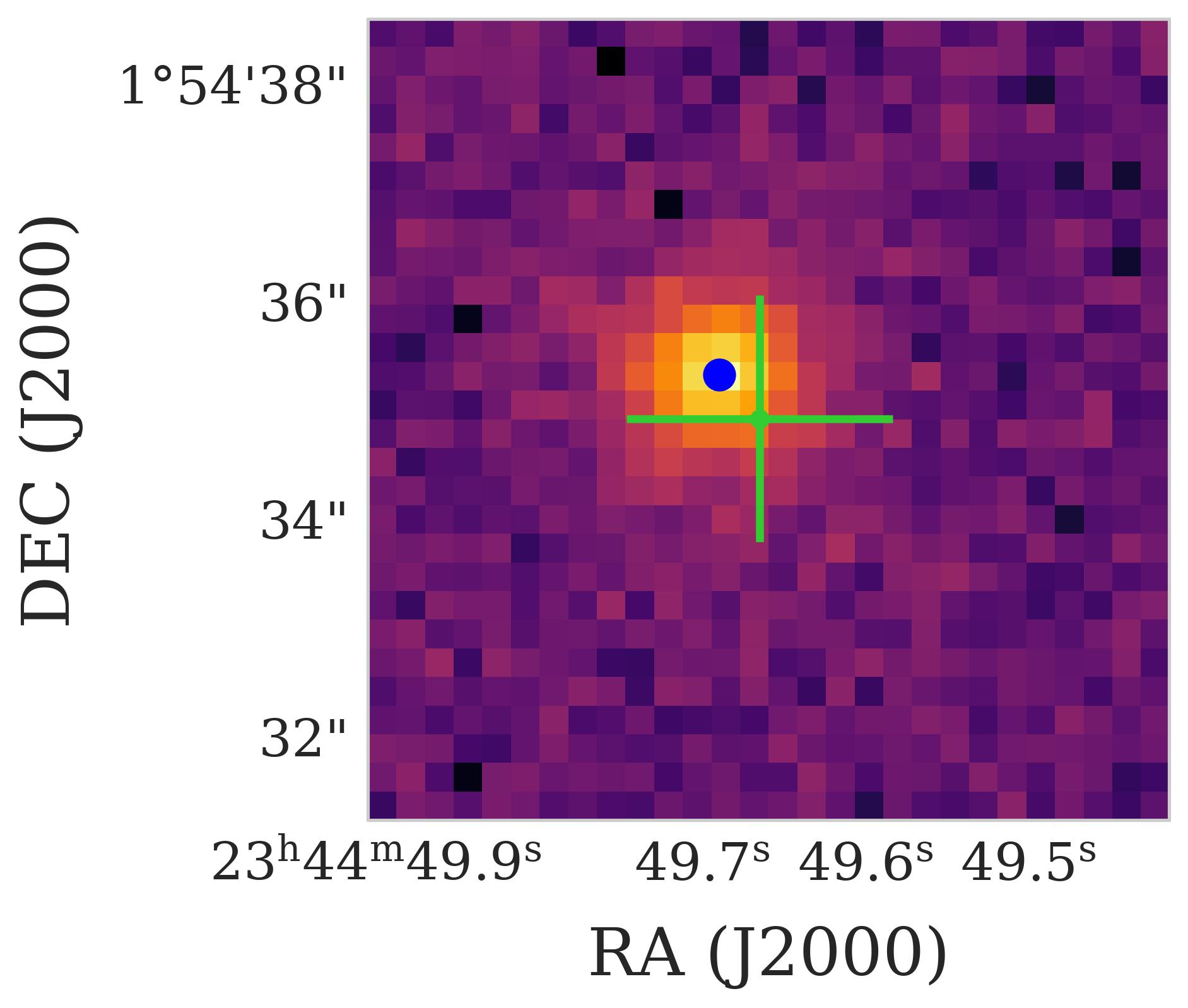}
\end{minipage}\hfil
\medskip
\caption{VAST J234449.6$+$015434}
\label{fig:J234449.6}
\end{figure*}

\clearpage
\newpage

\section{Host Galaxy ZTF Lightcurves}\label{appendix_ztf}
Here we show the difference image subtracted 420–650\,nm ATLAS lightcurves for the host galaxies of all twelve sources in our sample as well as the difference image subtracted g-band ZTF lightcurves for the host galaxies of the nine sources with ZTF detections. The vertical line indicates the time of the peak of the radio flare observed in the VAST Pilot. The horizontal line indicates the coverage in VAST for that particular source, the shaded maroon region indicates the start and end time of the flare; see Section \ref{sec:calculate_param}. For sources with a ZTF detection, an additional right panel shows three injected g-band flares from the ZTF observations of TDEs AT2020yue, AT2021qth, AT2020wey \citep{TDE_demographics}. The fluxes and timescales of the three TDEs are adjusted to the redshifts of the host galaxies.

\begin{figure*}[h] 
\centering
\includegraphics[width=\textwidth,keepaspectratio]{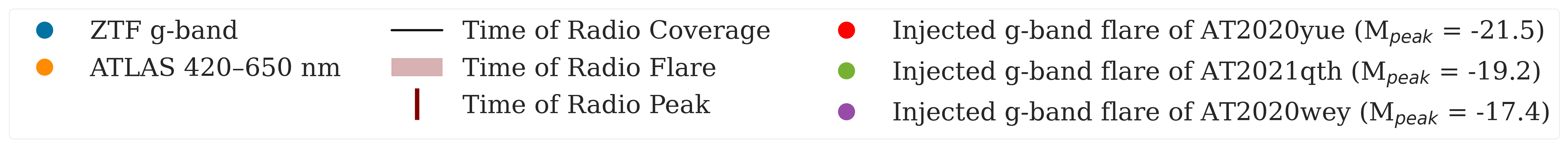}
\label{ztf_legend}
\end{figure*}

\begin{figure*}[h] 
\includegraphics[height=5cm,keepaspectratio]{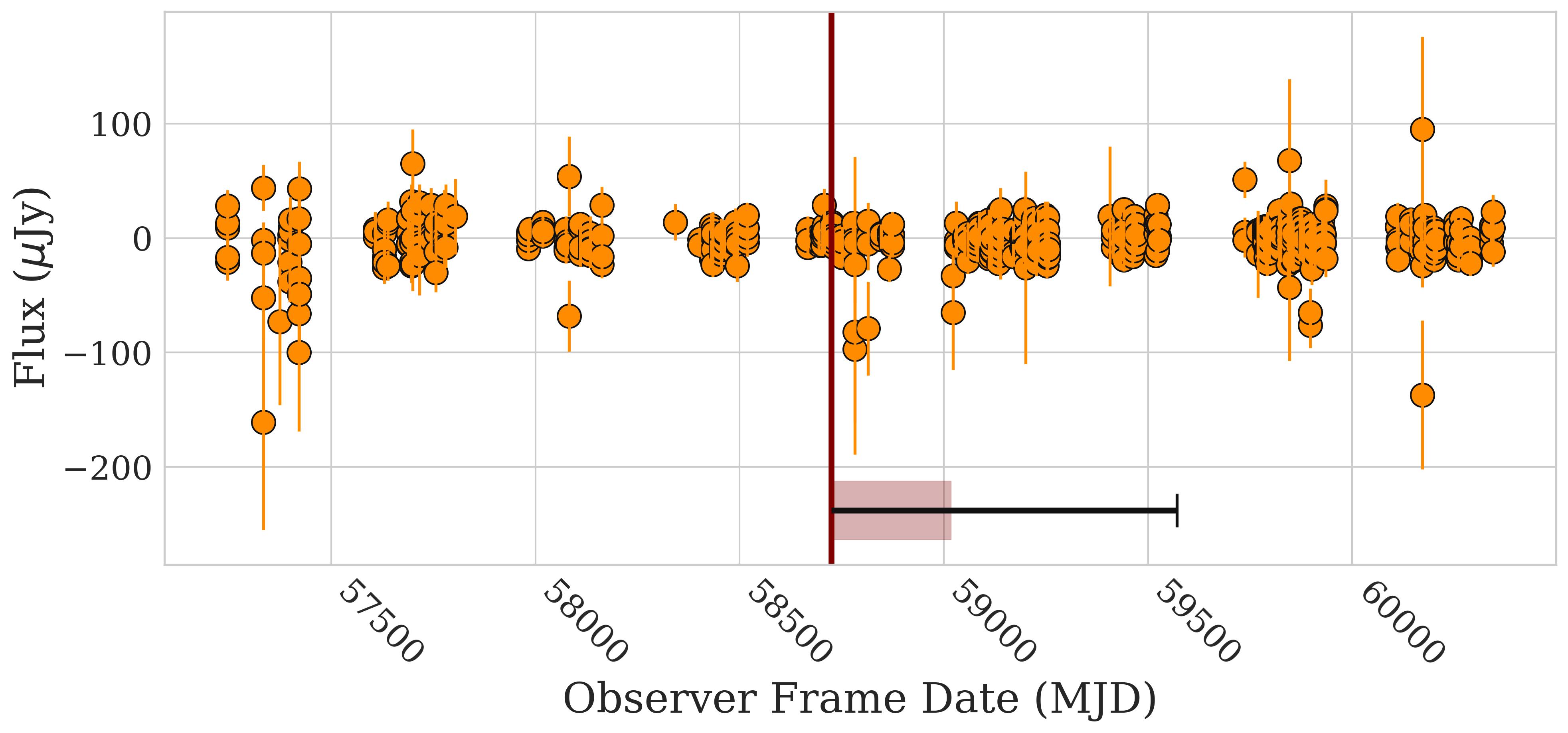}
\medskip
\caption{VAST J011148.1-025539 (photo z = 0.8$\pm$0.4)}
\label{fig:J011148.1-025539}
\end{figure*}

\begin{figure*}[h] 
\centering
\begin{minipage}{0.48\textwidth}
\centering
\includegraphics[height=5cm,keepaspectratio]{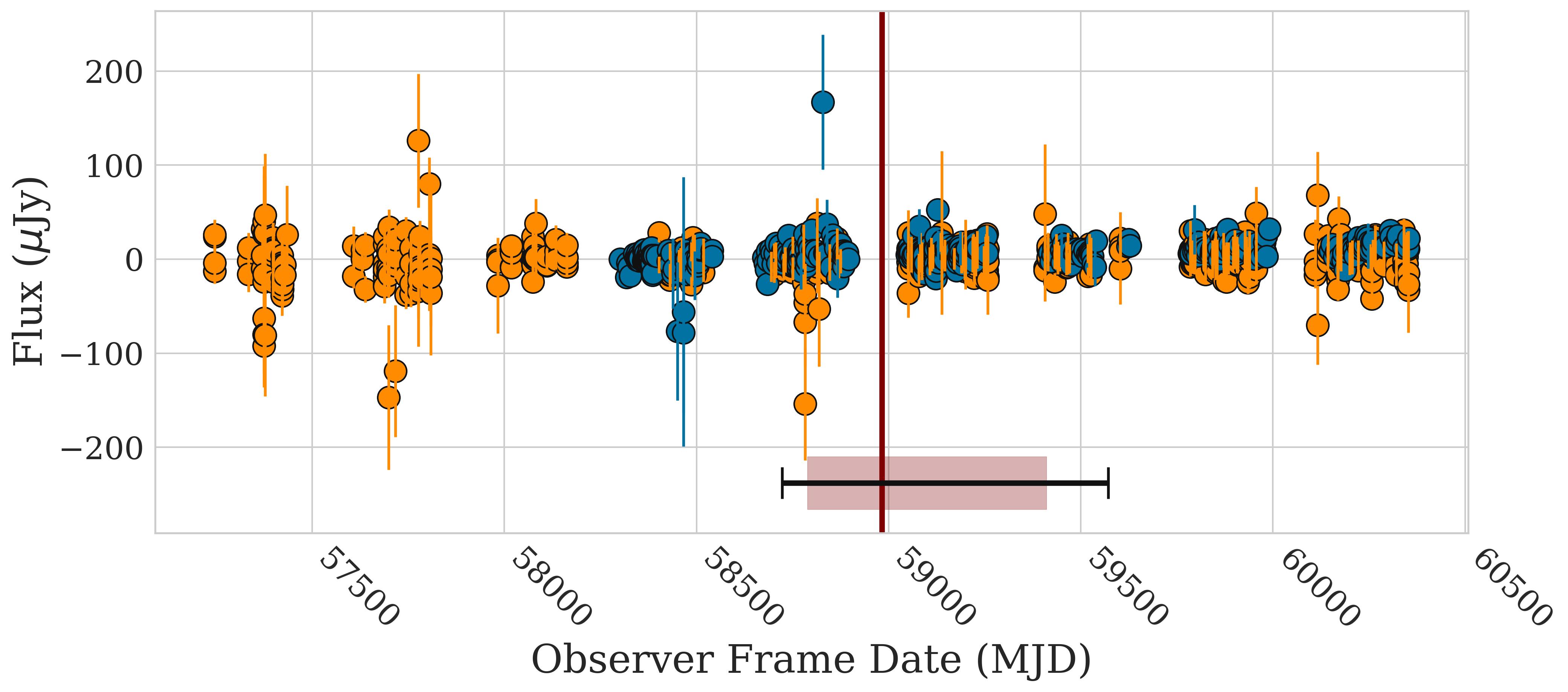}
\end{minipage}\hfil
\begin{minipage}{0.48\textwidth}
\centering
\includegraphics[height=5cm,keepaspectratio]{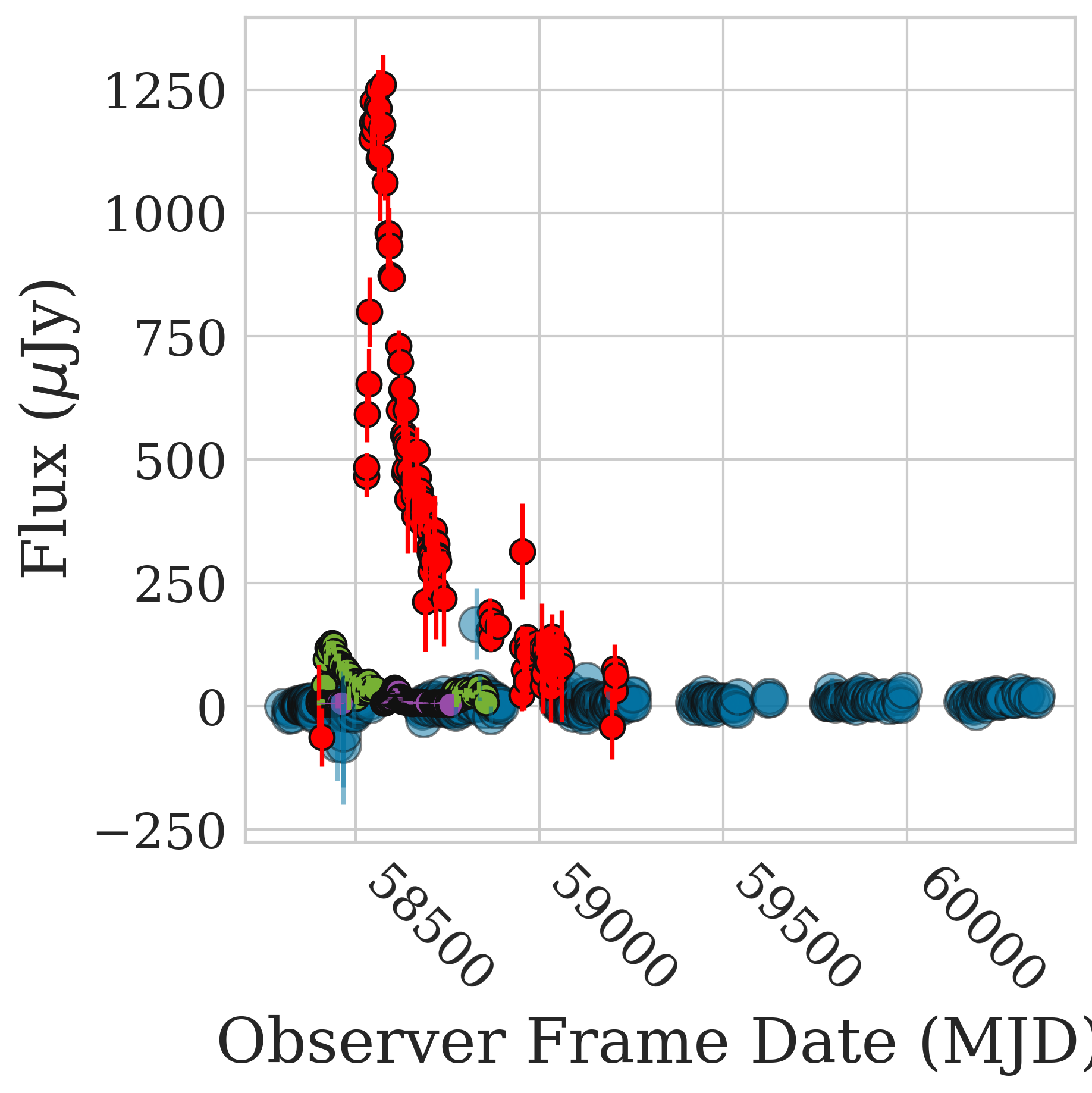}
\end{minipage}\hfil
\medskip
\caption{VAST J015856.8-012404 (photo z = 0.076$\pm$0.009)}
\label{fig:J015856.8-012404}
\end{figure*}

\begin{figure*}[h] 
\centering
\begin{minipage}{0.48\textwidth}
\centering
\includegraphics[height=5cm,keepaspectratio]{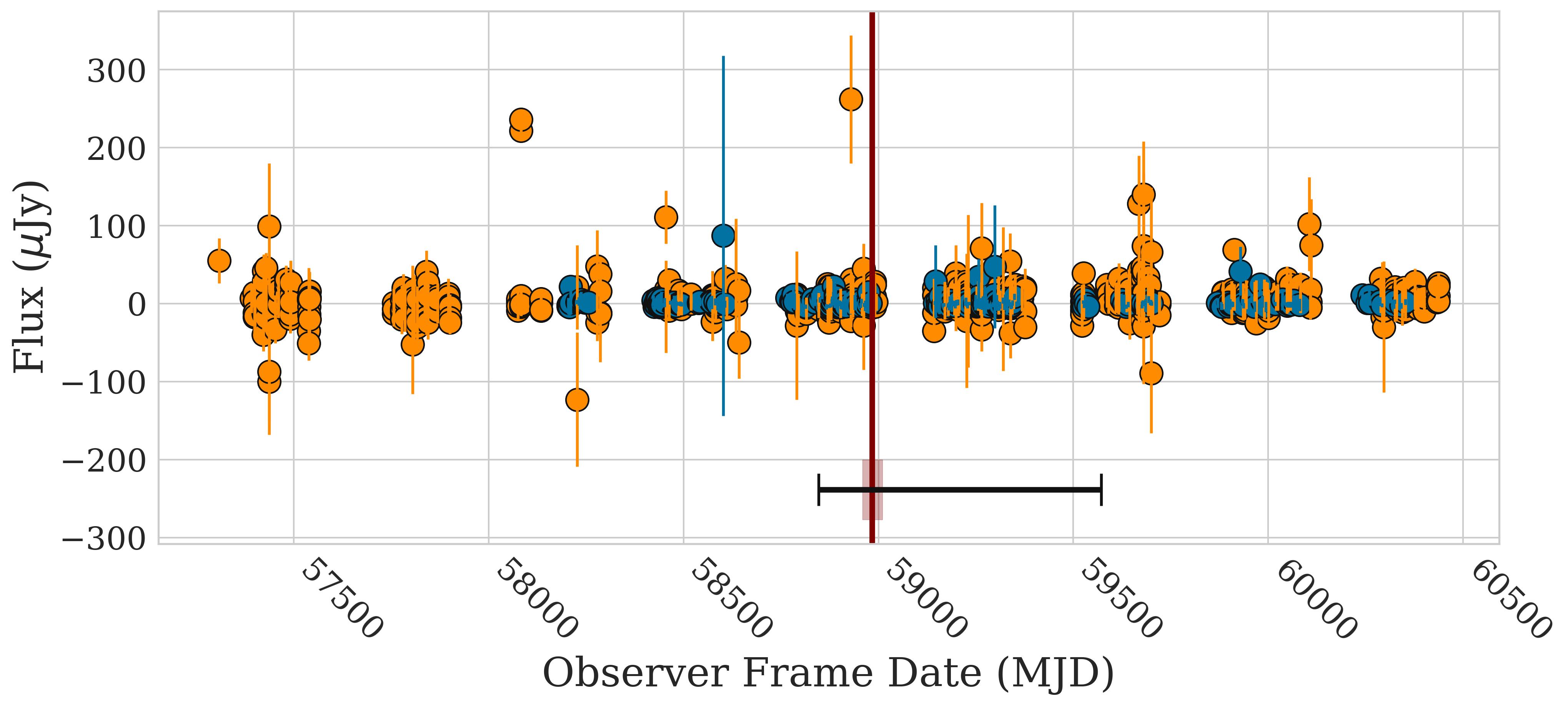}
\end{minipage}\hfil
\begin{minipage}{0.48\textwidth}
\centering
\includegraphics[height=5cm,keepaspectratio]{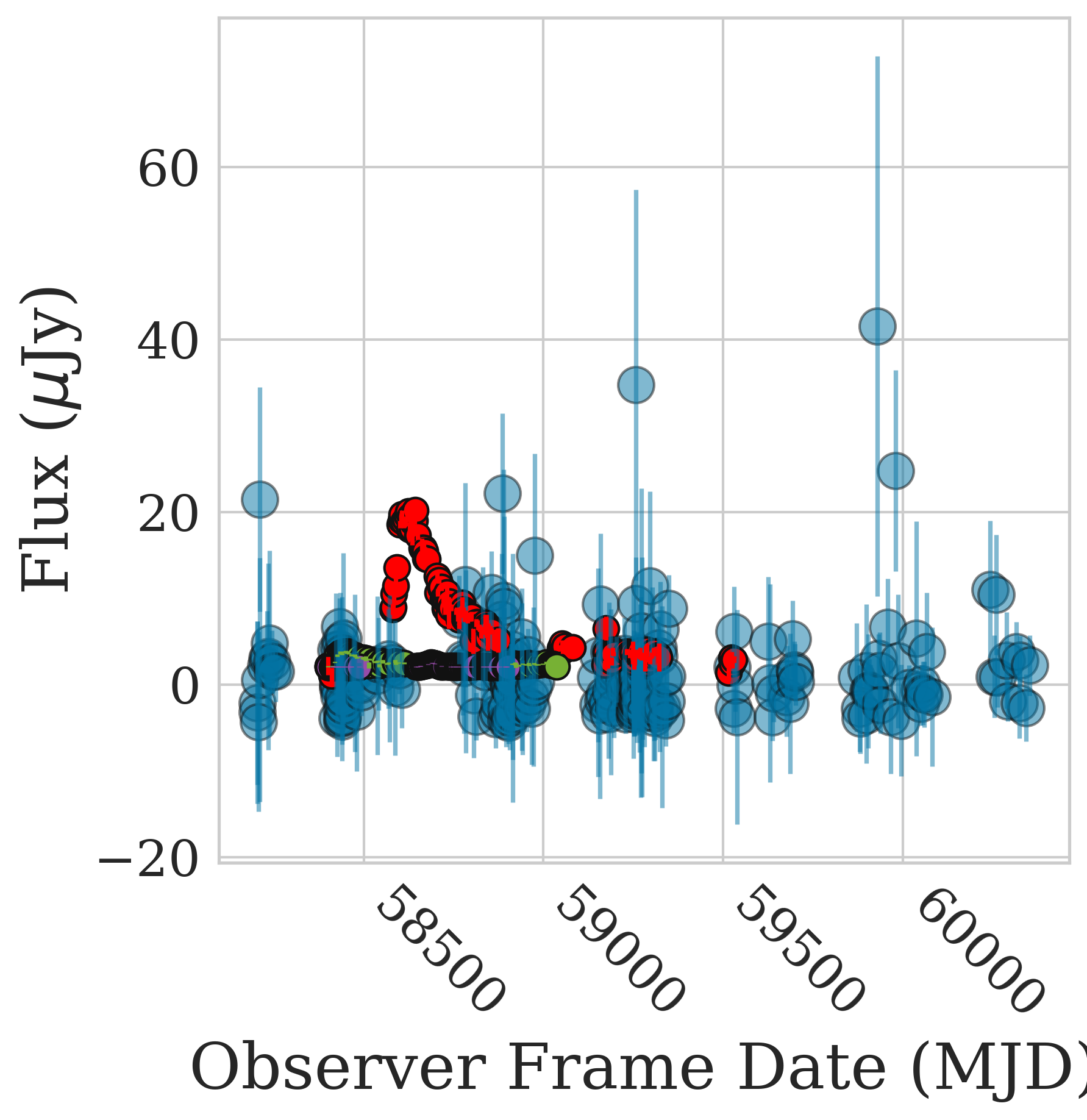}
\end{minipage}\hfil
\medskip
\caption{VAST J093634.7-054755 (photo z = 0.50$\pm$0.05)}
\label{fig:J093634.7-054755}
\end{figure*}

\begin{figure*}[h] 
\centering
\begin{minipage}{0.48\textwidth}
\centering
\includegraphics[height=5cm,keepaspectratio]{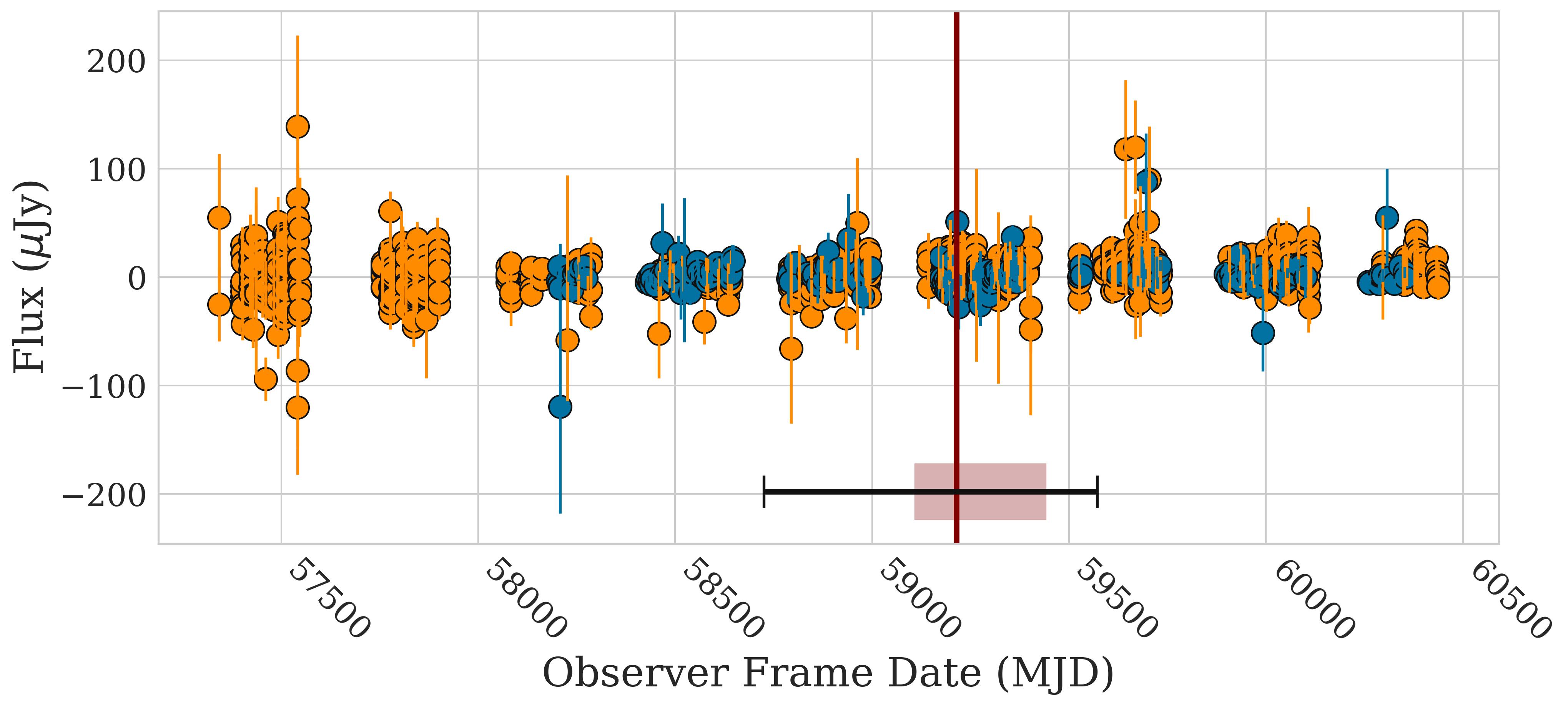}
\end{minipage}\hfil
\begin{minipage}{0.48\textwidth}
\centering
\includegraphics[height=5cm,keepaspectratio]{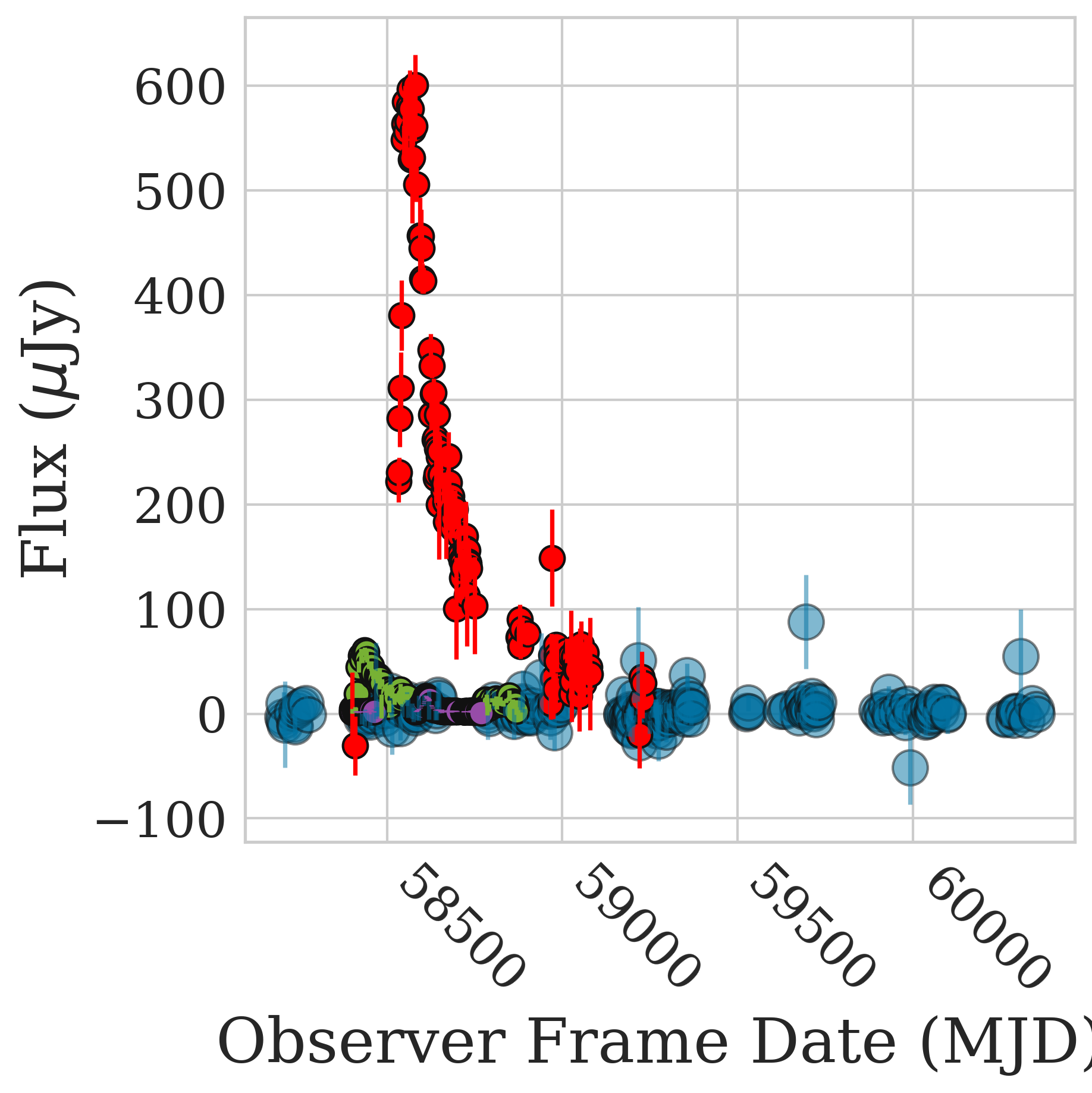}
\end{minipage}\hfil
\medskip
\caption{VAST J104315.9+005059 (photo z = 0.107$\pm$0.005)}
\label{fig:J104315.9+005059}
\end{figure*}

\begin{figure*}[h] 
\centering
\begin{minipage}{0.48\textwidth}
\centering
\includegraphics[height=5cm,keepaspectratio]{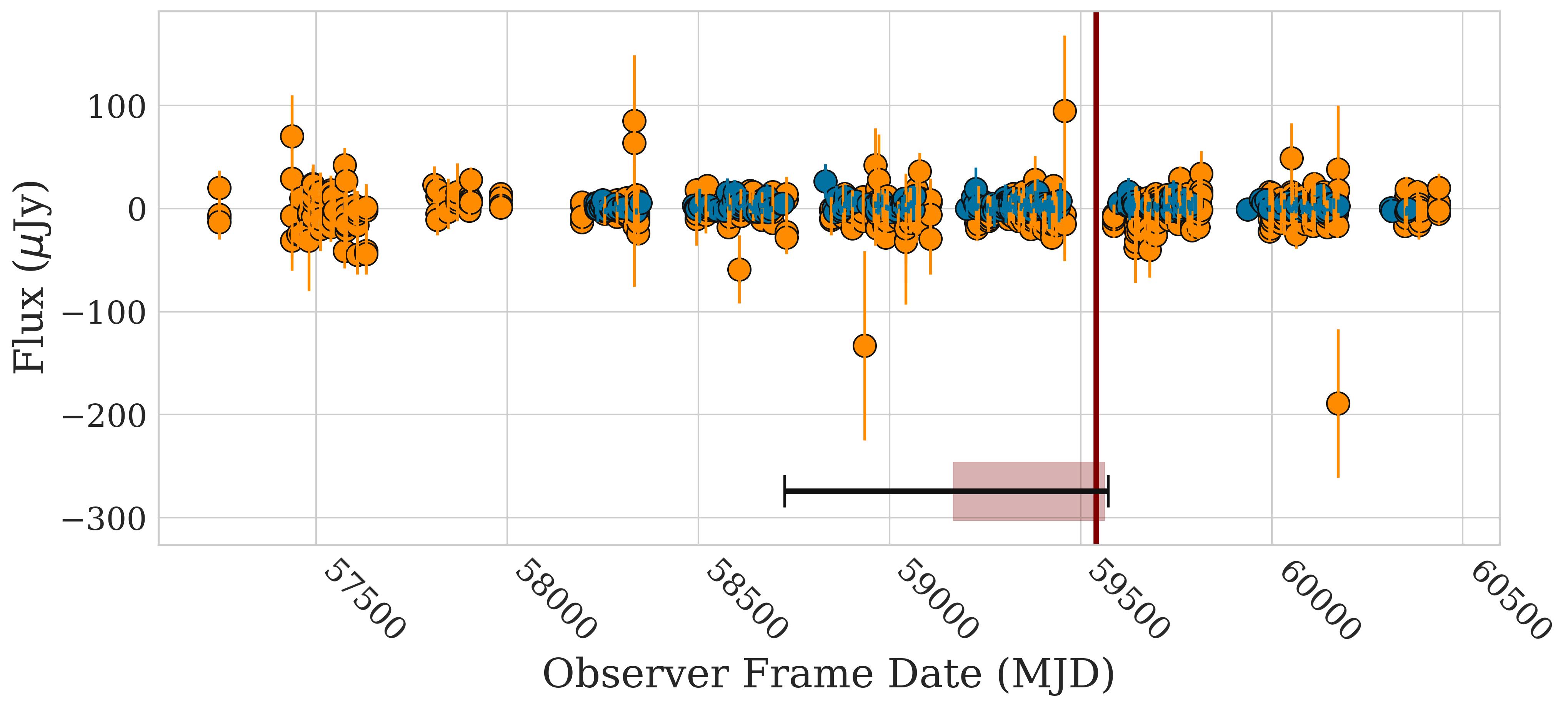}
\end{minipage}\hfil
\begin{minipage}{0.48\textwidth}
\centering
\includegraphics[height=5cm,keepaspectratio]{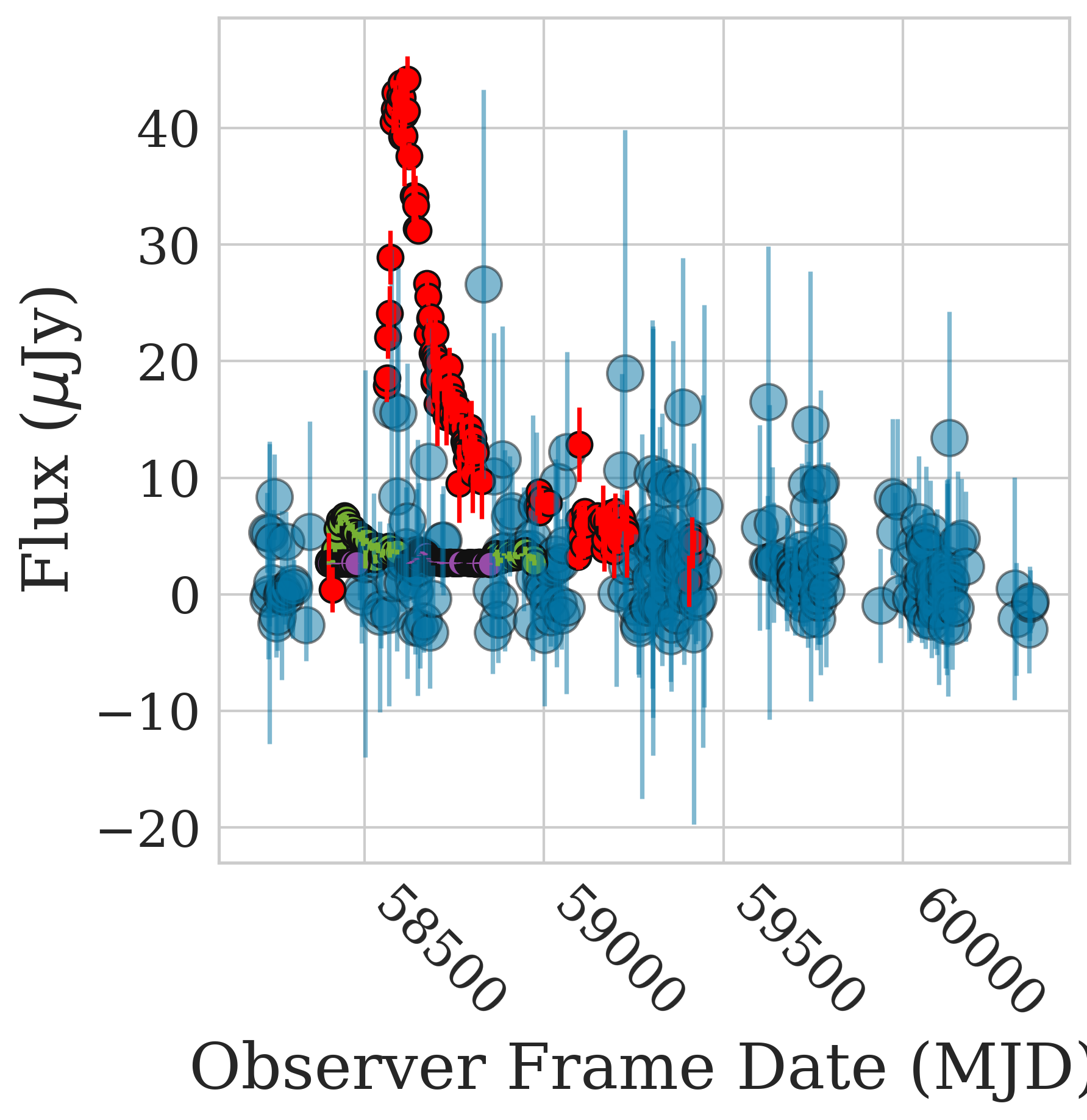}
\end{minipage}\hfil
\medskip
\caption{VAST J144848.2+030235 (photo z = 0.35$\pm$0.02)}
\label{fig:J144848.2+030235}
\end{figure*}

\begin{figure*}[h] 
\includegraphics[height=5cm,keepaspectratio]{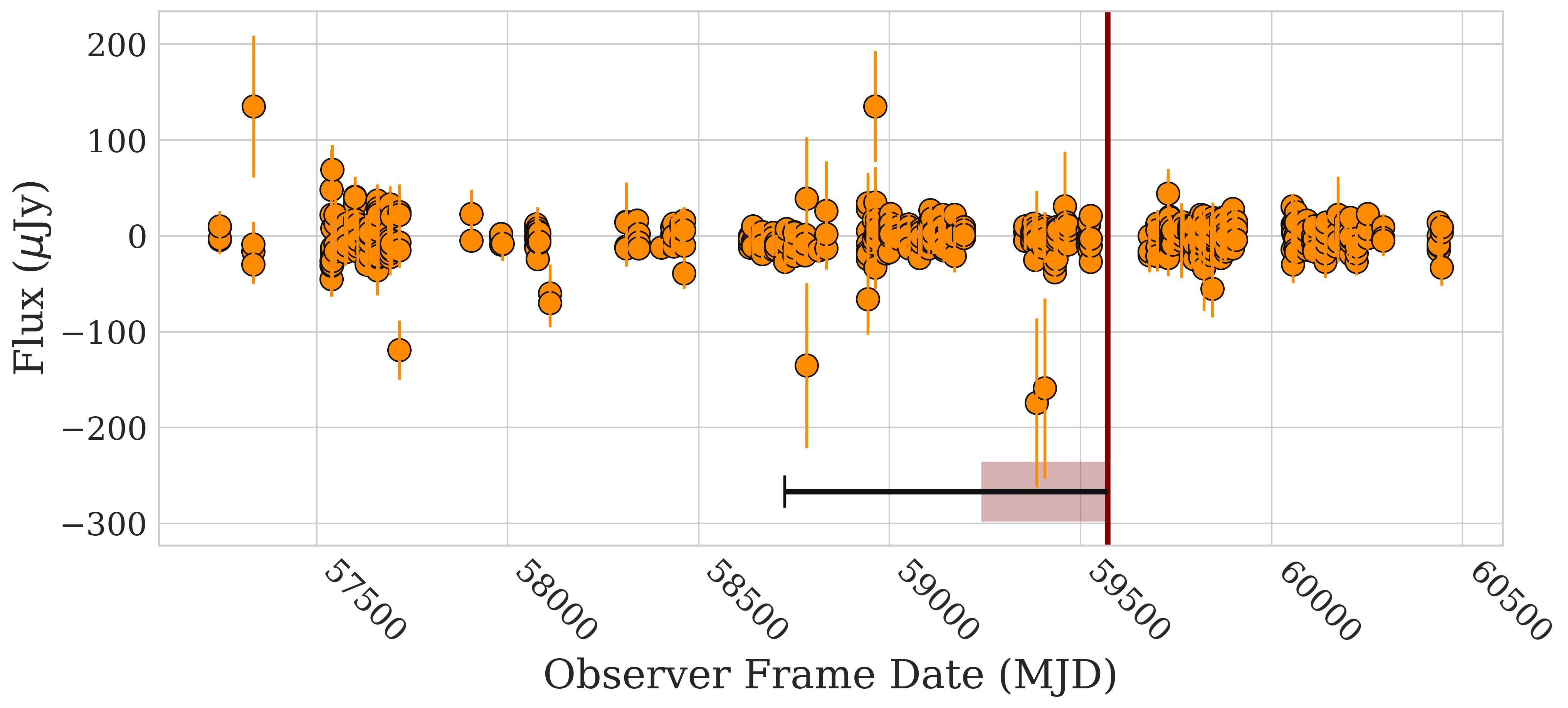}
\medskip
\caption{VAST J210626.2-020055 (photo z = 0.2$\pm$0.2)}
\label{fig:J210626.2-020055}
\end{figure*}

\begin{figure*}[h] 
\centering
\begin{minipage}{0.48\textwidth}
\centering
\includegraphics[height=5cm,keepaspectratio]{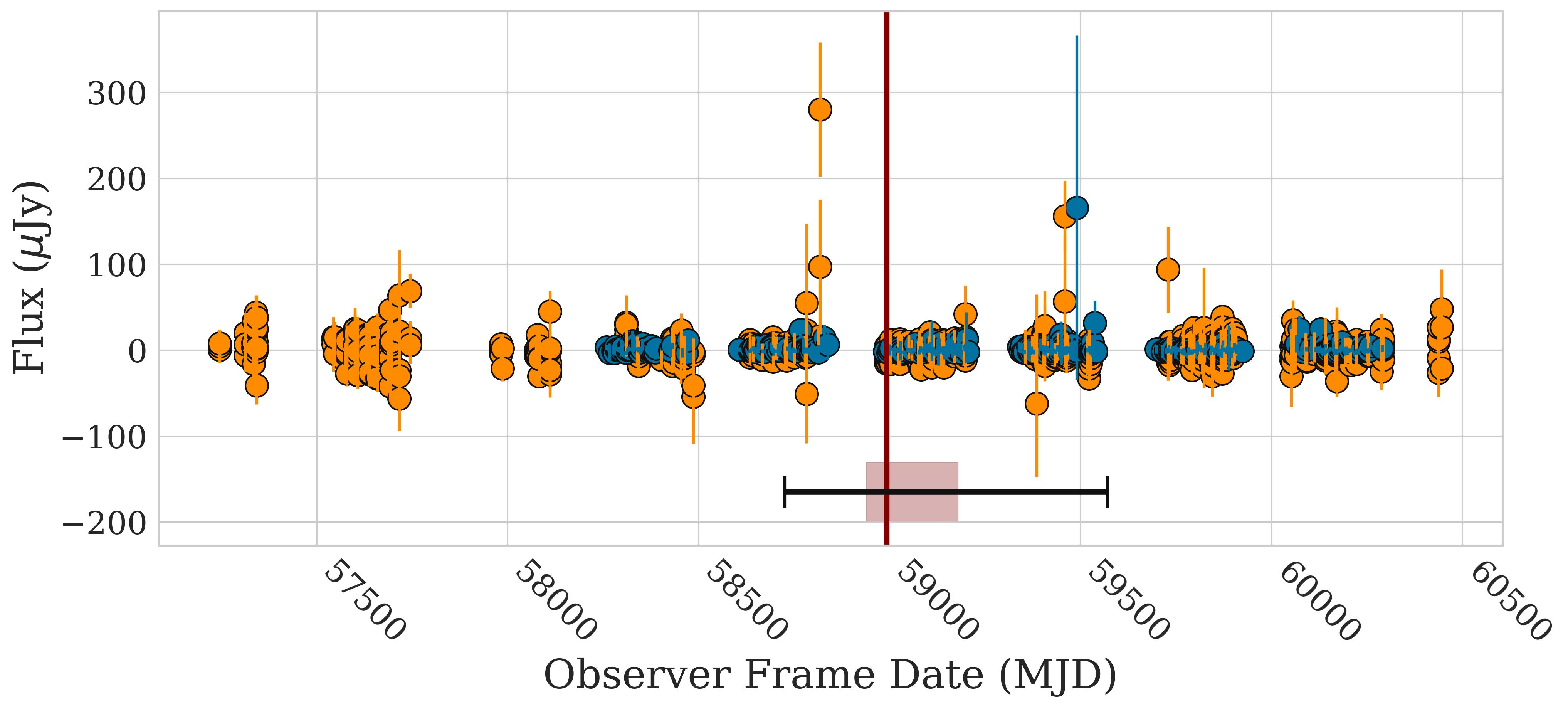}
\end{minipage}\hfil
\begin{minipage}{0.48\textwidth}
\centering
\includegraphics[height=5cm,keepaspectratio]{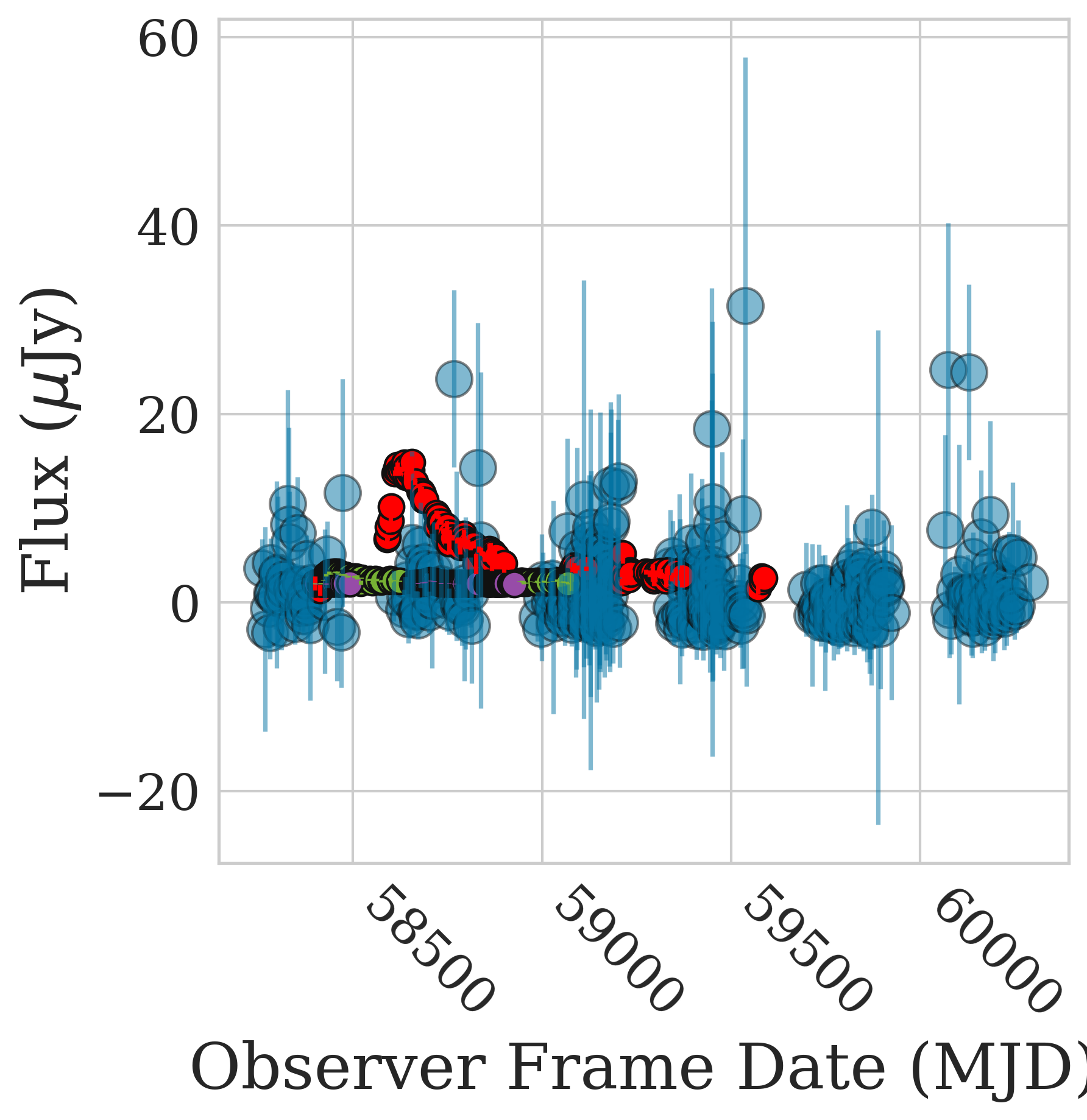}
\end{minipage}\hfil
\medskip
\caption{VAST J212618.5+022400 (photo z = 0.58$\pm$0.04)}
\label{fig:J212618.5+022400}
\end{figure*}

\begin{figure*}[h] 
\includegraphics[height=5cm,keepaspectratio]{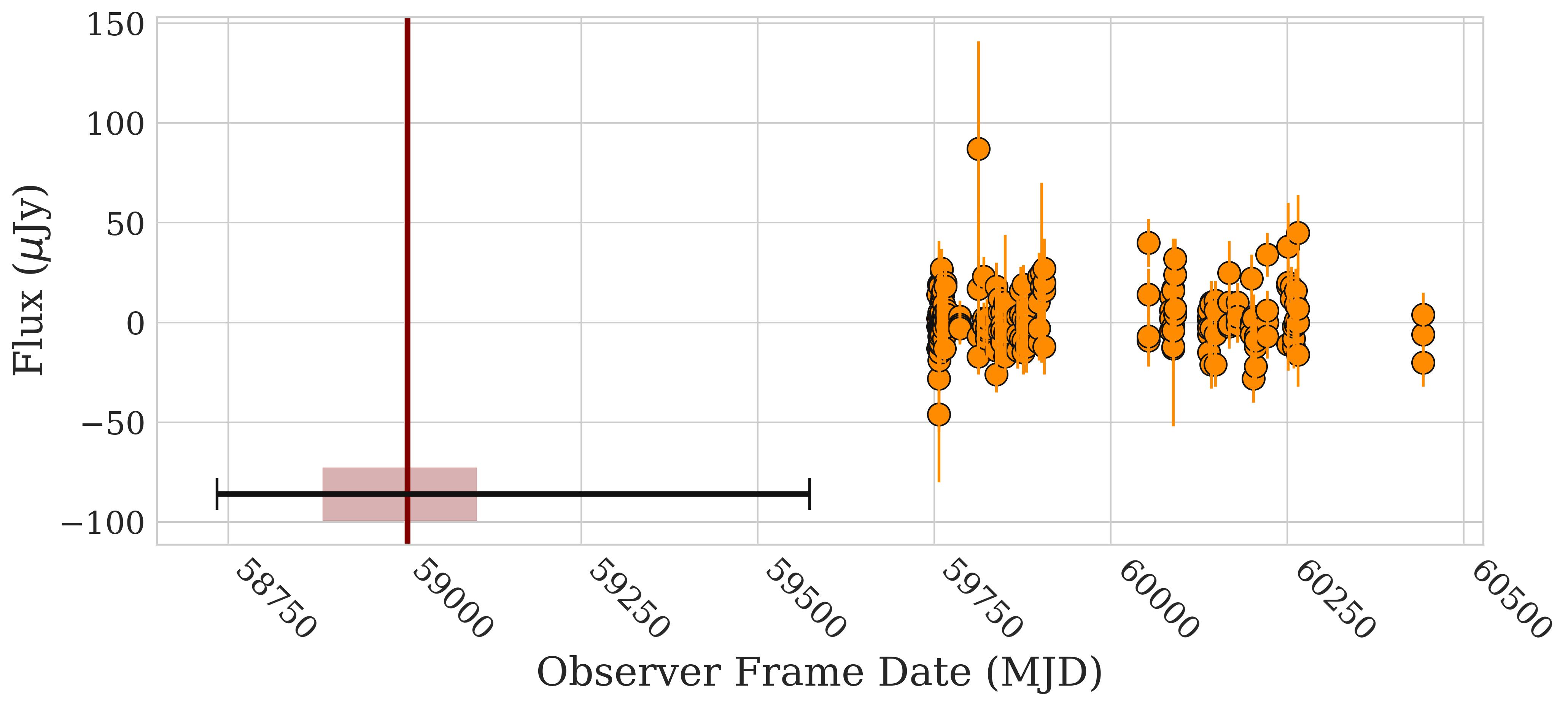}
\medskip
\caption{VAST J213437.8-620433 (photo z = 0.059$\pm$0.005)}
\label{fig:J213437.8-620433}
\end{figure*}

\begin{figure*}[h] 
\centering
\begin{minipage}{0.48\textwidth}
\centering
\includegraphics[height=5cm,keepaspectratio]{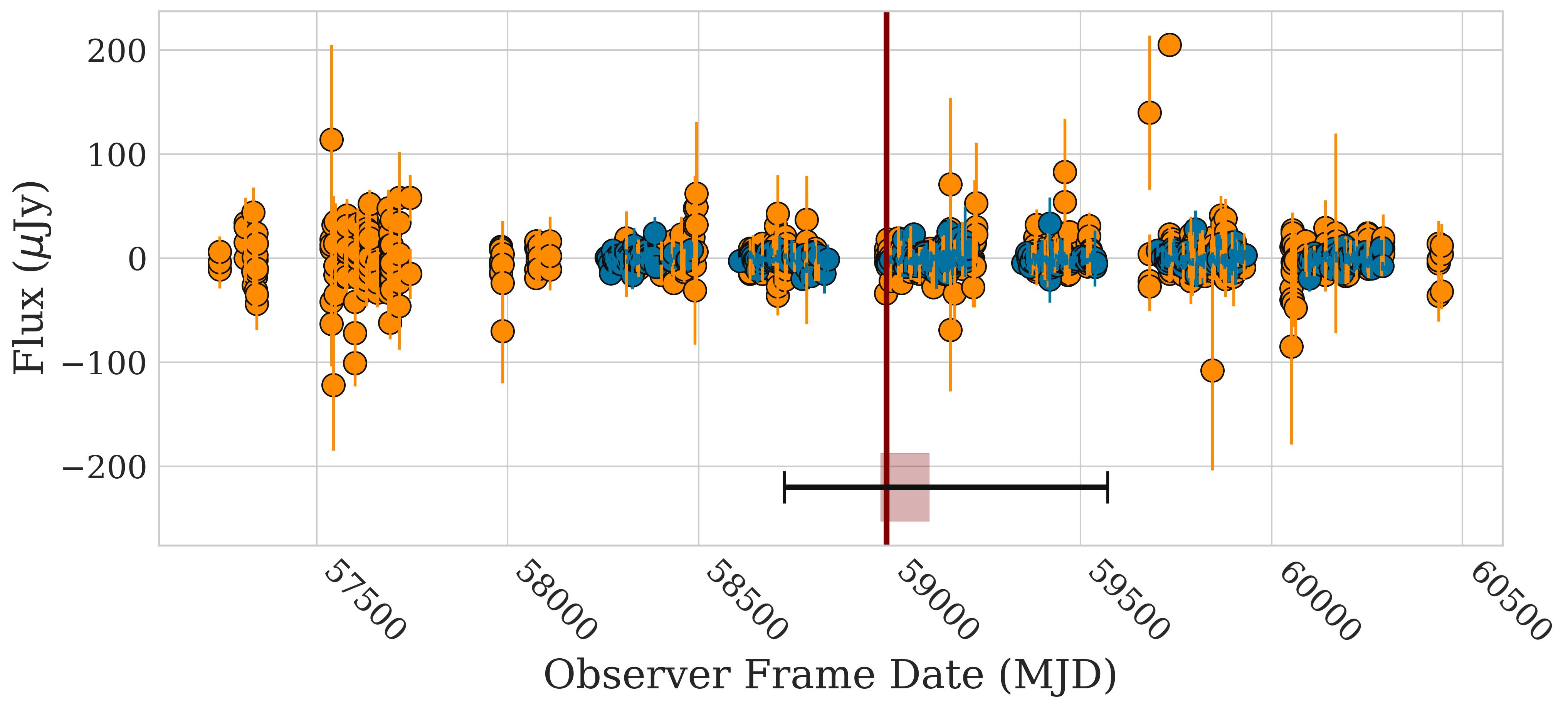}
\end{minipage}\hfil
\begin{minipage}{0.48\textwidth}
\centering
\includegraphics[height=5cm,keepaspectratio]{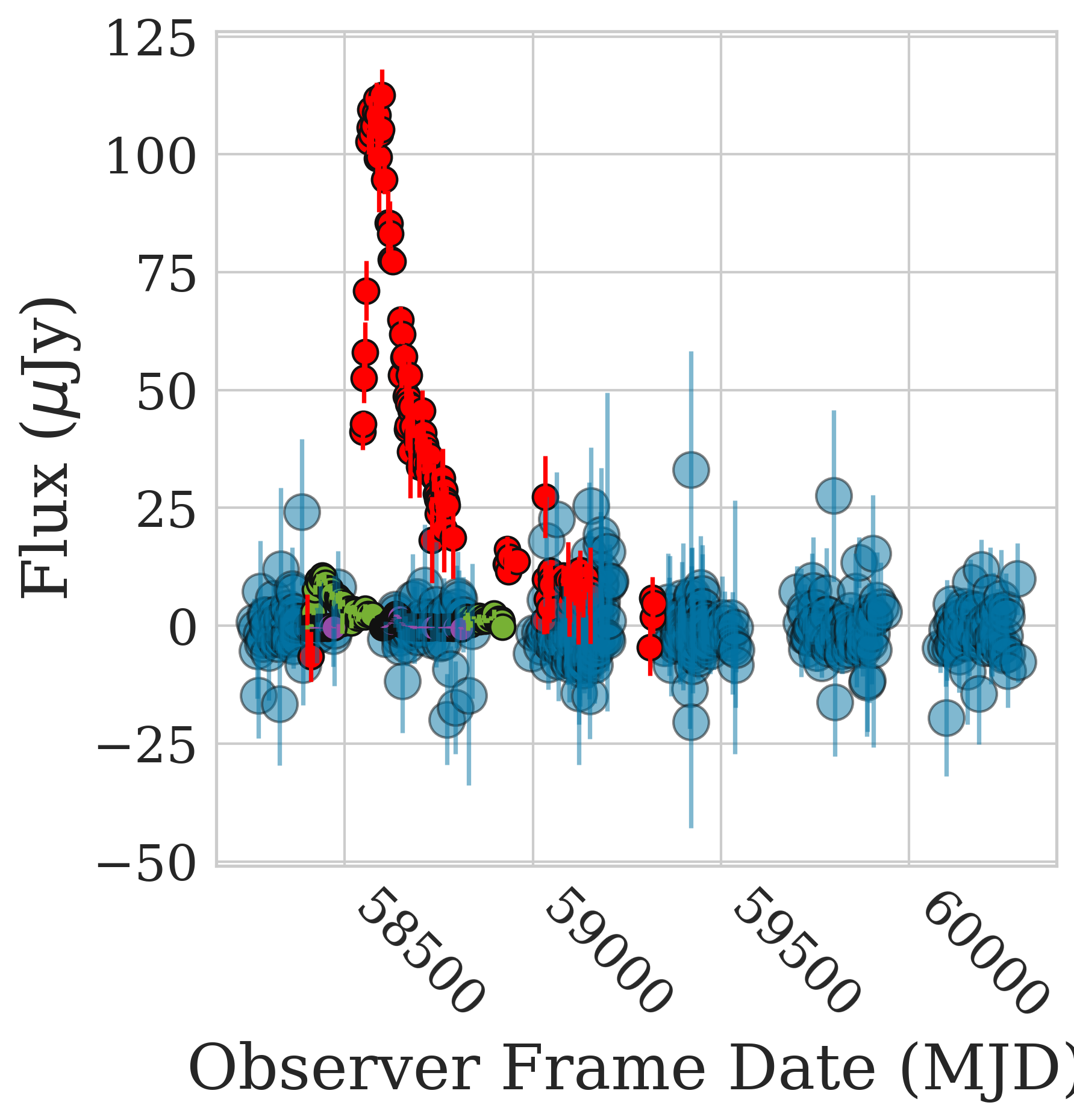}
\end{minipage}\hfil
\medskip
\caption{VAST J215418.2+002442 (photo z = 0.23$\pm$0.04)}
\label{fig:J215418.2+002442}
\end{figure*}

\begin{figure*}[h] 
\centering
\begin{minipage}{0.48\textwidth}
\centering
\includegraphics[height=5cm,keepaspectratio]{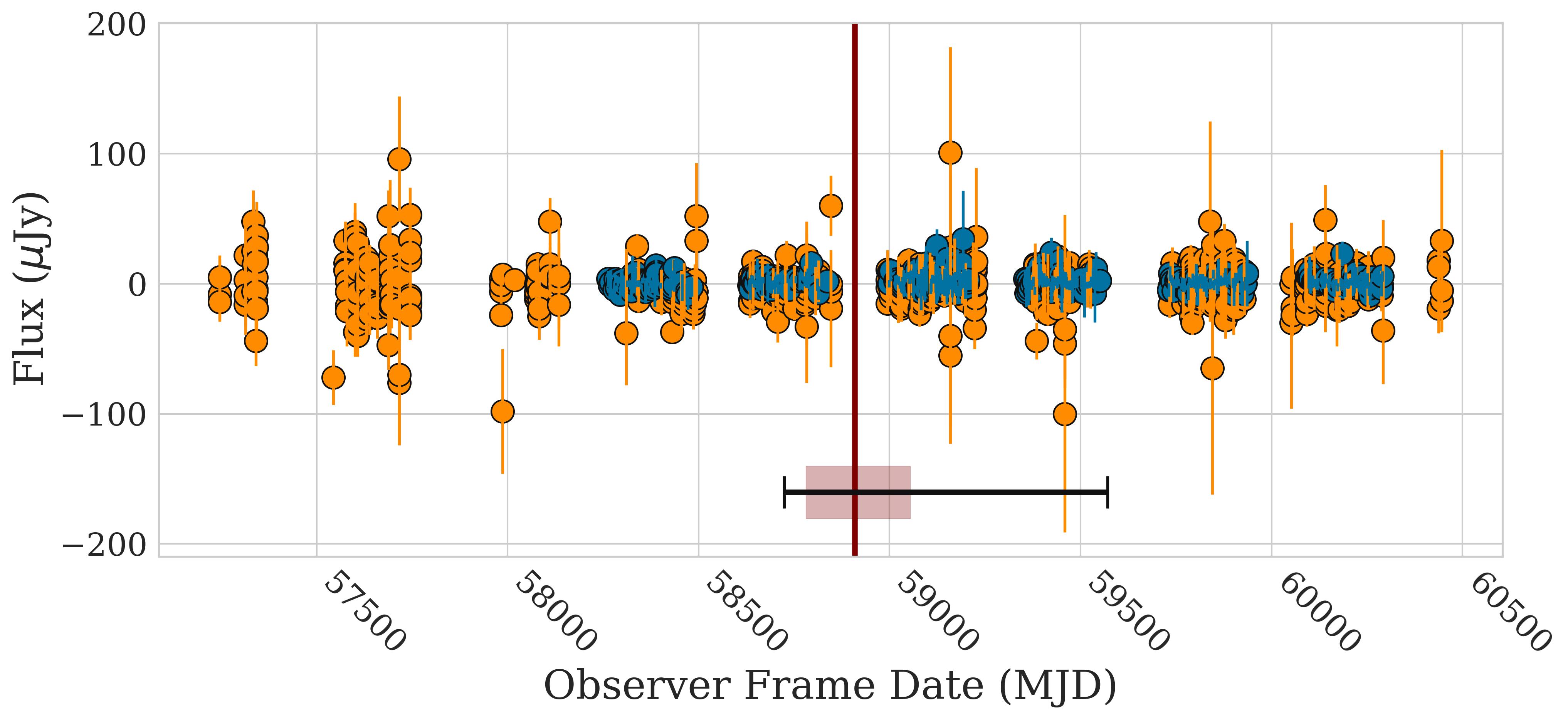}
\end{minipage}\hfil
\begin{minipage}{0.48\textwidth}
\centering
\includegraphics[height=5cm,keepaspectratio]{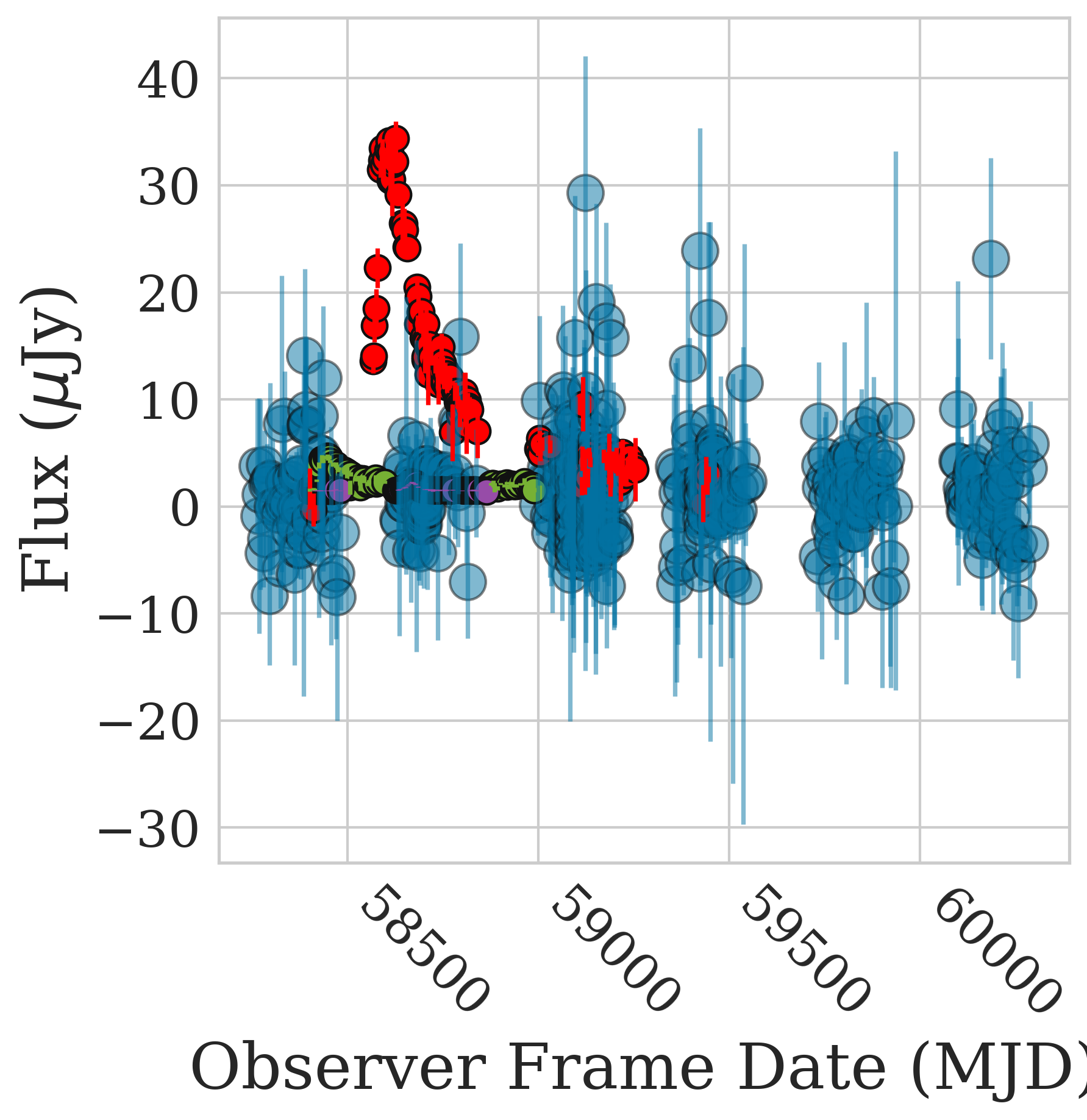}
\end{minipage}\hfil
\medskip
\caption{VAST J221936.0+004724 (photo z = 0.39$\pm$0.04)}
\label{fig:J221936.0+004724}
\end{figure*}

\begin{figure*}[h] 
\centering
\begin{minipage}{0.48\textwidth}
\centering
\includegraphics[height=5cm,keepaspectratio]{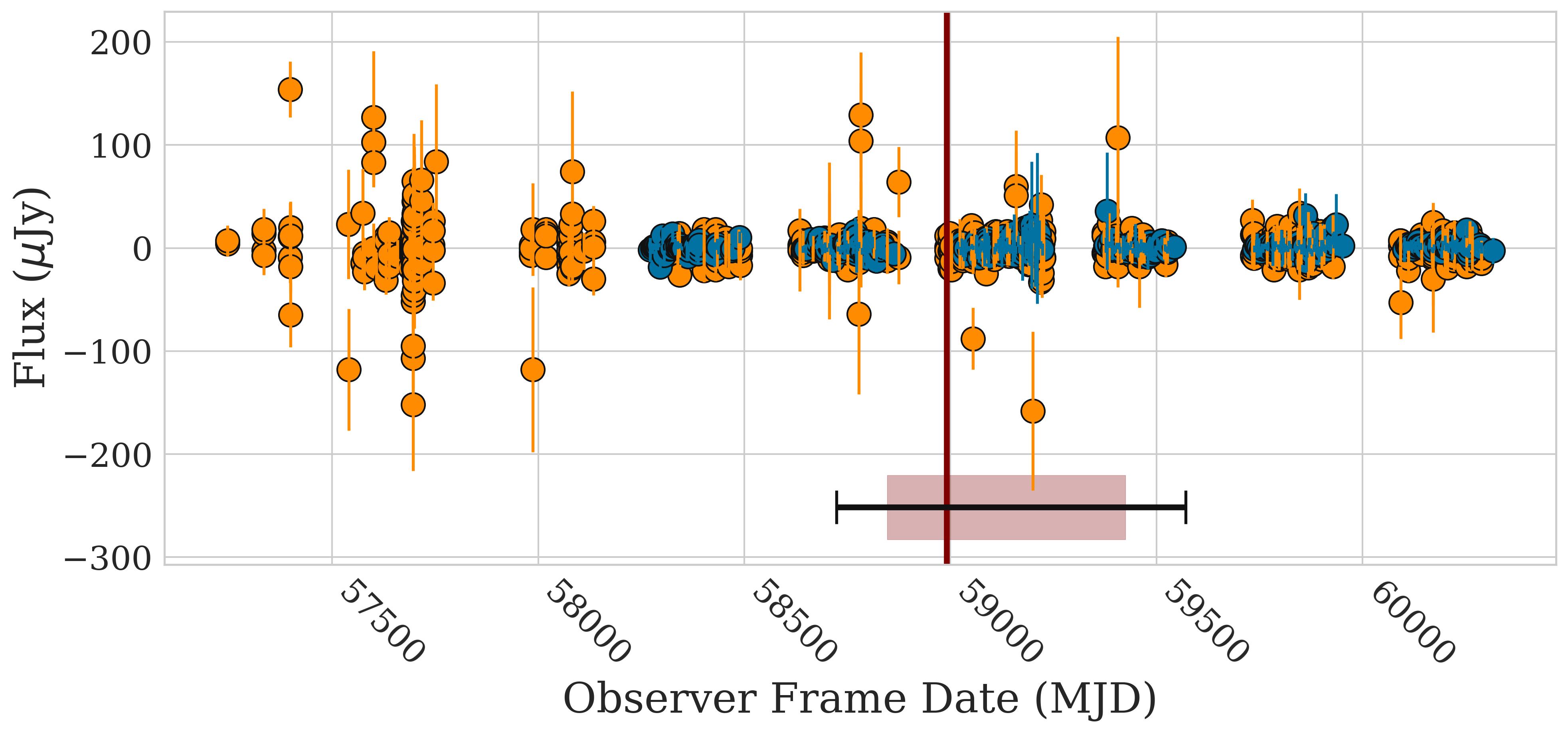}
\end{minipage}\hfil
\begin{minipage}{0.48\textwidth}
\centering
\includegraphics[height=5cm,keepaspectratio]{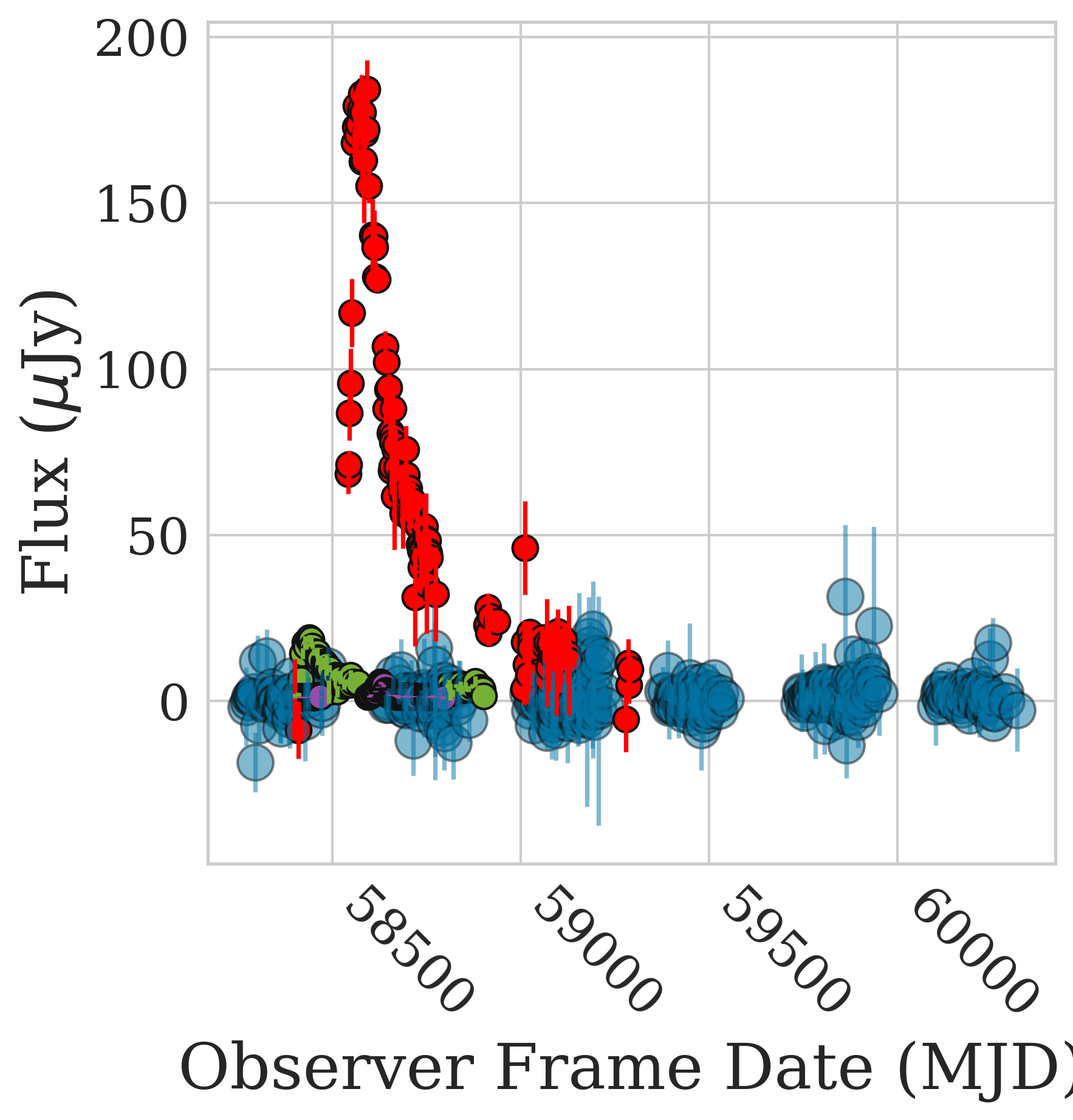}
\end{minipage}\hfil
\medskip
\caption{VAST J230053.0-020732 (photo z = 0.18$\pm$0.03)}
\label{fig:J230053.0-020732}
\end{figure*}

\begin{figure*}[h] 
\centering
\begin{minipage}{0.48\textwidth}
\centering
\includegraphics[height=5cm,keepaspectratio]{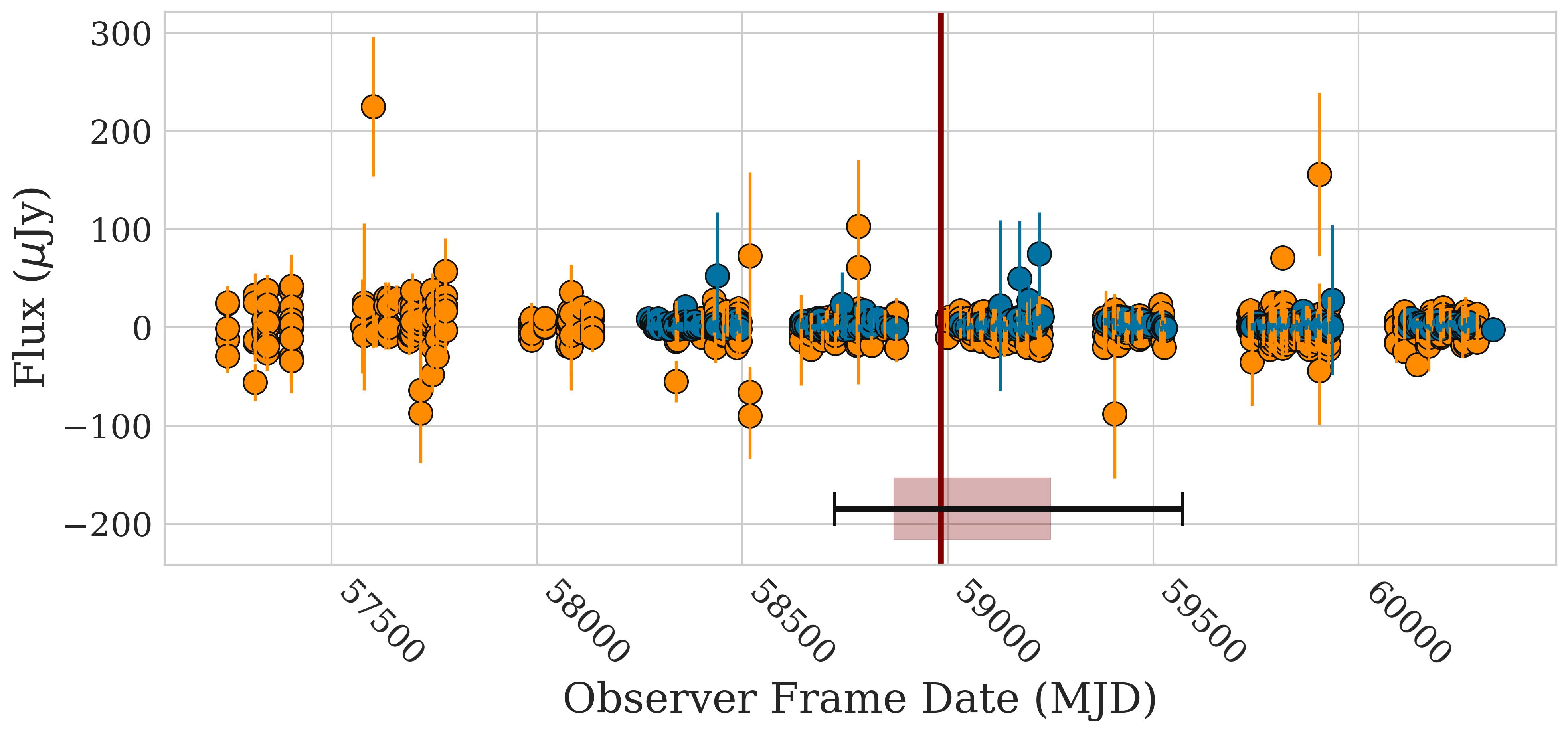}
\end{minipage}\hfil
\begin{minipage}{0.48\textwidth}
\centering
\includegraphics[height=5cm,keepaspectratio]{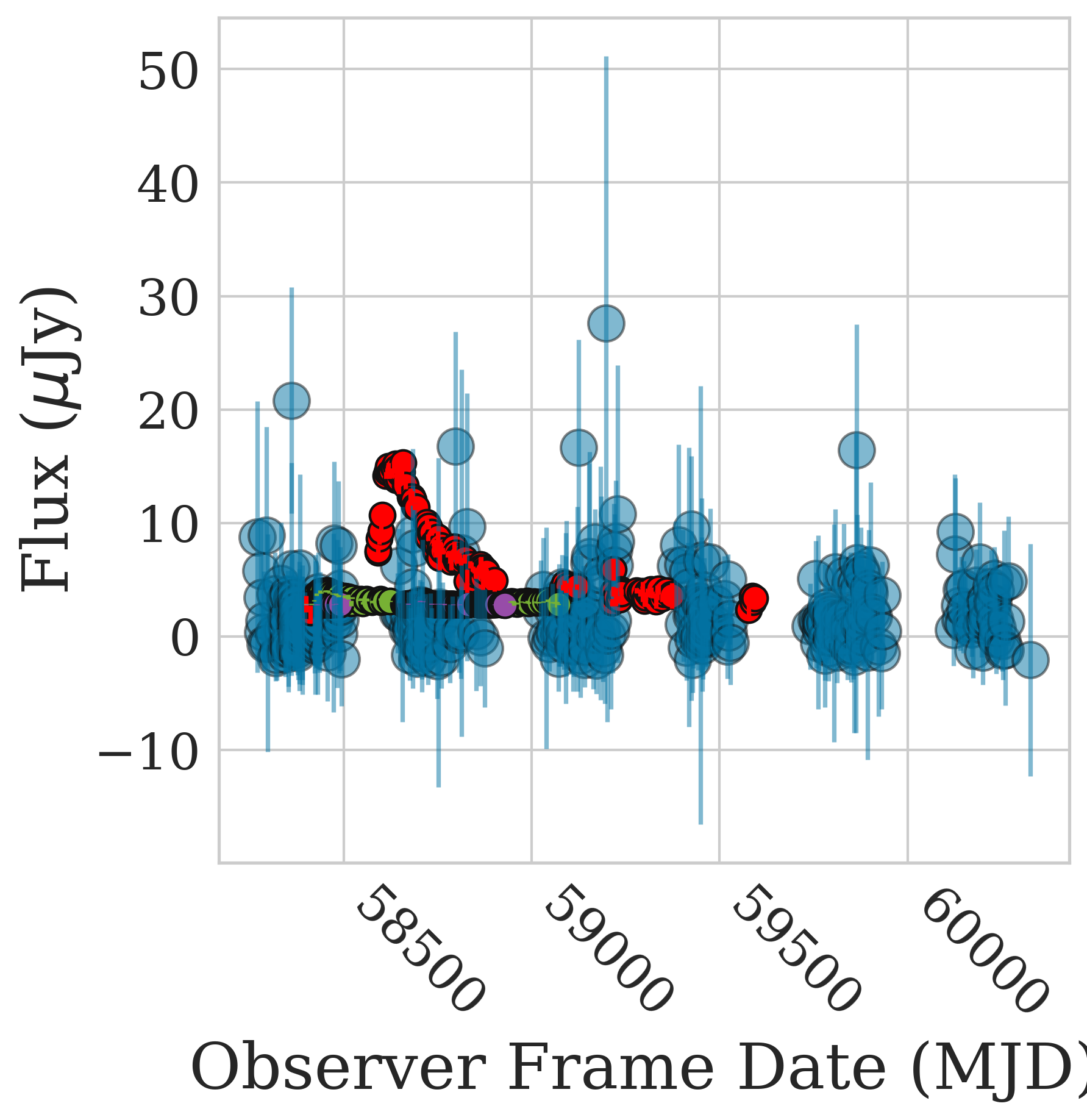}
\end{minipage}\hfil
\medskip
\caption{VAST J234449.6+015434 (photo z = 0.59$\pm$0.09)}
\label{fig:J234449.6+015434}
\end{figure*}

\clearpage

\end{document}